\documentclass[11pt,a4paper]{article}
\pdfoutput=1
\usepackage{jcappub}

\usepackage{bm}
\usepackage{epsfig}
\usepackage{bbold}
\usepackage{graphicx}
\usepackage{amsmath}
\usepackage{amssymb}
\usepackage{amsbsy}
\usepackage{color}
\usepackage{subfigure}
\usepackage{slashed}
\usepackage{afterpage}
\usepackage{psfrag}
\usepackage{hyperref}
\usepackage{dsfont}
\newcommand{\appropto}{\mathrel{\vcenter{
  \offinterlineskip\halign{\hfil$##$\cr
    \propto\cr\noalign{\kern2pt}\sim\cr\noalign{\kern-2pt}}}}}

\newcommand{\half}{\textstyle{\frac{1}{2}}}
\renewcommand\vec[1]{{\bf #1}}

\newcommand{\GG}{{\sf G}}
\newcommand{\PP}{{\sf P}}
\newcommand{\TT}{{\sf T}}
\newcommand{\MM}{{\sf M}}
\newcommand{\Ss}{{\sf S}}

\newcommand{\ma}[4]{ \left( \begin{array}{cc}   #1 & #2 \\ #3 & #4\end{array} \right)}

\newcommand{\be}{\begin{equation}}
\newcommand{\ee}{\end{equation}}
\newcommand{\bea}{\begin{eqnarray}}
\newcommand{\eea}{\end{eqnarray}}

\newcommand{\vv}[2]{ \left( \begin{array}{cc}   #1 \\ #2 \end{array} \right)}

\renewcommand\({\left(}
\renewcommand\){\right)}
\renewcommand\[{\left[}
\renewcommand\]{\right]}

\newcommand{\B}{\mathcal{B}}

\newcommand{\exclude}[1]{}

\begin{document}
\subheader{\hfill MPP-2016-329}

\title{Dielectric Haloscopes to Search for Axion Dark Matter:\\ Theoretical Foundations}
\author[a]{Alexander~J.~Millar,}
\author[a]{Georg~G.~Raffelt,}
\author[a,b]{Javier~Redondo,}
\author[a]{Frank~D.~Steffen}

\affiliation[a]{Max-Planck-Institut f\"ur Physik (Werner-Heisenberg-Institut),
F\"ohringer Ring 6,\\ 80805 M\"unchen, Germany}

\affiliation[b]{University of Zaragoza, P.\ Cerbuna 12, 50009 Zaragoza, Spain}

\emailAdd{millar@mpp.mpg.de}
\emailAdd{raffelt@mpp.mpg.de}
\emailAdd{jredondo@unizar.es}
\emailAdd{steffen@mpp.mpg.de}

\abstract{We study the underlying theory of dielectric haloscopes, a new way
to detect dark matter axions. When an interface between different dielectric
media is inside a magnetic field, the oscillating axion field acts as a
source of electromagnetic waves, which emerge in both directions
perpendicular to the surface. The emission rate can be boosted by multiple
layers judiciously placed to achieve constructive interference and by a large transverse area. Starting from
the axion-modified Maxwell equations, we calculate the efficiency of this new
dielectric haloscope approach. This technique could potentially search the
unexplored  high-frequency range of 10--100~GHz (axion mass 40--400~$\mu$eV),
where traditional cavity resonators have difficulties reaching the required
volume.}

\maketitle

\section{ Introduction}

The physical nature of dark matter remains one of the most intriguing
unsolved mysteries of the universe. A vast range of experimental efforts
worldwide aims to detect different kinds of postulated feebly-interacting
particles that could make up some or all of the galactic dark matter. One
long-standing candidate is the axion, a very low-mass pseudoscalar boson
required by the Peccei--Quinn mechanism. As the Peccei--Quinn mechanism is
the favoured explanation for CP conservation in quantum chromodynamics (QCD)~\cite{Peccei:2006as,
Kim:2008hd, Agashe:2014kda}, axions are very well motivated. The expected
mass of dark-matter axions depends on many details of the relaxation process
in the early universe after the axion field has been excited during the QCD
epoch \cite{Sikivie:2009fv, Kawasaki:2013ae}. We are particularly interested
in scenarios in which the Peccei--Quinn symmetry is broken after inflation
and in which both the vacuum realignment mechanism and decays of topological
defects contribute to the dark-matter density. Here the main production
processes can be decays of cosmic axion strings and domain walls. The desired
agreement of the resulting axion density with the observed dark-matter
density then suggests a relatively large axion mass of $m_a\sim 100~\mu{\rm
eV}$ \cite{Hiramatsu:2012gg, Kawasaki:2014sqa} which we will use as our
benchmark value.

Being bound to our galaxy, the velocity dispersion of the dark-matter axions
on Earth is described by the galactic virial velocity $v_a\sim 10^{-3}$. The
de~Broglie wavelength of these galactic axions is macroscopic
$\lambda_{\rm dB}=2\pi/(m_a v_a)=12.4~{\rm m}~(100~\mu{\rm eV}/m_a)\,(10^{-3}/v_a)$.
Axion dark matter is a highly degenerate Bose gas
and locally can be thought of as a classical field oscillating with frequency
$\nu_a=m_a/2\pi$, i.e., in the microwave range.%
\footnote{Notice that $\nu_a=m_a/2\pi$ is approximately 1~GHz for
$m_a=4~\mu{\rm eV}$. We use natural units with $\hbar=c=k_{\mathrm{B}}=1$ and
the Lorentz-Heaviside convention, where the fine structure constant is
$\alpha=e^2/4\pi\approx 1/137$. In these units, the permittivity and
permeability of vacuum are both unity: $\epsilon_{\rm vac}=1$ and $\mu_{\rm
vac}=1$. Accordingly, $\epsilon$ and $\mu$ specify respectively the {\em
relative\/} permittivity and {\em relative\/} permeability of a considered
medium.}
Moreover, axions have a generic two-photon interaction described by the
Lagrangian ${\cal L}_{a\gamma}=g_{a\gamma}{\bf E}\cdot{\bf B}\,a$, where
$g_{a\gamma}\sim\alpha/(2\pi f_a)$ is a coupling constant of dimension
(energy)$^{-1}$, $\alpha$ the fine-structure constant, $f_a$ the
Peccei--Quinn scale, and ${\bf E}$ and ${\bf B}$ the usual electric and
magnetic fields. In a space permeated by a strong external magnetic field
${\bf B}_{\rm e}$, the galactic axion field $a(t)$ provides an effective
oscillating electric current density ${\bf J}_a(t)=g_{a\gamma}{\bf
B}_{\rm e}\dot a(t)$ which appears in Maxwell's equations as a source term in
addition to the ordinary electromagnetic (EM) current. In a homogeneous
situation, the solution of the modified Maxwell's equations is simply a
homogeneous oscillating electric displacement field ${\bf
D}_a(t)=\epsilon{\bf E}_a(t)=-g_{a\gamma}{\bf B}_{\rm e}\,a(t)$, where
$\epsilon$ is the relative permittivity of a possible medium in the $B$-field
region. For a plausible laboratory field of 10~Tesla, the axion-dark-matter
induced oscillating electric field in vacuum is $|\vec E_a|\sim 10^{-12}~{\rm V/m}$,
independently of $m_a$, but oscillating with frequency $\nu_a$.

Instead of trying to measure this extremely small electric field directly one
may use ${\bf J}_a(t)$ to drive dynamical modes of the EM field. The
traditional approach is to use a microwave cavity placed in a strong
laboratory $B$-field \cite{Sikivie:1983ip}. If the resonance is tuned to the
axion mass, the cavity acts like a forced oscillator and on resonance
achieves a large axion-induced excitation. The quality factor of the cavity
is large ($Q\sim10^5$) but finite so that power is constantly dissipated.
This effect limits the amplitude of the excitations and implies that the
oscillating axion field continuously produces microwave power. The geometric
dimension of the cavity needs to be of the order of $\lambda/2=0.62~{\rm
cm}~(100~\mu{\rm eV}/m_a)$ where $\lambda$ is the photon wavelength
corresponding to the microwave frequency $\nu_a$. The longest
running ongoing effort, the ADMX experiment, currently begins with an axion
search mass in the few-$\mu$eV range where a single cavity on resonance is
big enough to produce a detectable signal~\cite{Rybka:2014xca}. If axions are
the galactic dark matter in this mass range, ADMX will likely find them.
ADMX-HF has recently started exploring higher mass ($\sim 20~\mu$eV) axions,
however it is yet to reach the most common axion benchmark
models~\cite{Kenany:2016tta}.

An $m_a$ range broader than the one accessible with ADMX is well motivated and must be
considered. For much smaller $m_a\sim10^{-3}$--$10^{-1}$~$\mu$eV,
one could use an LC circuit coupled to the
axion field with the help of a strong $B$-field~\cite{Cabrera:2010,
Sikivie:2013laa, Kahn:2016aff}. However, of greater interest to our discussion is the
``high-mass range'' of $m_a\gtrsim 10$~$\mu$eV
where one can consider multiple resonant cavity arrangements or
resonant cavities with unusual shapes or aspect ratios \cite{Baker:2011na}. Another
option is a linear resonator with a wiggled magnetic field, a first
pilot study being the Orpheus experiment~\cite{Rybka:2014cya}.

We pursue a different approach which in its simplest manifestation does not
invoke resonators at all. If one places an interface that separates two dielectric media
with different permittivity $\epsilon$ in a strong
external magnetic field, microwave photons will be emitted in both perpendicular
directions. The homogeneous axion-induced electric displacement field
${\bf D}_a(t)=\epsilon{\bf E}_a(t)=-g_{a\gamma}{\bf B}_{\rm e}\,a(t)$ is no longer
a self-consistent solution of Maxwell's equations because the total ${\bf E}$
(not~${\bf D}$) parallel to the interface needs to be continuous \cite{Jackson:1998nia}
and therefore requires microwave modes streaming in both directions away
from the interface. Physically, these microwaves are sourced by an
oscillating current ${\bf J}_a(t)$ in Maxwell's equations, but
in a homogeneous situation no propagating wave can be excited. Once translation invariance
is broken by an interface, ${\bf J}_a(t)$ leads to the emission of propagating
waves in both perpendicular directions. A sufficiently sharp change in ${\bf B}_{\rm e}$
has the same effect, i.e.,
either the source density must be inhomogeneous or the medium response
encoded in $\epsilon$.

In the most extreme case of a perfect mirror in vacuum immersed in a
$B$-field, microwaves are emitted only away from the mirror surface
and the electric field strength of the propagating mode equals
${\bf E}_a$---the two electric fields need to cancel at the mirror
surface.  The signal power can be enhanced by making the mirror
spherical to focus the microwaves in its center and by placing a
detector therein: the original axion dish antenna \cite{Horns:2012jf}.
The advantage of this approach is its simplicity and broadband nature,
but the signal boost that can be achieved in practice is not large
enough to detect axion dark matter with existing technology.  The dish
antenna is still useful to cover a hitherto unexplored parameter range
of more general axion-like particles or, without a magnetic field, of
hidden photons as in the FUNK and Tokyo experiments
\cite{Dobrich:2015tpa,Suzuki:2015sza}.

Another way of boosting the signal is to use multiple interfaces separating
different dielectric media, which are placed at suitable distances to achieve
constructive interference~\cite{Jaeckel:2013eha}. A minimal realization of
this concept would be two near-perfect facing mirrors. However, this setup
would be equivalent to a linear microwave cavity. One would need a large
transverse size (large mirror area $A$) and a distance of odd multiples of
$\lambda/2=0.62~{\rm cm}$ for our reference value $m_a=100~\mu{\rm eV}$. In
practice, such an arrangement suffers from the same problems as any other
cavity which tries to achieve a large volume for a small $\lambda$, i.e., the
difficulty to extract power in detectable amounts and to achieve the
resonance condition in a controlled and reproducible way.

A more encouraging realisation discussed in reference~\cite{Jaeckel:2013eha}
is the $\lambda/2$ dielectric mirror which consists of $N$ dielectric layers
with alternating high and low refractive indices $n\simeq\sqrt{\epsilon}$ and
thicknesses $d$ given by $n d=\lambda/2$, where $\lambda$ is the vacuum
wavelength determined by $m_a$ as specified above. This setting allows for
constructive interference such that the signal power is boosted by a factor
of $N^2$ with respect to the case of a planar perfect mirror. However, an
axion search will require a scan over $m_a$ and thereby over $\lambda/2=n d$.
This seems to be very challenging if not impossible from the practical point
of view because all layer thicknesses $d$ would have to be chosen accordingly
to explore each potential $m_a$ value.

Interestingly, with a somewhat related setting that requires a similar
adjustment of layer thicknesses, an axion-dark matter search in the region
with $m_a=4$--$400\,\mu$eV was claimed to be viable a long time ago in an
unpublished preprint~\cite{Morris:1984nu}. A special cavity design with
embedded dielectric plates was proposed to improve the cavity quality factor
by more than one order of magnitude to $Q\gtrsim 5\times 10^6$. The scanning
over frequencies would be accomplished by varying the spacings between and
the thicknesses of the plates and the width of the cavity in the direction
perpendicular to the plates. Two requirements were expressed for the spacings
and plate thicknesses: (i)~$\lambda/4$ spacings between the cavity walls and
the nearest plates with thickness $d\simeq\lambda/(4n)$ to reduce energy
density at the cavity walls and (ii)~$\lambda/2$ spacings between the other
plates with thickness $d\simeq\lambda/(2n)$ to optimize the phase coherence
of the electric field inside the cavity volume. As a drawback we expect a
significant adjustment time due to the required tuning to the cavity
resonance and again due to the requirements on the plate thicknesses. This
will slow down scans over a broad $m_a$ range significantly.

\begin{figure}[t]
\centering
\includegraphics[width=10cm]{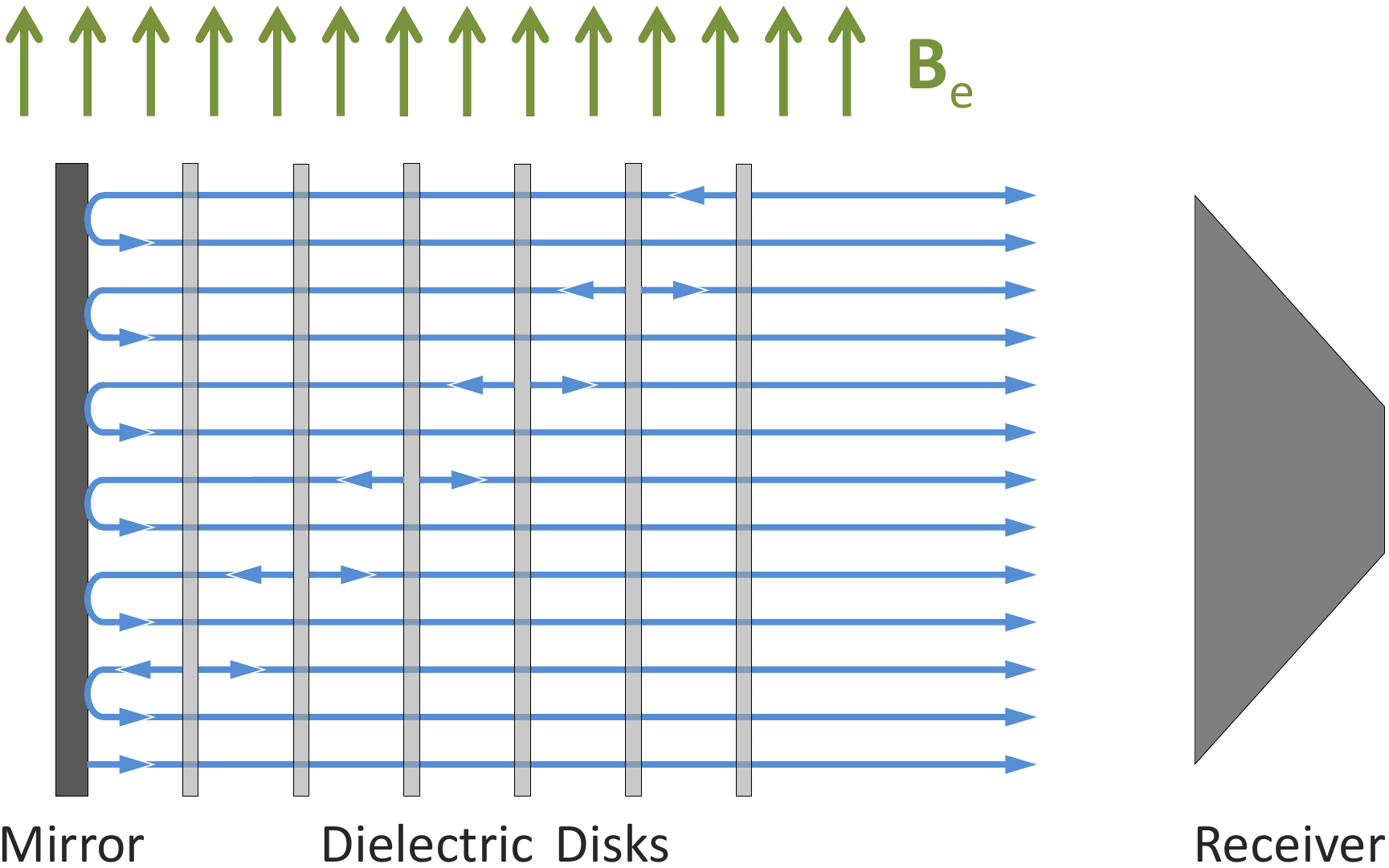}
\caption{A dielectric haloscope consisting of one mirror and several dielectric
disks placed in an external magnetic field ${\bf B}_{\rm e}$
and one receiver in the field-free region. The setup produces discontinuities
in the axion-induced electric field ${\bf E}_a$
at the various interfaces between empty space and either mirror or dielectric disk.
To satisfy the usual continuity requirements of the total electric and magnetic fields
in the directions parallel to the interfaces, ${\bf E}_{\parallel}$ and  ${\bf H}_{\parallel}$,
microwaves emerge in the perpendicular directions away from each interface
with a frequency $\nu_a=m_a/2\pi$ given by the axion mass.
These microwaves are illustrated by the horizontal blue arrows.
The signal strength depends on the thickness of the disks and the spacings
between them, which have to be varied to scan over the axions mass $m_a$.}
\label{fig:LayeredDielectricHaloscope}
\end{figure}

Dielectric haloscopes have recently been
proposed~\cite{TheMADMAXWorkingGroup:2016hpc} as practically viable
generalisation of the dielectric mirror setups presented in
reference~\cite{Jaeckel:2013eha} and as an alternative to the cavity setups
presented in reference~\cite{Morris:1984nu}. This would use a series of (up
to $N\sim 100$) dielectric disks each with m$^2$-scale transverse area and
fixed mm-scale thicknesses and precisely adjustable spacings in a strong
($\sim 10$~Tesla) magnetic field, open (no mirror) on one or both ends.
Figure~\ref{fig:LayeredDielectricHaloscope} shows an illustrative setup with
one mirror and six dielectric disks placed in a magnetic field ${\bf B}_{\rm
e}$ and one receiver in the $B$-field-free region.

With the dielectric haloscope approach, the thicknesses of the dielectric
disks will usually not satisfy $nd=\lambda/2$ (or $\lambda/4$) and are not
required to do so. In fact, when deviating from the $\lambda/2$ condition,
the disks become partially reflecting so that the stored power can be built
up as in a resonant cavity. With a judicious adjustment of the disk
separations, this will allow for a substantial signal boost over a sizeable
search frequency range~$\Delta\nu$. Moreover, by re-adjusting the disk
placement, one will be able to continuously shift~$\Delta\nu$ with the signal
boost and thereby to scan a sizeable range of possible $m_a$ values in a
realistic total measurement time. Indeed, with dielectric haloscopes, a
realistic galactic axion-dark matter search in the high-mass region of
$m_a=40$--$400\,\mu$eV seems to become
feasible~\cite{TheMADMAXWorkingGroup:2016hpc}.

One key advantage of our dielectric haloscope approach is that the frequency
dependent microwave emissivity can be adjusted in a flexible and varied way.
This is because the total power generated over a frequency range increases
linearly with the number of dielectric disks $N$, a finding which we call the
Area Law. For example, one can achieve a very large value of the signal boost
in a narrow range of frequencies, similar to a single resonator.
Alternatively one can adjust the spacings to achieve a relatively uniform
large signal boost over a broader range of frequencies. This freedom is very
important for a practical search for axions.

The challenge of every axion search experiment is to scan over a broad range
of frequencies. Once the axion mass has been found, measuring the signal in
detail will be easy. In search mode, however, one needs to take data for a
long enough time in a given search frequency range $\Delta\nu$ for a possible
signal to become significant relative to background fluctuations and, even
more critically, to avoid missing a true signal by a downward background
fluctuation. The background itself arises from ambient thermal emission and
from electronic noise of the microwave amplifiers. Therefore, the figure of
merit is not the signal power $P$ itself but $P^2$. As we will show, it is
possible to adjust the search frequency range $\Delta\nu$ with the large
signal boost by controlling the disk spacing. However, the Area Law implies
that $P\,\Delta\nu$ is roughly conserved. As the search rate per channel
$\Delta\nu/\Delta t$ scales with $P^2$, the overall search time required to
cover the mass range $\Delta m_a$ given by $(\Delta m_a/\Delta\nu)\Delta t$
decreases linearly with $\Delta\nu$, so working on a narrow search frequency
range would seem to be most advantageous. However, working with a narrow
signal boost can be impractical because of the difficulty to control and
stabilise this narrow boost and by the required adjustment time between such
configurations. Therefore, it can be more effective to work with a broader
frequency range where the signal boost is smaller, but still sufficiently
large to allow for a detection.

Fortunately, dielectric haloscopes are flexible enough to use a broadband search
strategy and, once a potential signal is found, to be reconfigured for a
narrower signal boost to quickly reach a high statistical significance.

Such a multi-parameter optimisation process, dictated by many practical
engineering issues, is not the subject of our discussion. Instead, we present
the theoretical background, extending the original proposal of
reference~\cite{Jaeckel:2013eha} in several ways and focussing directly on
axion-induced microwave production at interfaces instead of the more general
particle mixing approach. As a first step, we re-examine the axion-modified
Maxwell equations in the presence of dielectric and permeable media
(section~\ref{Sec:AxioEM}) and explain the basic physics of the axion-induced
microwave emission process at interfaces (section~\ref{1Dcase}). We then turn
to questions arising when many dielectric disks (perhaps up to around 100)
are used. We generalise the transfer matrix approach to include axions, which
allows one to calculate the frequency-dependent emission, transmission and
reflection functions for a chosen placement structure of the disks
(section~\ref{transfermatrixformalism}). We use this formalism to study
several generic setups which highlight both resonant and non-resonant
features of dielectric haloscopes (section~\ref{Genericexamples}). In
section~\ref{optimum} we consider more realistic and complicated setups,
showing that one can achieve a large degree of control over the response of
the system through disk placement, obtaining broadband, rectangular
responses. We then make projections for the potential reach of a realistic
axion dark-matter search experiment in
section~\ref{discoverypotential} and finally
conclude in section~\ref{conclusion}. Being a crucial feature of dielectric
haloscopes, a proof of the Area Law is provided in
appendix~\ref{sec:area-law}. We also show that a generalised overlap integral
formalism exists for large boost factor setups, which agrees with that of
Sikivie for a resonant cavity (appendix~\ref{Sikiviecomp}). Finally, in
appendix~\ref{scanoptimisation} we discuss optimal search strategies.

\section{Axion-photon mixing in an external B-field}
\label{Sec:AxioEM}

In this section we review the modified form of Maxwell's equations in
the presence of axions both in vacuum and in homogeneous media.
Focussing on a setting with a strong static external magnetic field
${\bf B}_{\rm e}$, we present the linearized macroscopic equations and
consider plane-wave solutions, including the option of lossy media
where the waves are damped.  For dark matter axions with negligible
velocity, we derive the axion-induced electric field ${\bf E}_a$
generated in the presence of ${\bf B}_{\rm e}$.

Throughout this paper, we will use a purely classical description of
the fields.  Our device can be seen as a linear mixer between ingoing
and outgoing axion and photon fields. In particular, we calculate the
linear response in the form of an outgoing propagating microwave
induced by the omnipresent oscillating axion field. On the level of
second quantization (amplitude quantization of the fields), the effect
can be pictured as a mixing of the fields on the operator level, i.e.,
a Bogoliubov transformation which mixes, for example, the creation
operator of a pure axion with that of a pure
photon~\cite{Raffelt:1991ck}.  Ultimately we do not measure an
amplitude but an average rate (average frequency-dependent microwave
power) which, when averaged over a long period of time, does not
reveal quantum fluctuations. Quantization issues may have to be faced
once axions have been found and we study temporal signal correlations.

\subsection{Equations of motion}
\label{Sec:EOM}

The interacting system of photons, axions, and EM currents is
described by the Lagrangian density
\begin{equation}\label{lagrangian}
{\cal L} = -\frac{1}{4}F_{\mu\nu}F^{\mu\nu}-J^\mu A_\mu+\frac{1}{2}\partial_\mu a \partial^\mu a
-\frac{1}{2}m_a^2a^2 -\frac{g_{a\gamma}}{4}F_{\mu\nu}\widetilde F^{\mu\nu}a ,
\end{equation}
where $a$ is the axion field, $m_a$ its mass, and $g_{a\gamma}$ a
coupling constant of dimension (energy)$^{-1}$.
$F_{\mu\nu}=\partial_\mu A_\nu-\partial_\nu A_\mu$ is the EM
field-strength tensor in terms of the vector potential
$A^\mu=(A_0,{\bf A})$ and $J^\mu=(\rho,{\bf J})$ is the electric
4-current.  \smash{$\widetilde F^{\mu\nu} =
  \frac{1}{2}\varepsilon^{\mu\nu\alpha\beta}F_{\alpha\beta}$} is the
dual tensor, where we use
$\varepsilon^{0123}=\varepsilon_{123}=+1$. The electric and magnetic
fields are explicitly
\begin{equation}
{\bf E} = -{\bm \nabla} A_0-\dot {\bf A}
\qquad\hbox{and}\qquad
{\bf B} =  {\bm \nabla}{\bm \times}  {\bf A}\,.
\end{equation}
We work in natural units with $\hbar = c = 1$ and the Lorentz-Heaviside convention
$\alpha=e^2/4\pi$, as mentioned above.
In this case, the energy density of the EM field
is $\frac{1}{2}({\bf E}^2+{\bf B}^2)$ and the conversion of units is
$1~{\rm V}/{\rm m}=6.5162\times 10^{-7}~{\rm eV}^2$ and
$1~{\rm T}=1~{\rm Tesla}=195.35~{\rm eV}^2$.

For a general axion-like particle (ALP), the mass and photon coupling
are independent phenomenological parameters. However, in the dark-matter context we
are interested in QCD axions where both parameters are given in terms of
the Peccei--Quinn scale or axion decay constant $f_a$ by
$m_af_a\sim m_\pi f_\pi$ and $g_{a\gamma}\sim\alpha/(2\pi f_a)$, where $m_\pi$ and
$f_\pi$ are respectively the pion mass and decay constant. A recent detailed study
yields the numerical values~\cite{diCortona:2015ldu}
\begin{subequations}
\begin{eqnarray}
m_a&=&5.70(6)(4)\, {\rm \mu eV}\,\left(\frac{10^{12}\rm\,GeV}{f_a}\right)\,,\label{eq:ma}\\
g_{a\gamma}&=&-\frac{\alpha}{2\pi f_a}\,C_{a\gamma}
=-2.04(3)\times10^{-16}~{\rm GeV}^{-1}\,\left(\frac{m_a}{1\,\mu{\rm eV}}\right)\,C_{a\gamma}\,,\label{eq:gag}\\
C_{a\gamma}&=&\frac{{\cal E}}{{\cal N}}-1.92(4)\label{eq:cag}\,,
\end{eqnarray}
\end{subequations}
where numbers in brackets denote the uncertainty in the last digit.
For $m_a$ the first error is from quark-mass uncertainties
and the second one from higher-order corrections.

In expression~(\ref{eq:cag}), ${\cal E}$ is the EM anomaly and
${\cal N}$ the color anomaly, or equivalently domain wall number, of the Peccei-Quinn symmetry. In models
where ordinary quarks and leptons do not carry Peccei--Quinn charges,
the axion-photon interaction arises entirely from $a$-$\pi^0$-$\eta$
mixing and ${\cal E}/{\cal N}=0$, the KSVZ model \cite{Kim:1979if,
  Shifman:1979if} providing a traditional example. In more general
models, ${\cal E}/{\cal N}$ is a ratio of small integers, the DFSZ
model \cite{Dine:1981rt, Zhitnitsky:1980tq} with
${\cal E}/{\cal N}=8/3$ being an often-cited example, although there exist many
other cases \cite{Kim:2014rza}.
While $g_{a\gamma}$ can be either positive or negative
via the model-dependent ${\cal E}/{\cal N}$ value,
we will see that the detectable power will depend on $g_{a\gamma}^2$.

The Euler--Lagrange equations of motion for the axion and photon fields following
from the Lagrangian density~(\ref{lagrangian}) are
\begin{subequations}
\begin{eqnarray}
\label{GaussAmpere}
\partial_\mu F^{\mu\nu}  &=&J^\nu -g_{a\gamma}\widetilde F^{\mu\nu}\partial_\mu a\,, \\
\(\partial_\mu\partial^\mu  +m_a^2\) a&=& -\frac{g_{a\gamma}}{4}F_{\mu\nu}\widetilde F^{\mu\nu}\,.
\end{eqnarray}
\end{subequations}
The first equation is a modification of the laws of Gauss and Amp\`ere
in the presence of axions, which leads to the extra current 
$J_a^\nu\equiv -g_{a\gamma}\partial_\mu (\widetilde F^{\mu\nu} a)=-g_{a\gamma}\widetilde F^{\mu\nu}\partial_\mu a$ 
on the rhs of this equation. The laws of Gauss for magnetism and of Faraday derive from
a geometric property of electrodynamics, the Bianchi identity
$\partial_\mu \widetilde F^{\mu\nu}=0$, which does not get modified by
including the axion.  In terms of electric and magnetic fields, one
finds \cite{Sikivie:1983ip, Wilczek:1987mv}
\begin{subequations}
\begin{eqnarray}
{\bm\nabla}\cdot {\bf E} &=& \rho -	g_{a\gamma} {\bf B}\cdot{\bm\nabla} a\,,
\label{eq:Maxwell-a}\\
{\bm\nabla}{\bm \times}{\bf B}- \dot{\bf E}  &=& {\bf J}
+g_{a\gamma}\({\bf B}\, \dot a -{\bf E}{\bm \times}{\bm\nabla} a\)\,,\label{eq:Maxwell-b}\\
{\bm\nabla}\cdot{\bf B}&=& 0\,,
\label{eq:Maxwell-cone}\\
{\bm\nabla}{\bm \times}{\bf E}+\dot{\bf B}&=&0\,,
\label{eq:Maxwell-d}\\
\ddot a-{\bm\nabla}^2 a +m_a^2 a &=&
g_{a\gamma} {\bf E}\cdot {\bf B}\,.\label{eq:Maxwell-c}
\end{eqnarray}
\end{subequations}
While axions do not enter the homogeneous equations, we need the latter
to derive the EM boundary conditions for interfaces
such as those considered in section~\ref{1Dcase} below.

\subsection{Macroscopic form of Maxwell's equations}
\label{Sec:MacroscopicEOM}

As discussed in the Introduction, dielectric materials will be essential for the proposed
new axion dark matter haloscope. Accordingly, we now reformulate these equations
in terms of macroscopic fields to account for the EM response of the background medium.

The form of the homogeneous equations~(\ref{eq:Maxwell-cone})
and~(\ref{eq:Maxwell-d}) does not change.  Nevertheless, the fields
${\bf E}$ and ${\bf B}$ are here and henceforth understood to be the
macroscopic ones obtained from the microscopic ones by macroscopic
smoothing.  This applies, in particular, to the rhs of
equation~(\ref{eq:Maxwell-c}), which keeps its form as well. We
mention in passing that there are other cases of axion-electrodynamics
where one would rather consider the microscopic fields.  The Bragg
technique to search for solar axions uses their conversion to $X$-rays
in the very strong microscopic electric fields in a crystal.  In our
case of extremely small momentum transfers, the microscopic structure
of the medium is not resolved.

Following a standard treatment of electrodynamics in
media~\cite{Jackson:1998nia}, we consider macroscopic charge and
current densities and split them into a free part and a bound part:
$\rho=\rho_{\rm f}+\rho_{\rm b}$ and ${\bf J}={\bf J}_{\rm f}+{\bf J}_{\rm b}$.
The bound parts are tied to the molecules of the medium
and responsible for the EM response.
They can be expressed in terms of the macroscopic polarization ${\bf P}$
(given by the macroscopic average over the molecular dipole moments)
and the macroscopic magnetization ${\bf M}$
(given by the macroscopic average over the molecular magnetic moments)
\begin{equation}
	\rho_{\rm b} = -{\bm\nabla}\cdot{\bf P}
	\quad {\rm and} \quad
	{\bf J}_{\rm b} = {\bm\nabla}{\bm \times}{\bf M}+\dot {\bf P}\, .
\label{eq:boundchargecurrent}
\end{equation}
The free charge and current densities $\rho_{\rm f}$ and ${\bf J}_{\rm f}$ are simply
the remaining parts, which satisfy the continuity equation
\begin{equation}
\dot {\rho}_{\rm f}+{\bm\nabla}\cdot {\bf J}_{\rm f}=0 \, .
\label{eq:ContinuityEq}
\end{equation}
Moreover, we introduce the usual macroscopic electric displacement field ${\bf D}$ and
the macroscopic magnetic field ${\bf H}$ which absorb
effects of the polarisation ${\bf P}$ and the magnetisation ${\bf M}$, respectively,
\begin{equation}
{\bf D}={\bf E}+{\bf P}
\quad\hbox{and}\quad
{\bf H}={\bf B}-{\bf M}\, .
\end{equation}
(Following usual practice, we have ignored possible effects of
molecular quadrupole moments.)
The inhomogeneous equations~(\ref{eq:Maxwell-a}), (\ref{eq:Maxwell-b}) and
(\ref{eq:Maxwell-c}) become
\begin{subequations}
\bea
\label{coulombaxion}
{\bm\nabla}\cdot {\bf D} &=& \rho_{\rm f} - g_{a\gamma} {\bf B}\cdot {\bm\nabla} a		\, ,
\label{eq:Maxwell-a-matter}\\
{\bm\nabla}{\bm \times} {\bf H} - \dot {\bf D}   &=& {\bf J}_{\rm f}  +g_{a\gamma}\({\bf B}\dot a  -{\bf E}{\bm \times} {\bm\nabla} a \)\, ,
\label{eq:Maxwell-b-matter}\\
\ddot a-{\bm\nabla}^2 a +m_a^2 a &=& g_{a\gamma} {\bf E}\cdot {\bf B}.
\label{eq:Maxwell-c-matter}
\eea
\label{eq:Maxwell-x-matter}
\end{subequations}
In our context, the important point is that on the rhs of
equations~(\ref{eq:Maxwell-a-matter})--(\ref{eq:Maxwell-c-matter}) it remains
${\bf E}$ (not {\bf D}) and ${\bf B}$ (not {\bf H}) that appear in the axion--photon
interaction terms, which remain unaffected by the medium response.

To solve these equations, additional constitutive relations
${\bf D}={\bf D}[{\bf E},{\bf B}]$, ${\bf H}={\bf H}[{\bf E},{\bf B}]$, and
${\bf J}_{\rm f}={\bf J}_{\rm f}[{\bf E},{\bf B}]$ are needed that characterize
the medium. Later we will assume simple isotropic media with a linear response.
In this case, the third of these relations is Ohm's law which in our context
applies, for example, to a metallic mirror.

\subsection{Linearization}
\label{linearization}

In our cases of practical interest, the macroscopic Maxwell equations
with axions can be linearized both in the fields and concerning the
medium response. To linearize in the fields, we notice that our
situation of interest is one where a static external magnetic field
${\bf B}_{\rm e}$ has been set up by means of an external current
${\bf J}_{\rm e}$.  As we will see, this field is much stronger than
any other field in our problem. In particular, it is much larger than
the time-varying part corresponding to the axion-induced EM waves.
Keeping only the leading terms in the static field ${\bf B}_{\rm e}$
on the rhs of
equations~(\ref{eq:Maxwell-a-matter})--(\ref{eq:Maxwell-c-matter}) we
find\footnote{Note that we do not consider static axion field
  contributions such as those associated with a static domain
  wall~\cite{Huang:1985tt} on which photons can be reflected and
  transmitted.}
\begin{subequations}
\begin{eqnarray}
{\bm\nabla}\cdot {\bf D} -  \rho_{\rm f}
&=&
- g_{a\gamma} {\bf B}_{\rm e}\cdot{\bm\nabla} a\,,
\label{eq:Maxwell-aa}\\
{\bm\nabla}{\bm \times} {\bf H}-\dot{\bf D}-{\bf J}_{\rm f}
&=&
g_{a\gamma}{\bf B}_{\rm e} \dot a\,,\label{eq:Maxwell-bb}
\\
\ddot a-{\bm\nabla}^2 a +m_a^2 a &=&
g_{a\gamma}{\bf E}\cdot{\bf B}_{\rm e}\,,\label{eq:Maxwell-cc}
\end{eqnarray}
\label{eq:Maxwell-xx}
\end{subequations}
where ${\bf D}$, ${\bf E}$, ${\bf H}$, $\rho_{\rm f}$, ${\bf J}_{\rm f}$, and $a$
all depend on $t$ and ${\bf x}$.

Here ${\bf H}$ stands for the small time-varying part of the magnetic
field, not including ${\bf H}_{\rm e}$ associated with
${\bf B}_{\rm e}$, and ${\bf J}_{\rm f}$ for the current density
without
${\bf J}_{\rm e}$.  Indeed, ${\bf H}_{\rm e}$ and ${\bf J}_{\rm e}$ do not
appear since they fulfill equation~(\ref{eq:Maxwell-b-matter})
separately.  We have now written ${\bf J}_{\rm f}$ on the lhs of the
equation because the only remaining current exists in response to the
electric field.  Likewise, in the modified Gauss
law~(\ref{eq:Maxwell-aa}), $\rho_{\rm f}$ is written on the lhs
because we will always consider neutral media.  Indeed, since we do
not consider settings with a static charge density, we do not lose
information by applying a time derivative on
equation~(\ref{eq:Maxwell-aa}) to obtain
\begin{equation}
{\bm\nabla}\cdot(\dot{\bf D}-{\bf J}_{\rm f})= - g_{a\gamma} {\bf B}_{\rm e}\cdot{\bm\nabla} \dot a\, ,
\label{eq:TimeDerivedGauss}
\end{equation}
where the continuity equation~(\ref{eq:ContinuityEq}) has been used. 
In the following we will use equation~\eqref{eq:TimeDerivedGauss}
instead of equation~\eqref{eq:Maxwell-aa} 
since this will allow us to include the
case of conductors in a straightforward way.

The equations are now linear in all space-time dependent quantities. 
It is therefore convenient to expand them in plane waves
proportional to $e^{-i(\omega t-{\bf k}\cdot{\bf x})}$, leading to
\begin{subequations}
\begin{eqnarray}
{\bf k}\cdot  (\omega{\bf{\hat D}}+i {\bf{\hat J}}_{\rm f})&=& -g_{a\gamma}\omega{\bf k}\cdot{\bf B}_{\rm e}{\hat a}\,,
\label{eq:Maxwell-aaaa}
\\
{\bf k}{\bm \times} {\bf\hat H}+\omega {\bf\hat D}+i{\bf\hat J}_{\rm f}  &=&-g_{a\gamma}\omega{\bf B}_{\rm e}\hat a\,,
\label{eq:Maxwell-bbbb}
\\
\(\omega^2-{\bf k}^2-m_a^2\) \hat a &=&
-g_{a\gamma} {\bf\hat E}\cdot{\bf B}_{\rm e}\,,
\label{eq:Maxwell-cccc}
\end{eqnarray}
\end{subequations}
where now $\hat a$, ${\bf\hat D}={\bf\hat E}+{\bf\hat P}$,
${\bf\hat H}$, ${\bf\hat E}$, and ${\bf\hat J}_{\rm f}$ are complex amplitudes
depending on $\omega$ and ${\bf k}$.  

Let us now focus on homogeneous and isotropic materials which may be conductors and let
us assume that the medium response is linear. In this case, the magnetic field
and the magnetic induction are related in terms of the magnetic permeability $\mu$ in the form
\begin{equation}
{\bf\hat H}={\bf\hat B}/\mu\,.
\end{equation}
Notice that all quantities
depend on $\omega$ and ${\bf k}$, encoding temporal and spatial
dispersion.
In configuration space, this relation is generally neither local in time nor in space.
Thus, one cannot use the same simple form because ${\bf H}(t,{\bf x})$ is given
by a convolution of ${\bf B}(t,{\bf x})$ with the $(t,{\bf x})$-dependent magnetic permeability.
In practice, of course, the response functions may vary only slowly as functions of
$\omega$ and ${\bf k}$ and one may not always need to worry about this issue.

If the medium is a conductor, the electric field drives a current given by Ohm's law
in terms of the conductivity $\sigma$ in the form
${\bf\hat J}_{\rm f}=\sigma{\bf\hat E}$. Moreover,
in the linear response domain, the polarization is given as ${\bf\hat P}=\chi\,{\bf\hat E}$ where $\chi$
is the electric susceptibility. Therefore, we may combine the two terms and write on the
lhs of equations~(\ref{eq:Maxwell-aaaa}) and~(\ref{eq:Maxwell-bbbb}) all electric effects in the form
$\omega {\bf\hat D}+i{\bf\hat J}_{\rm f}=\omega\epsilon {\bf\hat E}$, where the
total effective dielectric permittivity is
\begin{equation}
		\epsilon = 1+\chi + \frac{i\sigma}{\omega}\,.
\label{Eq:epsilontotal}
\end{equation}
In this form we can include both dielectric and conductive properties of the chosen material.
Assuming that $\sigma$ is real-valued for the frequency range of interest, its contribution to
the imaginary part of $\epsilon$ derives from Ohm's law
and accounts for absorption due to the conduction current.
The additional contribution to $\text{Im}(\epsilon)$ from $\text{Im}(\chi)$
accounts for losses caused by the displacement current.
Accordingly, the imaginary part of the permittivity is used to classify materials, e.g.,
as a perfect conductor for $\text{Im}(\epsilon)\simeq \sigma/\omega\to\infty$
or as a perfect dielectric for  $\text{Im}(\epsilon)=0$.
In general, both $\epsilon$ and $\mu$ have real and imaginary parts;
we comment on lossy media explicitly in section~\ref{sec:LossyMedia} below.
Moreover, in general $\epsilon$ like $\mu$ depends on $\omega$ and ${\bf k}$,
causing temporal and spatial dispersion.

With these definitions, the linearized macroscopic axio-electrodynamic
equations in Fourier space are
\begin{subequations}
\begin{eqnarray}
\epsilon{\bf k}\cdot {\bf\hat E} &=& -g_{a\gamma}{\bf k}\cdot{\bf B}_{\rm e}\hat a\,,
\label{eq:Maxwell-aaa}\\
{\bf k}{\bm \times} {\bf\hat H}+\omega \epsilon{\bf\hat E}  &=&
-g_{a\gamma}\omega{\bf B}_{\rm e} \hat a\,,
\label{eq:Maxwell-bbb}\\
{\bf k}\cdot{\bf\hat B}&=& 0\,,
\label{eq:Maxwell-cccone}\\
{\bf k}{\bm \times}{\bf\hat  E}-\omega{\bf\hat B}&=&0\,,
\label{eq:Maxwell-ddd}\\
\(\omega^2-{\bf k}^2-m_a^2\) \hat a &=&
-g_{a\gamma} {\bf\hat E}\cdot{\bf B}_{\rm e}\, .
\label{eq:Maxwell-ccc}
\end{eqnarray}
\label{eq:Maxwell-xxx}
\end{subequations}
For completeness, we have written out the homogeneous equations in Fourier space as well,
in which ${\bf B}_{\rm e}$ does not appear because it is assumed to be stationary
and ${\bm\nabla}\cdot{\bf B}_{\rm e}=0$.
For the case of a perfect dielectric, $\rho_{\rm f}={\bf J}_{\rm f}=0$
in equation~(\ref{eq:Maxwell-xx})
so that one obtains equation~(\ref{eq:Maxwell-xxx}) directly from~(\ref{eq:Maxwell-xx})
without having to apply a time derivative on equation~(\ref{eq:Maxwell-aa}).

In light of the smallness of $g_{a\gamma}$, the axion coupling to ${\bf B}_{\rm e}$ can be treated as a perturbation.
To lowest order in $g_{a\gamma}$, the dispersion relation for propagating EM fields remains
\begin{equation}
	k_{\gamma}^2=n^2\omega_{\gamma}^2
\label{Eq:DispersionRelGamma}
\end{equation}
as in the absence of axions,  with the refractive index $n$ given by
\begin{equation}
	n^2=\epsilon\mu .
\label{Eq:RefractiveIndex}
\end{equation}
The corresponding dispersion relation for axions is
\begin{equation}
	k_a^2=\omega_a^2-m_a^2
\label{Eq:DispersionRelAxion}
\end{equation}
as in the absence of ${\bf B}_{\rm e}$.

Henceforth we will simplify our notation and write the Fourier components of fields without
carets. A quantity like ${\bf E}$ will simply denote the complex
amplitude whenever no confusion can arise. 
Moreover, as detailed below, we will approximate the axion velocity to be zero 
which implies a vanishing wave vector ${\bf k}_a=0$. 
For simplicity, we thus will use ${\bf k}$ (without $\gamma$ subscript)
to denote a non-vanishing wave vector of a propagating EM field.

\subsection{Lossy media}
\label{sec:LossyMedia}

In general, dielectrics are not perfect, i.e., not all the EM energy that propagates in the form of waves
is transferred or reflected but absorbed as well.
For us, this means that the full signal generated by axions (described in detail below) will not be available
and stacking too many lossy dielectric layers may be counter productive.
To describe the effect of losses, we allow $\epsilon$ to be complex valued
as discussed earlier.
We quantify losses in terms of the usual dielectric loss tangent
\begin{equation}
\tan\delta=\frac{\text{Im}\,\epsilon}{\text{Re}\,\epsilon}\,.
\label{eq:loss-tangent}
\end{equation}
For the dielectric layers in realistic experimental situations,
we would use materials with $\tan\delta\ll 1$. Typical values may be of the order of $10^{-5}$
so that with excellent approximation $\tan\delta\simeq\delta$.
Assuming a non-permeable material with $\mu=1$,
the refractive index can then be written as $n\simeq |n|(1+i\delta/2)$.

EM waves are damped in lossy media.
For our specific case, it is shown below that all time variations
are driven by the external axion field with frequency $\omega=m_a=\omega_a=\omega_{\gamma}$,
leading to a steady-state situation with no damping as a function of time. Therefore, when we
study propagating plane-wave solutions which are proportional to $e^{-i(\omega t-{\bf k}\cdot{\bf x})}$
and which obey the dispersion relation~(\ref{Eq:DispersionRelGamma}),
it is the wave number $k$ which acquires an imaginary part and not the frequency.
For example, a linearly polarized electric field of a plane wave propagating in the positive $x$-direction
is proportional to
\be
  e^{-i\omega\left (t-nx\right )}=
  e^{-{\text {Im}}(n)\omega x}\, e^{-i\omega\left[t-{\text {Re}}(n)x\right]}.
\ee
For ${\rm Im}\,n>0$ this wave is exponentially damped as a function of $x$.
In the opposite case, for example in an inverted medium in the context of laser physics,
the wave grows exponentially.

Waves which are exponentially damped as a function of spatial coordinates cannot exist
in a (semi-)infinite medium, i.e., when we discuss plane-wave solutions in a homogeneous medium,
we are having in mind piecewise homogeneous situations with a plane boundary of the medium.
Of course, interfaces between different media, or between a medium and empty space, is exactly
what we study in our paper.
Notice also that a solution of the form $e^{-i(\omega t-{\bf k}\cdot{\bf x})}$ does
not represent a spatial Fourier mode of the electric-field distribution. Because
${\bf k}$ has now an imaginary part commensurate with the complex dispersion relation~(\ref{Eq:DispersionRelGamma}),
this plane wave is one component of a Laplace transform.

In the following parts of our paper, all formulas will be consistent with complex dielectric constants $\epsilon$
and with plane waves which are damped along their direction of propagation.
One should remember, however,
that the electric and magnetic fields
of a propagating EM wave are not in phase
if the wave number $k$ has an imaginary part.

\subsection{Dark matter axions}
\label{Sec:DMaxions}

Our case of interest is very special in that the galactic dark matter axions
are highly nonrelativistic ($v_a\lesssim 10^{-3}$). With the small axion mass~(\ref{eq:ma}) and
the large value of the corresponding de Broglie wavelength 
\begin{equation}
	\lambda_{\rm dB}
	=\frac{2\pi}{m_a v_a}
	=12.4~{\rm m}~\left(\frac{100~\mu{\rm eV}}{m_a}\right)\left(\frac{10^{-3}}{v_a}\right),
\label{eq:lambdadeBroglie}
\end{equation}
the axion field can be considered to be essentially homogeneous over the relevant laboratory scales.
Therefore, we approximate the axion field as homogeneous (${\bf k}_a=0$)
leaving the description of modifications related to a small but non-zero $v_a$ for future work.

In the zero-velocity limit for axions, the local dark-matter axion field
is simply described by an amplitude $a_0$ and frequency $\omega=m_a$ in the form
\begin{equation}
	a(t)= a_0 e^{-i m_a t},
\label{Eq:axionDMfield}
\end{equation}
where the physical axion field is the real part of this expression.
The axion field strength $a_0$ governs the local axion dark matter density according to
\begin{equation}
\rho_a=\frac{m_a^2|a_0|^2}{2}=f_{\rm DM}\,\frac{300~{\rm MeV}}{{\rm cm}^3}\,.
\label{eq:axionDMdensity}
\end{equation}
Here we have introduced $f_{\rm DM}$ as a fudge factor to express the uncertainty in the local
dark-matter density%
\footnote{The local dark matter density has been estimated by various
authors using different data and assumptions \cite{Catena:2009mf, Strigari:2009zb, Weber:2009pt, Bovy:2012tw, Nesti:2013uwa, Bozorgnia:2013pua, Read:2014qva}. The value
$300~{\rm MeV}~{\rm cm}^{-3}$ is considered to be the
canonical one, which the particle data group \cite{Agashe:2014kda} quotes to come with an uncertainty of a factor of 2--3.
In the axion literature, $400~{\rm MeV}~{\rm cm}^{-3}$ is frequently used, which is compatible, e.g., with the findings of
the often-cited reference~\cite{Catena:2009mf}.}
as well as the uncertainty of the dark-matter fraction consisting of axions relative to possible other forms of dark matter
or relative to the fraction gravitationally bound in axion mini clusters.

For ${\bf k}_a=0$, the homogeneous axion-induced electric field can be read directly from equation~(\ref{eq:Maxwell-bbb}):
\begin{equation}
\label{eq:Ea}
{\bf E}_a(t)=-\frac{g_{a\gamma}{\bf B}_{\rm e}}{\epsilon}\,a(t)\,.
\end{equation}
Thus, to linear order in $g_{a\gamma}$, the axion-induced electric field ${\bf E}_a(t)$
oscillates with frequency $\omega=m_a$ and is the real part  of
\begin{equation}
\label{eq:EaE0}
{\bf E}_a(t)=-\frac{{\bf E}_0}{\epsilon}\,e^{-i m_a t}\,,
\end{equation}
where we have introduced the definition
\begin{equation}
\label{eq:vecE0}
{\bf E}_0\equiv g_{a\gamma}{\bf B}_{\rm e} a_0\,.
\end{equation}
Notice that all effects driven by the axion field vary with frequency $\omega=m_a$,
but they do not need to be in phase.
In lossy media, where the dielectric permittivity $\epsilon$ has a non-vanishing imaginary part,
${\bf E}_a(t)$ and $a(t)$ are out of phase
by a shift corresponding to the loss tangent defined in equation~(\ref{eq:loss-tangent}).

In most of our paper we will use
$E_0=|{\bf E}_0|=|g_{a\gamma}{B}_{\rm e}a_0|$  with $B_{\rm e}=|{\bf B}_{\rm e}|$
as a scale for all axion-induced EM fields.
With the axion-photon coupling~(\ref{eq:gag}) and~(\ref{eq:axionDMdensity}), we find
\begin{equation}
E_{0}
=1.3\times10^{-12}~{\rm V}/{\rm m}~\frac{B_{\rm e}}{10~{\rm T}}~
|C_{a\gamma}|f_{\rm DM}^{1/2}\,.
\end{equation}
This result is independent of the axion mass because $| a_0|\propto \rho_a^{1/2}/m_a$ whereas
$g_{a\gamma}\propto m_a$.

For comparison we mention that the energy density associated with the
external magnetic field is
$\frac{1}{2}\,B_{\rm e}^2=2.5\times10^{14}~{\rm MeV}/{\rm cm}^3~(B_{\rm e}/10~{\rm T})^2$,
i.e., around 12 orders of magnitude larger than the local dark matter energy density.
On the other hand, the energy density associated with the axion-induced electric field
$\frac{1}{2}\,E_{a}^2=\frac{1}{2}\,E_0^2/|\epsilon|^2$
is found to be some 31 orders of magnitudes 
(or more depending on $\epsilon$)
smaller than $\rho_a$.

Corrections which arise from the slow axion motions with $v_a \sim 10^{-3}$ will be
negligible in our context. For example, the magnetic field ${\bf H}_a$
associated with the axion field is smaller than ${\bf E}_a$ by an approximate factor $v_a$.
Indeed, $v_a=0$ and thus ${\bf k}_a=0$ requires ${\bf B}_a=0$
to satisfy the homogeneous Maxwell equation~(\ref{eq:Maxwell-ddd}).
At this level of approximation, the impact of the axion field derives entirely from
equation~(\ref{eq:Ea}). The experimental challenge is to measure this extremely
small electric field which oscillates with the frequency $m_a$.

\section{Axion-induced electromagnetic radiation at an interface}
\label{1Dcase}
After setting up a static magnetic field ${\bf B}_{\rm e}$ in the
laboratory, we can measure an electric field ${\bf E}_{a}(t)$ given by
equation~(\ref{eq:Ea}) which is driven by the nonrelativistic
dark-matter axion field and oscillates with the frequency $\omega=m_a$
corresponding to the axion mass.  In practice, to couple a detector
efficiently to this EM signal requires amplification
which we propose to achieve with layered dielectrics. To explain this
concept, we first derive the EM response of the simplest
element of our overall system: a single interface between two media,
which could be a boundary between regions with either different dielectric
properties, with different a applied field strength ${\bf B}_{\rm e}$, or
both.

\subsection{Radiation from an interface}

\begin{figure}[bt]
\centering
\includegraphics[width=10cm]{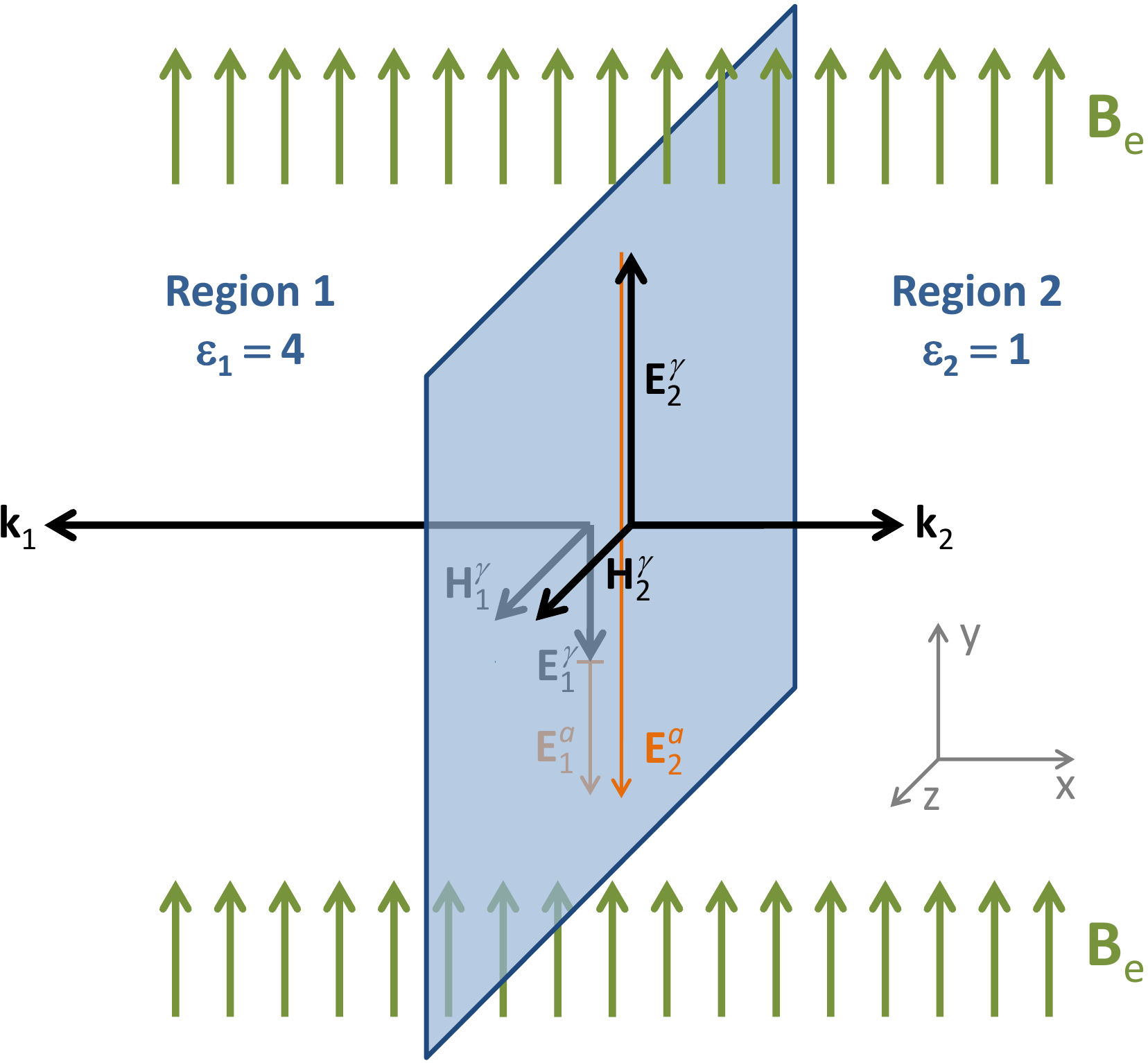}
\caption{Interface between two regions 1 and 2 with equal
  ${\bf B}_{\rm e}$ and chosen material properties $\mu_1=\mu_2=1$,
  $\epsilon_1=4$ and $\epsilon_2=1$, implying $n_1=\sqrt{\epsilon_1}=2$
  and $n_2=1$ so that $k_1=2\omega$ and $k_2=\omega$.  The
  EM waves propagating away from
  both sides of the interface ensure
  continuity of ${\bf H}_\parallel$ and ${\bf E}_\parallel$, here
  implying $H_{1}^\gamma=H_{2}^\gamma=1/2$,
  $E_{1}^a=E_{1}^\gamma=-1/4$, $E_{2}^a=-1$ and
  $E_{2}^\gamma=1/2$, where all fields are given in units of
  $E_0=|g_{a\gamma}B_{\rm e} a_0|$.}
\label{fig:interface}
\end{figure}

We begin with the simplest possible configuration, where the interface
between two regions~1 and 2 is an infinite plane with a parallel
applied magnetic field ${\bf B}_{\rm e}$ as shown in
figure~\ref{fig:interface}.  At the interface, there is a jump of the
dielectric constant $\epsilon$ or the magnetic field ${\bf B}_{\rm e}$ or
both, and so the axion-induced electric field ${\bf E}_a$ jumps at the
interface as well. Note that in each region Maxwell's equations in the presence of axions \eqref{eq:Maxwell-xxx} are inhomogeneous differential equations: the general solution is given by adding solutions of the homogeneous equations (regular EM waves) to an inhomogeneous solution (i.e, ${\bf E}_a$). The global solution to Maxwell's equations then comes from choosing EM waves in each region such that they match
the boundary conditions. In
other words, the presence of the interface couples 
the axion-induced electric field to propagating waves. We
ignore possible EM waves which are causally unrelated to
the axion field in this section. In particular, we exclude waves moving toward the
interface from far away. Therefore, the symmetries of our
configuration allow only waves moving away from the interface on
either side in perpendicular directions.

The axion-induced electric field and the electric and magnetic fields
associated with these propagating waves are all parallel to the
interface. The usual continuity requirements at an interface between
different media are~\cite{Jackson:1998nia}
\begin{equation}\label{eq:continuity}
{\bf E}_{\parallel,1} ={\bf E}_{\parallel,2}
\quad\hbox{and}\quad
{\bf H}_{\parallel,1} ={\bf H}_{\parallel,2}\,,
\end{equation}
where the first condition follows from Faraday's law
(\ref{eq:Maxwell-d}), which remains unchanged by axions. The second
condition follows from Amp\`ere's law (\ref{eq:Maxwell-b}) in the
usual way because the presence of axions amounts to an external volume
current density, which vanishes for a surface.

One consequence of these boundary conditions is that ${\bf B}_{\rm
  e}$, assumed to be parallel to the interface, must jump if the
static (DC) permeability $\mu_{\rm DC}$ is different between the two
media. Recall that in our equations, usually the symbols $\mu$ and
$\epsilon$ represent the material properties at angular frequency
$\omega=m_a$, not the DC quantities, although in practice, the
frequency dependence may be small. In our conceptual discussion here
we will mostly ignore a possible ${\bf B}_{\rm e}$ discontinuity at the
interface caused by a jump of $\mu_{\rm DC}$ because we are primarily
concerned with such dielectric media that have only a negligible magnetic
response.  This issue is one of many small effects to be studied in a
realistic experimental design.

Turning to propagating waves, equation~(\ref{eq:Maxwell-bbb})
without the source term on the rhs and using ${\bf H}={\bf B}/\mu$
implies the usual condition
${\bf k}\times{\bf H}_\gamma+\omega\epsilon{\bf E}_\gamma=0$,
where the subscript $\gamma$ indicates that these are the fields of a
propagating EM wave. With the wave number satisfying $k=n\omega$ (refractive index~$n$) one finds
$H_\gamma=\pm(\epsilon/n)\,E_\gamma$.  Notice that $n=\sqrt{\epsilon\mu}$
so that $\epsilon/n=\sqrt{\epsilon/\mu}$. However, because the medium
is described by two material constants $\mu$ and $\epsilon$ we prefer
to use instead the pair of parameters $n$ and $\epsilon$ which avoids
the appearance of many square-root symbols. In most practical cases,
$\mu\approx 1$ so that $n$ becomes essentially a notation for
$\sqrt{\epsilon}$. For the EM waves, ${\bf H}_\gamma$ is
perpendicular to ${\bf E}_\gamma$ which itself is collinear with
${\bf E}_a$ and thus with ${\bf B}_{\rm e}$. Therefore, the continuity of
${\bf H}_\parallel$ does not involve ${\bf B}_{\rm e}$ and
implies\footnote{We put the symbols $a$ and $\gamma$ as subscripts or
  superscripts depending on typographical convenience.}
${\bf H}_{1}^\gamma={\bf H}_{2}^\gamma$.  Because ${\bf k}_1$ and
${\bf k}_2$ point in opposite directions, also ${\bf E}^{\gamma}_1$ and
${\bf E}^{\gamma}_2$ must be oriented in opposite
directions as shown in figure~\ref{fig:interface}.
Moreover, the continuity of ${\bf E}_\parallel$
must involve the axion-induced field ${\bf E}_a$ so that
at the interface, which is defined by a jump of ${\bf E}_a$, the total
${\bf E}_\parallel$ is continuous. Therefore, the continuity conditions of
equation~(\ref{eq:continuity}) imply for the field values at the interface
\begin{subequations}
\begin{eqnarray}
  \hbox to 9em{Continuity of ${\bf H}_\parallel$\hfil}&&
  -\frac{\epsilon_1}{n_1}\,E_1^\gamma=\frac{\epsilon_2}{n_2}\,E_2^\gamma,\\[1ex]
  \hbox to 9em{Continuity of ${\bf E}_\parallel$\hfil}&&
        {E}_{1}^\gamma+{E}_{1}^a={E}_{2}^\gamma+{E}_{2}^a,
\end{eqnarray}
\end{subequations}
where a positive $E$-field means that it is oriented along
${\bf B}_{\rm e}$, i.e., in the positive $y$-direction in the geometric
setup of figure~\ref{fig:interface}.
 Overall we find
\begin{subequations}
\begin{eqnarray}
  {E}_{1}^\gamma&=&+\left({E}_{2}^a-{E}_{1}^a\right)\,
  \frac{\epsilon_2n_1}{\epsilon_1n_2+\epsilon_2n_1}\,,
\\[1ex]
  {E}_{2}^\gamma&=&-\left({E}_{2}^a-{E}_{1}^a\right)\,
  \frac{\epsilon_1n_2}{\epsilon_1n_2+\epsilon_2n_1}\,,
  \\[1ex]
  {H}_{1,2}^\gamma&=&-\left({E}_{2}^a-{E}_{1}^a\right)\,
  \frac{\epsilon_1\epsilon_2}{\epsilon_1n_2+\epsilon_2n_1}\,,
\end{eqnarray}
\end{subequations}
where the direction of the $H$ fields is orthogonal to
${\bf B}_{\rm e}$ and parallel to the interface as shown in
figure~\ref{fig:interface}, i.e., defining the $z$-direction.

In the lossless limit in which $n_1$ and $n_2$ are real,
these values can be taken as real numbers  
because in this case
all $E$ and $H$ fields at the interface are in phase. 
As a function of time $t$ and
distance $x$ from the interface, the physical axion-induced electric 
and propagating electric and magnetic fields are respectively the real
parts of the expressions
\begin{equation}
E_{1,2}^a\,e^{-i\omega t}\,,
\quad
E_{1,2}^\gamma\,e^{-i(\omega t - k_{1,2} x)}\,,
\quad \mathrm{and} \quad
H_{1,2}^\gamma\,e^{-i(\omega t - k_{1,2} x)}\,,
\label{eq:xtdependentfields}
\end{equation}
where
$\omega=m_a$,
$k_{1}=-n_1\omega$ (wave moving in the negative $x$-direction), and
$k_{2}=+n_2\omega$ (wave moving in the positive $x$-direction).

If the media are lossy, the refractive indices $n_1$ and $n_2$ are complex numbers 
with ${\rm Im}(n_{1,2})>0$ 
and the fields at the interface are not in phase. In particular, the axion-induced fields $E_a(t)$ on both sides of the interface are no longer in phase with each other or with the driving galactic dark-matter axion field $a(t)$. In this case, $E_a(t)$ jumps at the interface even if $|n_1|=|n_2|$, but if the media have different loss tangents~(\ref{eq:loss-tangent}).
One can convince oneself that all derivations and results of this section remain unchanged if $\epsilon_{1,2}$ and $n_{1,2}$ are complex numbers. In particular, the axion-induced electric fields $E_{1,2}^a(t)$ remain homogeneous in the two half-spaces separated by the interface, whereas the amplitudes of the propagating EM waves decrease in proportion to $e^{-{\rm Im}(n_{1,2})\omega |x|}$ as a function of distance $|x|$ from the interface.

\subsection{Simple examples}

Let us now consider three simple examples.
For each we will provide the explicit expressions 
of $E_{1,2}^a$, $E_{1,2}^\gamma$, and $H_{1,2}^\gamma$ at the interface.
The corresponding fields away from the interface and as a function of time 
are given by~(\ref{eq:xtdependentfields})
and used to determine the energy flux density.

\subsubsection*{$\bullet$ \em Sudden change of external magnetic field}

As a first simple example we consider a magnetic field region which
ends abruptly at the interface so that ${\bf B}_{\rm e}=0$ in
region~2. While such a situation is not strictly possible, 
we simply assume that ${\bf B}_{\rm e}$ falls off very quickly,
i.e., on a scale much faster than the wavelength of the produced
EM waves. The medium itself is taken to be vacuum
everywhere so that
$\epsilon_1=\epsilon_2=n_1=n_2=1$.
Moreover, as a scale for all fields we use $E_0=g_{a\gamma}{B}_{\rm e}a_0$
as previously defined in equation~(\ref{eq:vecE0}).
If the ${\bf B}_e$ direction is taken to define
the electric field direction, the $y$-direction in figure~\ref{fig:interface},
the axion-induced electric field in a medium with dielectric permittivity $\epsilon$ is
\begin{equation}
E_a=-\frac{E_0}{\epsilon}\,.
\end{equation}
In our example this means ${E}_1^a=-E_0$ and $E_{2}^a=0$ so that
\begin{equation}
   E^\gamma_1=+\half\,{E}_{0}, 
\quad
    E^\gamma_2=-\half\,{E}_{0},
\quad\hbox{and}\quad
    H_{1,2}^\gamma=-\half\,{E}_{0}\,.
\end{equation}
The propagating waves both have half the electric field strength of
the homogeneous axion-induced field-strength in the magnetic field
region.

\subsubsection*{$\bullet$ \em Dielectric interface}

Our main example, however, is a dielectric interface between two media
with $\epsilon_1$ and $\epsilon_2$. We assume negligible magnetic
responses at our frequencies, i.e., $\mu_1=\mu_2=1$ so that
$\epsilon_1=n_1^2$ and $\epsilon_2=n_2^2$ and we express the material
properties in terms of the refractive indices. ${\bf B}_{\rm e}$
is taken to be homogeneous, not jumping at the interface.
The $E_a$-field discontinuity is
$E^a_2-E^a_1=-(\epsilon_2^{-1}-\epsilon_1^{-1})E_0$ and the
interface-values of the wave electric fields are
\begin{equation}\label{eq:dielectricE}
  E_1^\gamma=-\frac{E_0}{n_1}\,\left(\frac{1}{n_2}-\frac{1}{n_1}\right)
\quad\hbox{and}\quad
  E_2^\gamma=+\frac{E_0}{n_2}\,\left(\frac{1}{n_2}-\frac{1}{n_1}\right)\,,
\end{equation}
whereas
\begin{equation}\label{eq:dielectricH}
  H_{1,2}^\gamma=E_0\,\left(\frac{1}{n_2}-\frac{1}{n_1}\right)
\end{equation}
is the magnetic field associated with both waves. Notice that the positive
$H$ direction is the $z$-direction in the convention of figure~\ref{fig:interface}.

The energy flux density (energy transfer per unit area and per unit time or equivalently power per unit area)
of an EM field configuration is given by Poynting's theorem as
\begin{equation}\label{eq:Poynting}
{\bf S}={\bf E}\times{\bf H}\,.
\end{equation}
If these vectors are complex
Fourier components, we must use the real parts of ${\bf E}$ and of
${\bf H}$ to obtain the physical energy flux.
When considering ${\bf S}$ in lossless media,
the cycle average introduces a factor $1/2$
so that our two waves carry the cycle-averaged power per unit surface
\begin{equation}\label{eq:Poynting-2}
\bar S_i^{\gamma}=\mp\frac{E_0^2}{2n_i}\,\left(\frac{1}{n_2}-\frac{1}{n_1}\right)^2\,,
\end{equation}
for $i=1$ (upper sign) or 2 (lower sign). A positive sign means energy flowing in the
positive $x$-direction, i.e., the energy flows away in opposite directions from the
interface. In absolute terms, the medium with the smaller refractive index carries
the larger energy in produced EM waves.

\subsubsection*{$\bullet$ \em Perfect mirror}

As an extreme case, we may assume that medium~2 is vacuum ($n_2=1$)
whereas medium~1 has a huge dielectric constant so that
$|n_1|\to\infty$. In this case $E_1^\gamma=0$
(and $H_{1,2}^\gamma=0$),
whereas the wave emitted
into vacuum is $E_2^\gamma=E_0$. Notice that $E_2^a=-E_0$, so
the electric field of the wave
simply has to cancel the ambient axion-induced field such that the
total electric field vanishes at the mirror surface.

A mirror could be a purely dielectric medium with a very large real value of $\epsilon$, 
but in practice more realistically is a metallic surface with a large real-valued conductivity $\sigma\to\infty$
so that $\epsilon\sim 1+i\sigma/\omega$.
Put another way, the value of the mirror's loss tangent does not affect the production of axion-induced
EM waves from its surface; in our context the only defining property of a perfect mirror is $|n_1|\to\infty$.

\subsection{Energy flux}

In our example for a dielectric interface, we have stated the power per unit area
carried by the produced EM waves based on the Poynting vector ${\bf S}$
which we applied to the propagating solution alone. The result is somewhat
paradoxical because power seems to pour from the
interface in both directions, carried by the EM waves
which are generated by the presence of the interface. On the other hand,
the Poynting vector in equation~(\ref{eq:Poynting}) is constructed
from ${\bf E}$ and ${\bf H}$ which are both parallel to the interface
and so continuous. Therefore, the Poynting vector itself must
be continuous as well, so the interface cannot be a source of energy.

The apparent paradox is resolved if we note that the energy flux given
in terms of the fields of the generated EM wave is
meaningful only at a large distance.  If we assume that ${\bf
  B}_{\rm e}$ adiabatically decreases as a function of distance, far
away the only EM fields are those of the propagating
waves.  However, in the ``near-field region'' around the interface,
this interpretation is not complete. The Poynting vector involves all
EM fields, and especially the axion-induced field
${\bf E}_a$, whereas the external field ${\bf B}_{\rm e}$ does not appear
in any of the cross products because all electric fields are parallel
to ${\bf B}_{\rm e}$. Overall we thus have
${\bf S}=({\bf E}_a+{\bf E}_\gamma)\times{\bf H}_\gamma$,
correcting what looked like an
inconsistency earlier. In other words, we cannot think of the
``background axion-induced field'' and the propagating waves as
incoherent phenomena. Rather, they are phase-locked as the
oscillating axion field drives them all. Both ${\bf E}_a(t)$ and
${\bf E}_\gamma(t)$ are part of the overall solution of Maxwell's
equations.

\begin{figure}[t]
\centering
\includegraphics[width=10cm]{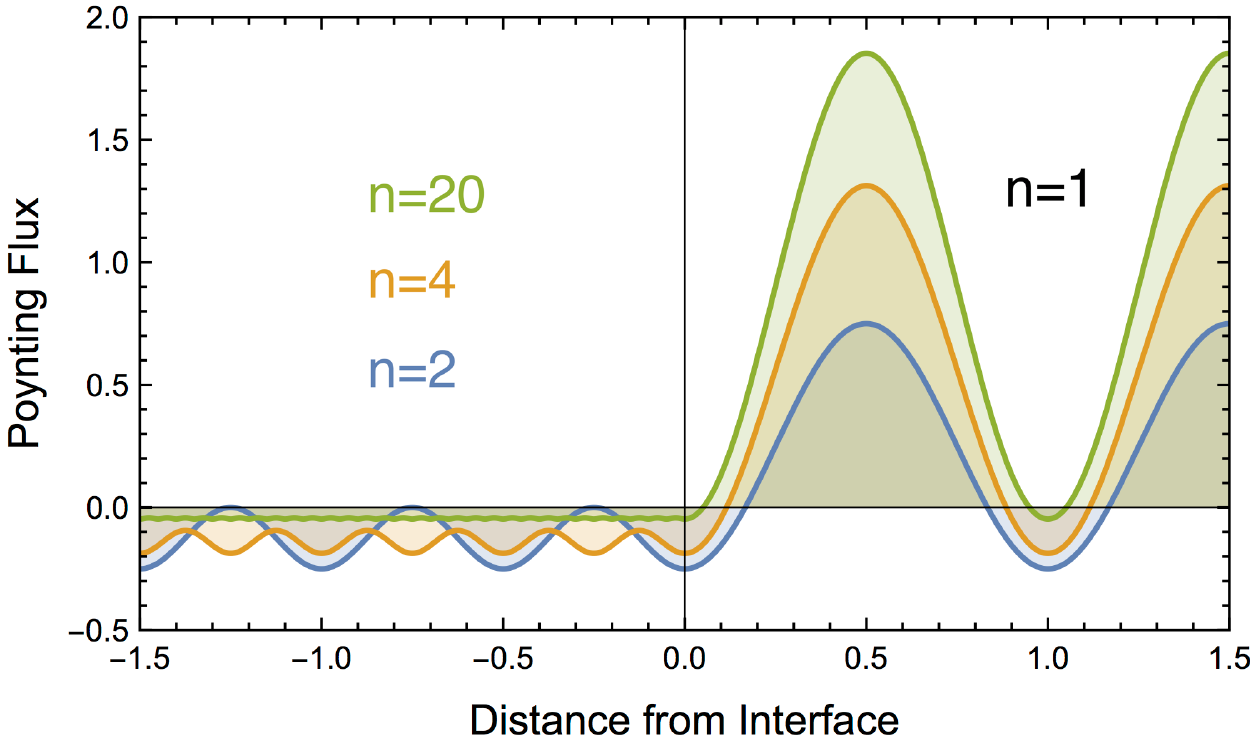}
\caption{Energy flux density (cycle-averaged value of Poynting vector) in the $x$-direction~(\ref{eq:poynting13})
  in units of $E_0^2/2$
  for the setup of figure~\ref{fig:interface} 
  with region~1 (region~2) at negative (positive) distance values, except that here we
  use $n_1=2$, 4, and 20 (blue, orange, and green), always keeping $n_2=1$.
\label{fig:Poynting}}
\end{figure}

Considering explicitly the overall solution of Maxwell's equations
in the presence of an interface as given in equation~(\ref{eq:xtdependentfields}),
the cycle-averaged energy flux density is found to be
\begin{equation}
\bar S_i=\half\bigl[E_i^\gamma+E_i^a\cos(k_i x)\bigr]\,H_i^\gamma\,,
\end{equation}
where $i=1$ or 2 and we assume real $n$. Notice that a positive value of $\bar S$ means
energy flowing in the positive $x$-direction in the arrangement
of figure~\ref{fig:interface}. For our main example, a dielectric
interface where $B_{\rm e}$ is continuous and $\mu=1$ everywhere
(no magnetic response), the explicit solutions for the fields
at the interface were given in equations~(\ref{eq:dielectricE})
and~(\ref{eq:dielectricH}). In this case, the explicit cycle-averaged energy
flux density is
\begin{equation}
\bar S_i=-\frac{E_0^2}{2n_i}
\left(\frac{1}{n_2}-\frac{1}{n_1}\right)
\left[\pm\left(\frac{1}{n_2}-\frac{1}{n_1}\right)
+\frac{\cos(n_i\omega x)}{n_i}\right]\,,\label{eq:poynting13}
\end{equation}
where $i=1$ (upper sign) or 2 (lower sign). The first term reproduces
our previous result of equation~(\ref{eq:Poynting-2})
for the energy carried by the propagating EM waves alone. The second term
arises from the interference of $H_\gamma$ with the homogeneous
but time-varying electric field $E_a$. The Poynting fluxes on
the two sides of the interface are explicitly
\begin{subequations}
\begin{eqnarray}
\bar S_1&=&-\frac{E_0^2}{2n_1^2}\left(\frac{1}{n_2}-\frac{1}{n_1}\right)
\left[\frac{n_1}{n_2}-2\sin^2\left(\frac{n_1\omega x}{2}\right)\right]\,,
\\[1ex]
\bar S_2&=&+\frac{E_0^2}{2n_2^2}\left(\frac{1}{n_2}-\frac{1}{n_1}\right)
\left[2\sin^2\left(\frac{n_2\omega x}{2}\right)-\frac{n_2}{n_1}\right]\,.
\end{eqnarray}
\end{subequations}
Let us assume that $n_1\gg n_2$, having in mind that region~1 is a
dielectric and region~2 vacuum or air. We immediately
glean from these equations that the term in square
brackets of $\bar S_1$ is positive for any $x$ if $n_1>2 n_2$,
meaning that for any $x<0$ energy flows in the negative $x$-direction.
On the
other hand, in $\bar S_2$ this term has a positive
spatial average, but it varies between $-n_2/n_1<0$ and $2-n_2/n_1>0$,
i.e., between positive and negative values.
For $n_2=1$ and several examples of $n_1>1$ we show the Poynting flux
in figure~\ref{fig:Poynting}.

Therefore, in the medium with smaller $\epsilon$, there are surfaces
parallel to the interface through which no energy flux passes. Of
course, as $B_{\rm e}$ falls off adiabatically in the $x$-direction,
the interference term disappears and we are left with an energy flux
carried by the propagating EM wave alone. The purpose of the ``layered
dielectric haloscope" to be discussed in the next section is, of course, to boost
$E_\gamma$ to much larger values by multiple dielectric surfaces. In
this case the interference term with $E_a$ is small and we can think
of the produced EM wave independently from $E_a$ in any
practically relevant sense.

However, this exercise shows that the EM power transferred from the
oscillating axion field to a propagating EM wave does not pour from
the interface into space in both directions.  Rather, it is
transferred within the magnetized volume in a spread-out region,
commensurate with the picture that neither the axion nor the
propagating EM wave can be exactly localized.

\subsection{Detection}
\label{detection}

To get a first grasp of the detection requirements we consider the EM radiation
emitted by a perfect mirror ($n_1\to \infty$) into vacuum ($n_2=1$).
From equation~\eqref{eq:Poynting-2} we find for the cycle-averaged energy flux 
density (power per unit surface $A$) in the $x$-direction
\begin{equation}
\frac{P_\gamma}{A}
=\bar S_2^\gamma
=\frac{1}{2}\,\overline{\left[{\rm Re}\({\bf E}^\gamma_2\)\times {\rm Re}\({\bf H}^\gamma_2\)\right]}_x
=\frac{E_0^2}{2}=2.2\times 10^{-27}\,\frac{\rm W}{\rm m^2}
\(\frac{B_{\rm e}}{10~{\rm T}}\)^2 C_{a\gamma}^2 f_{\rm DM}\,.
\end{equation}
The corresponding photon flux, 
\begin{equation}
F_\gamma=\frac{P_\gamma}{A\,\omega}
=\frac{12}{{\rm m}^{2}~{\rm day}}\,
\(\frac{100~\mu{\rm eV}}{m_a}\)
\(\frac{B_{\rm e}}{10~{\rm T}}\)^2 C_{a\gamma}^2 f_{\rm DM} \, , 
\end{equation}
is extremely tiny.

To illustrate the detection challenge we recall that virialised galactic axions have
a velocity dispersion of around $v_a\sim 10^{-3}$. The kinetic energy of a nonrelativistic axion is
$m_a v_a^2/2$ so that 
these axions have an energy spread around $\omega_a=m_a$ of
$\Delta\omega_a\sim m_a v_a^2/2\sim 10^{-6}\,m_a$, 
corresponding to a signal bandwidth given by the frequency range
$\Delta\nu_a=\Delta\omega_a/2\pi$. If
the power in this frequency band
is detected with a linear amplifier with system noise $T_{\rm sys}$,
the Gaussian noise power fluctuations after a measurement time $\Delta t$
have a standard deviation of 
$T_{\rm sys}\Delta\nu_a/\sqrt{\Delta\nu_a\,\Delta t}$. 
The signal-to-noise ratio in this band is given by the Dicke radiometer equation~\cite{Dicke}

\begin{equation}
\frac{\rm S}{\rm N} 
= \frac{P_\gamma}{T_{\rm sys}}\sqrt{\frac{\Delta t}{\Delta \nu_a}} 
=1.0\times 10^{-4}\(\frac{A}{1\,\rm m^2}\) \sqrt{\frac{100\,\mu{\rm eV}}{m_a}}\,
\sqrt{\frac{\Delta t}{\rm week}}\(\frac{8\,{\rm K}}{T_{\rm sys}}\)
\(\frac{{B}_{\rm e}}{10\,{\rm T}}\)^{\!\!2} C_{a\gamma}^2 f_{\rm DM}\,.
\end{equation}
For plausible parameters $B_{\rm e}\sim10~{\rm T}$ and $A\sim 1~{\rm m}^2$ this is
far too small for a realistic experiment as noted earlier~\cite{Jaeckel:2013eha}.
To be able to detect axion dark matter in a reasonable time scale, the signal power
must be enhanced by a factor of $10^4$ or more.

The main point of our paper is that the signal can be amplified by using multiple
interfaces. We will define a boost factor $\beta$ for the electric-field amplitude
of the produced EM wave, i.e., the power is amplified by a factor $\beta^2$,
\begin{equation}
\frac{P_\gamma}{A}=\frac{E_0^2}{2}
\quad\to\quad
\frac{P_\gamma}{A}=\frac{\beta^2 E_0^2}{2}\,\eta
\label{eq:Pbeta2eta}
\end{equation}
where also $\eta$ is introduced to account for the efficiency of the detector. 
If we require a certain signal-to-noise ratio S/N in a given
search channel $\Delta\nu_a$, the scanning speed per channel~is
\begin{equation}
\frac{\Delta t}{1.3\,\rm days}\sim
\(\frac{{\rm S}/{\rm N}}{5}\)^2\(\frac{400}{\beta}\)^4
\(\frac{1\,\rm m^2}{A}\)^2\(\frac{m_a}{100~\mu{\rm eV}}\)
\(\frac{T_{\rm sys}}{8\,\rm K}\)^2
\(\frac{10\,\rm T}{B_{\rm e}}\)^4 \(\frac{0.8}{\eta}\)^2 C_{a\gamma}^{-4} f_{\rm DM}^{-2} \,. \label{eq:scan}
\end{equation}
Of course, the main problem is that we do not know the axion mass 
and thus have to scan over a range of search masses that is as broad as possible.
In a given run, we are not limited to covering a single channel 
defined by the natural line width 
$\Delta\nu_a$
of the virialised axion field.
Instead we may achieve a large boost factor over a broader frequency range
and in this way achieve a reasonable overall scanning speed. 
The option of a broadband search is one of the attractions 
of the dielectric haloscope approach.

\section{Boost from multiple layers}
\label{transfermatrixformalism}

The main point of our paper is that one can gain a boost relative to
the axion-induced EM wave produced by a mirror in a strong $B$-field
by augmenting it with a series of parallel interfaces between
different media. The outgoing wave is a coherent superposition of the
amplitudes produced at all interfaces, including the effects of
transmission and reflection at each interface. Here we develop the
required theoretical tools in terms of the transfer matrix formalism
to calculate both the emerging waves as well as the transmission and
reflection coefficients of the full ensemble.

\subsection{General setup of the layered system}

\begin{figure}
\centering
\includegraphics[width=14cm]{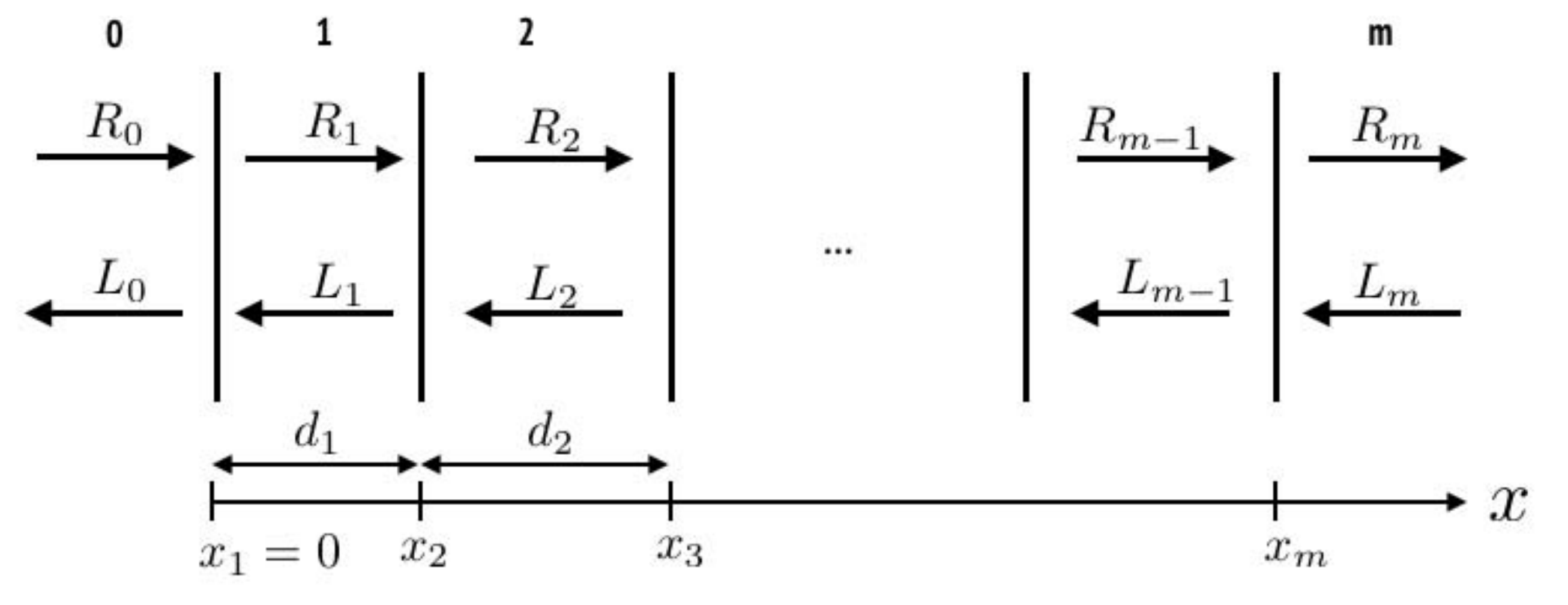}
\caption{Several dielectric regions with labels $r=0,1,\ldots,m$. The amplitudes $L_r$ and $R_r$ denote
the electric-field amplitudes of left and right moving EM waves in each homogeneous region.}
\label{scheme}
\end{figure}

We consider the idealised case of a sequence of plane parallel regions
labelled $0,1, \ldots, m$, with different indices of refraction
$n_0,\ldots,n_m$ as sketched in figure~\ref{scheme}.  Here 0 and $m$
are semi-infinite regions, which will be typically taken as either a
perfect reflector or vacuum. In practice, each pair of interfaces will
delimit a dielectric layer or a vacuum gap, but for the moment we keep
our setup completely general. We align the interfaces along the
$y$-$z$-directions, using the same coordinate system that was shown in
figure~\ref{fig:interface}. The boundaries between media are located
along the $x$ direction at $x_1,\ldots,x_m$. We arrange an external
magnetic field along the $y$-direction, i.e., parallel to the
boundaries in the same way as in figure~\ref{fig:interface}, although
we allow it to change in magnitude and sign in each region. For
example, if the regions have different DC permeabilities,
${\bf B}_{{\rm e},\parallel}$ can jump because ${\bf H}_{{\rm e},\parallel}$
is continuous. Likewise, we can model a variation
of $B_{\rm e}$ as a function of $x$, notably its finite extension, by
treating it as a sequence of sufficiently thin $B_{\rm e}$-field
layers.

In each magnetised region, $r$, the solution of Maxwell's equations in the presence of the external axion current ${\bf J}^a_r(t) = g_{a\gamma} {\bf B}_{{\rm e},r} \dot a(t)$ consists of a homogeneous electric field ${\bf E}_r^a(t)$, as discussed earlier, and two EM waves, left and right moving.
The former are fixed by the dielectric properties and the $B$-field in each region and are in general discontinuous across the boundaries.
The latter must be determined to have continuous ${\bf E}_{\parallel}$ and ${\bf H}_{\parallel}$ across each interface in analogy to the single interface in the previous section. However, previously we only considered EM waves moving away from the interface, whereas now we also need to include waves moving toward each interface from both sides. Besides photon production by axions this also includes the effects of transmission and reflection at each interface.

In analogy to the definition in equation~(\ref{eq:vecE0}) we use the notation $E_0=g_{a\gamma} B_{\rm e,max}a_0$ as a scale for the axion-induced electric field, where now $B_{\rm e,max}$ is the largest value of the external field. Moreover, we assume again that the axion field is perfectly homogeneous and at rest. All fields bear a common time variation $e^{-i\omega t}$ with $\omega=m_a$ which henceforth is implicit everywhere. We then write the axion-induced electric field, homogeneous in each region $r$, in the form
\begin{equation}\label{eq:Ardef}
E_a^r=-A_r E_0
\quad\hbox{where}\quad
A_r=\frac{1}{\epsilon_r}\,\frac{{B}_{{\rm e},r}}{B_{\rm e,max}}\,.
\end{equation}
The $x$-dependent electric and magnetic fields of the propagating EM waves are
\begin{subequations}
\begin{eqnarray}
E^{R}_r(x) =  R_r e^{+i \omega n_r \Delta x}
&\quad\hbox{and}\quad&
H^{R}_r(x) = +\frac{\epsilon_r}{n_r} R_r e^{+i \omega n_r \Delta x},
\\[1ex]
E^{L}_r(x) = L_r e^{-i \omega n_r \Delta x}
&\quad\hbox{and}\quad&
H^{L}_r(x) = - \frac{\epsilon_r}{n_r} L_r e^{-i \omega n_r \Delta x},
\end{eqnarray}
\end{subequations}
where $\Delta x = x-x_r$ measures $x$ relative to the left boundary of every region $r$.
Notice that the electric-field amplitudes $L_r$ and $R_r$ of the left and right moving EM waves are defined at the left boundary of every region, except for $L_0$ and $R_0$ which denote the amplitudes in region $r=0$ on the left of the interface at $x_1$. Recall also that the field orientations are analogous to
figure~\ref{fig:interface}.

In each region, the total electric and magnetic fields are the superposition of these components, leading to
\begin{subequations}
\begin{eqnarray}
E_r^{{\rm tot}}(x)&=&-E_0A_r+R_re^{+i \omega n_r \Delta x}+L_re^{-i \omega n_r \Delta x},
\\[1ex]
H_r^{{\rm tot}}(x)&=&\frac{\epsilon_r}{n_r} \(R_r e^{+i \omega n_r \Delta x} - L_r e^{-i \omega n_r \Delta x}\).
\end{eqnarray}
\end{subequations}
The continuity of ${\bf E}_\parallel$ and ${\bf H}_\parallel$ at $x_{r+1}$ implies
$E_{r}(x_{r+1})={E}_{r+1}(x_{r+1})$ and $H_{r}(x_{r+1})=H_{r+1}(x_{r+1})$. These conditions are explicitly
\begin{subequations}\label{eq:boundary-conditions}
\begin{eqnarray}
- E_0A_r  +R_r e^{i \delta_r} + L_r e^{-i \delta_r} &=& - E_0A_{r+1}+R_{r+1} + L_{r+1},
\\[1ex]
\frac{\epsilon_r}{n_r}\Bigl(R_re^{i \delta_r} - L_re^{-i \delta_r}\Bigr)
&=& \frac{\epsilon_{r+1}}{n_{r+1}} \Bigl(R_{r+1} -L_{r+1} \Bigr).
\end{eqnarray}
\end{subequations}
We have introduced the optical thickness
\begin{equation}
\delta_r=\omega n_r\(x_{r+1}-x_r\)
\label{eq:deltar}
\end{equation}
for the phase difference between the boundaries of region $r$. Notice that the media can be lossy, leading to waves which are damped along their direction of propagation, an effect consistently included in these formulas. In this case $\delta_r$ is a complex number. Notice also that $\delta_r$ is not to be confused with the loss tangent of the medium in region $r$.

The continuity conditions at all $m$ interfaces have provided us with $2m$ linear equations~(\ref{eq:boundary-conditions}) for the $2(m+1)$ unknown field amplitudes $L_r$ and $R_r$.
To solve them, we need to specify the two incoming waves $R_0$ and $L_m$. When we ask only for the axion-produced outgoing waves $L_0$ and $R_m$ these must be set to zero. When we ask for the global transmission and/or reflection coefficients of the overall ensemble we can specify non-vanishing incoming amplitudes.

\subsection{Transfer matrix formalism}

To solve equations~(\ref{eq:boundary-conditions}) we use the transfer matrix formalism \cite{Sanchez-Soto} which simply relates, at every interface, the amplitudes in region $r+1$ to those in region $r$. Unlike with standard transfer matrices, we will include the axion-produced amplitudes at each interface. For simplicity we now specialize to non-permeable media, i.e., $\mu_r=1$ in every region. In this case we have only one medium response parameter $n_r=\sqrt{\epsilon_r}$ in every region. This choice is motivated for practical reasons: dielectric disks with large $\mu$ would be very difficult to move accurately in strong magnetic fields. The pair of equations~(\ref{eq:boundary-conditions}) is equivalent to
\begin{equation}
\vv{R_{r+1}}{L_{r+1}}=
\GG_{r} \PP_r \vv{R_r}{L_r} + E_0\,\Ss_{r}\vv{1}{1}\,,
\end{equation}
where sans-serif letters denote the $2{\times}2$ matrices
\begin{subequations}
\begin{eqnarray}
\GG_r &=& \frac{1}{2n_{r+1}}\(\begin{array}{cc}
n_{r+1}{+}n_r &~~ n_{r+1}{-}n_r\\
n_{r+1}{-}n_r &~~ n_{r+1}{+}n_r  \end{array}\)\,,
\\[2ex]
\PP_r &=& \(\begin{array}{cc}
e^{+i \delta_r} & 0 \\
0 & e^{-i \delta_r}  \end{array}\)\,,
\\[2ex]
\Ss_r &=& \frac{A_{r+1}-A_{r}}{2}\,\(\begin{array}{cc}
1 &~ 0 \\
0 &~ 1  \end{array}\)\label{Sdef}\,.
\end{eqnarray}
\end{subequations}
Note that $\GG_r$ is the regular EM transfer matrix for the interface at $x_{r+1}$ between regions $r$ and $r+1$, whereas $\PP_r$ gives us the phase accrued by propagation in region $r$. The diagonal matrix
$\Ss_r$ describes the source of EM waves induced by axions at the interface between regions $r$ and $r+1$.

For multiple layers we iterate this equation to relate the amplitudes of the $R$ and $L$ waves in the external media, 0 and $m$:
\begin{eqnarray}
\vv{R_m}{L_m}
&=& \GG_{m-1} \PP_{m-1} \vv{R_{m-1}}{L_{m-1}} + E_0\Ss_{m-1}\vv{1}{1}
\nonumber\\[1ex]
&=& \GG_{m-1} \PP_{m-1}
\biggl[\GG_{m-2} \PP_{m-2} \vv{R_{m-2}}{L_{m-2}} + E_0\Ss_{m-2}\vv{1}{1}\biggr] + E_0\Ss_{m-1}\vv{1}{1}
\nonumber\\[1ex]
&=&\ldots
\nonumber\\[1ex]
&=& \TT^m_0 \vv{R_{0}}{L_{0}}
+E_0\sum_{s=1}^m  \TT^{m}_s \Ss_{s-1}\vv{1}{1}\,. \label{transfereq}
\end{eqnarray}
Here we have defined a transfer matrix from region $b$ to $a$ which for $b<a$ is of the form
\begin{equation}
\TT^a_b = \GG_{a-1}\PP_{a-1}\GG_{a-2}\PP_{a-2}\,\ldots\, \GG_{b+1}\PP_{b+1}\GG_{b}\PP_{b}\label{eq:emtransfer}
\end{equation}
and $\TT^a_a=\mathbb{1}$. Moreover, $\PP_0=\mathbb{1}$ because the left-most external amplitudes $L_0$ and $R_0$ are the amplitudes on the left of the interface at $x_1$, whereas all other amplitudes are defined on the right of each interface.

The relation~(\ref{transfereq}) has been split into two parts. The first accounts for the usual transfer matrix between in and out EM waves in a 1D system. The second part is the sum over all axion-induced source terms $E_0 \Ss_r$, one for each layer, and then propagated to the right external region of the ensemble at $x_m$. We write equation~(\ref{transfereq}) in even more compact form as
\begin{equation}
\vv{R_m}{L_m} = \TT \vv{R_{0}}{L_{0}} + E_0 \MM \vv{1}{1},
\label{transfereq2}
\end{equation}
where we have defined
\begin{equation}\label{eq:T-M-matrices}
\TT=\TT_0^m
\quad\hbox{and}\quad
\MM=\sum_{s=1}^m  \TT^{m}_s \Ss_{s-1}.
\end{equation}
We may also solve for the left-most amplitudes in region $0$ in terms of those in region $m$ by multiplying equation~(\ref{transfereq2}) from the left with $\TT^{-1}$. The inverse matrix involves the determinant of $\TT$ for which one finds
\begin{equation}\label{eq:DetT}
{\rm Det}[\TT]=\frac{n_0}{n_m}\,.
\end{equation}
This is because the determinant of a product of matrices is the product of the determinants and the determinants are
${\rm Det}[\PP_r]=1$ and ${\rm Det}[\GG_r]=n_r/n_{r+1}$.
In the special case when one external medium is a perfect mirror, e.g., $|n_0|\to\infty$, one must be careful. The correct results can be obtained by taking the limit of a large but finite $n$.

\subsection{Transmission and reflection coefficients}

Ignoring for the moment such special cases, we consider two instructive limits of these relations. In the absence of axion dark matter we have $E_0=0$ and the transfer matrix simply relates incoming and outgoing EM waves. For the usual transmission and reflection amplitude coefficients we find
\begin{subequations}\label{eq:T-R-Coefficients}
\begin{eqnarray}
{\cal T}_L &=& \frac{R_m}{R_0}\Big{|}_{L_m=0} = +\frac{\rm Det[\TT]}{\TT[2,2]}\,, \\[1ex]
{\cal T}_R &=& \frac{L_0}{L_m}\Big{|}_{R_0=0}  =  +\frac{{1} }{\TT[2,2]}\,,\label{eq:transmisivityb}  \\[1ex]
{\cal R}_L &=&  \frac{L_0}{R_0}\Big{|}_{L_m=0}  = -\frac{\TT[2,1]}{\TT[2,2]}\,, \\[1ex]
{\cal R}_R &=&  \frac{R_m}{L_m}\Big{|}_{R_0=0} = +\frac{\TT[1,2]}{\TT[2,2]}\,,
\end{eqnarray}
\end{subequations}
where $\TT[i,j]$ denotes the matrix component in the $i$th row and $j$th column. In the presence of axion-induced photon production ($E_0\not=0$) we can still use these definitions for the transmissivity and reflectivity.
In practice one would use reflection and transmission to test an experimental setup: when non-zero incoming photon waves are present they would typically be much larger than any contribution from axions.

\subsection{Boost amplitude and boost factor}
\label{sec:Boost-Amplitude}

The second limit is when there are no incoming waves ($R_0=0$
and $L_m=0$) and all outgoing waves arise from axion-induced
production. The outgoing amplitudes are
\begin{subequations}\label{eq:outgoing-waves}
  \begin{eqnarray}
L_0  &=& -E_0\,\frac{\MM[2,1]+\MM[2,2]}{\TT[2,2]}\,,   \\
R_m &=& E_0\,\(\MM[1,1]+\MM[1,2] -\frac{\MM[2,1]+\MM[2,2]}{\TT[2,2]}\,\TT[1,2]\).
 \end{eqnarray}
 \end{subequations}
As motivated in section~\ref{detection}, we may express the outgoing
amplitudes in terms of a boost amplitude $\B$ which gives us the
outgoing amplitude in units of the one produced by a single perfect
mirror. For a mirror the outgoing wave simply has the electric field
amplitude $E_0$, so the boost amplitudes are simply the outgoing
electric field amplitudes of equation~(\ref{eq:outgoing-waves}) in
units of $E_0$. Our multi-layer system therefore produces
\begin{equation}\label{eq:boost-factor}
\hbox{Boost amplitude:}\qquad
\B_L = \frac{L_0}{E_0}
\quad\hbox{and}\quad
\B_R = \frac{R_m}{E_0}\,.
\end{equation}
Often we will not be interested in the phase of the outgoing wave
relative to the original axion field. In this case we will use the
term
 \begin{equation}\label{eq:boost-factor-2}
\hbox{Boost factor:}\qquad
\beta=|\B|
\end{equation}
to denote the modulus of the boost amplitude.  Usually we will
not distinguish between $\B_L$ and $\B_R$
without danger of confusion because our examples are either
left-right-symmetric or have a perfect mirror on one end. Note that in the case of a perfect mirror, one can either use the above equations in a limiting sense, or simply use (for $n_0=\infty)$
\be
\mathcal B_R=\MM[1,1]+\MM[1,2].
\ee
The axion-produced EM power is proportional to~$|\B|^2=\beta^2$, cf.~\eqref{eq:Pbeta2eta}.
For arrangements which are open on both ends (no
mirror) and left-right-symmetric, the boost amplitude and boost factor
applies only to one of the emerging waves. The produced power refers
to the one emerging from one side only.

The boost amplitude is our main quantity of interest because it
describes the response of the haloscope to the axion
field. By construction, $\B$ is a function of the refractive
indices $n_0,\ldots,n_m$, the distances between the interfaces
$d_1,\ldots,d_{m-1}$, and the chosen frequency $\omega$. The latter
quantities together provide us with the phase depths
$\delta_r=n_r\omega d_r$ with $r=1,\ldots,m-1$. It is really the phase
depths and refractive indices which determine $\B$.

In practice, a given haloscope will consist of a series of dielectric
plates with fixed properties that can be shifted relative to each
other. Keeping all refractive indices fixed, we may think of
$\B({\bf d},\omega)$
as a function of an $(m{-}1)$-dimensional ``configuration vector''
\hbox{${\bf d}=(d_1,\ldots,d_{m-1})$} of distances and of the frequency
$\omega$. If all $n_r$ are real numbers (no losses),
it is easy to show (appendix~\ref{sec:area-law}) that
\begin{equation}\label{eq:general-area-law}
  \langle |\B|^2\rangle_{\bf d}=\langle |\B|^2\rangle_{\omega}\,.
\end{equation}
Here the average $\langle\ldots\rangle_{\bf d}$ is the average over
all configurations, whereas $\langle\ldots\rangle_{\omega}$ the
average over all frequencies. In particular, the configuration average
does not depend on frequency and the frequency average not on
configuration.

In the latter interpretation, this result means that the power emitted by the
haloscope, averaged over a flat spectrum of $\omega$, does not depend
on configuration. In particular, if $|\B|^2$ shows a resonance as a
function of $\omega$, small modifications of the disk spacings
will change the resonance structure, perhaps shift it a bit, but
leave the ``area'' $\int d\omega\, |\B|^2$ unchanged. We refer to this
rule as the Area Law. For a simple resonance described by
a Lorentzian response, this result means that the integral over the
Lorentzian does not depend on the resonance width $\Gamma$ as
expected.

\subsection{Overlap integral formalism}

Here we have derived a formalism using transfer matrices, 
but many readers are probably more familiar with the overlap integral formalism 
introduced by Sikivie for studying cavity haloscopes~\cite{Sikivie:1985yu}.
The power extracted from the cavity is~\cite{Sikivie:1985yu}
\be
P_{\rm cav} = \kappa {\cal G} V \frac{Q}{m_a} \rho_{a} g_{a\gamma}^2B_{\rm e}^2, \label{eq:sikivieoverlap1}
\ee
where $V$ is the volume, $Q$ the loaded quality factor and $\kappa$ the cavity coupling factor
(cf.~appendix~\ref{sec:3Dcavity} for more details).
The geometry factor $\cal G$ is defined as,
\be
{\cal G} = \frac{\(\int dV {\bf E}_\alpha\cdot {\bf B}_{\rm e}\)^2}{V B_{\rm{e}}^2\int dV {\bf E}_\alpha^2},
\ee
where ${\bf E}_\alpha$ is the $E$-field of a given mode $\alpha$ of the cavity.

The systems studied in this paper are explicitly open ones, and in general are non-resonant. Thus it is not immediately obvious what could constitute a ``cavity mode" or ``quality factor". However, as we show in appendix~\ref{Sikiviecomp} one can find analogous quantities for the power emitted from the lhs and rhs of the device,
\begin{subequations}
\begin{eqnarray}
P_{L,R}&\simeq &  {\cal G}_{{\rm d},L,R}V\frac{Q_{{\rm d},L,R}}{m_a}\rho_a g_{a\gamma}^2B_{\rm e}^2, \label{eq:dielectricoverlap}\\
{\cal G}_{{\rm d},L,R}&=&\frac{\left|\int dx E_{L,R} B_{{\rm e}}\right
  |^2}{L B^2_{{\rm e}} \int dx |E_{\rm in}|^2}\,,
\\
Q_{{\rm d},L,R}&=&\frac{1}{4}\frac{\int dx |E_{L,R}|^2}{E_0^2/m_a},
\end{eqnarray}
\end{subequations}
where $E_{L,R}$ is the complex $E$-field given by shining a wave of magnitude $|E_{\rm in}|$ in from the $0,m$th layer (lhs, rhs) in the absence of axions.
In this expression, $L$ is the length of the haloscope with $V=AL$. As it factors out in equation~\eqref{eq:dielectricoverlap}, the choice of $L$ is irrelevant.
Note that $Q_{{\rm d},L,R}$ are not true quality factors. The cavity coupling $\kappa$ is built into our formalism without requiring an explicit term.

Unfortunately, while equation~\eqref{eq:dielectricoverlap} generalises \eqref{eq:sikivieoverlap1} in the 1D limit to non-resonant and open setups, the physical interpretation of the $E$-field one integrates over does not generalise. In the Sikivie picture, one thinks of the axion field exciting ${\bf E}_\alpha$, but for dielectric haloscopes this is generally not the case---the $E_{L,R}$-fields seem to be simply mathematical tools to calculate the electric fields produced due to axions. As transfer matrices are able to encode all the information of this integral without approximation (including two terms which, while small, are generically missed by this integral, as shown in appendix~\ref{Generalised_Overlap_Integral_Formalism}) and are computationally much more efficient, we will use them throughout this paper.

\section{Generic examples}
\label{Genericexamples}
The EM radiation emitted by a single interface can be boosted in two
generic ways.  One is the coherent superposition of the radiation from
many surfaces, the other is by creating a resonance between two
reflecting surfaces. A realistic layered haloscope takes advantage of
both effects in what can be a complicated arrangement of dielectric
disks. In this section we consider several generic examples which
illustrate these effects: a single dielectric disk, a resonator or
``cavity'' consisting of a dielectric disk at some distance in front
of a perfect mirror, and a series of equally spaced dielectric disks.

\subsection{Single dielectric disk}
\label{slab}

\begin{figure}[t]
\centering
\includegraphics[width=5.5cm]{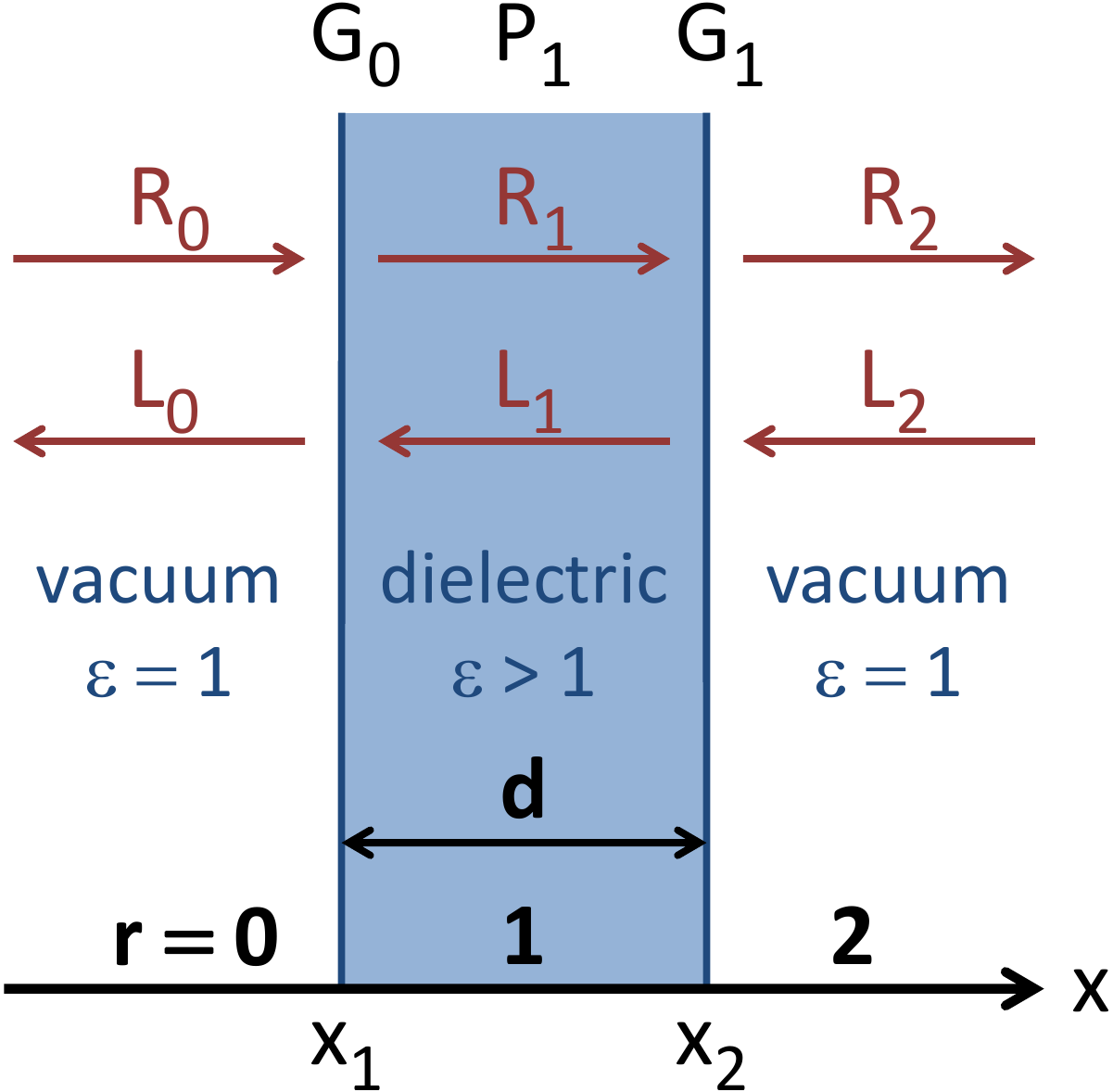}
\caption{Schematic picture of a single dielectric disk.}
\label{fig:scheme-single-disk}
\end{figure}

The first generic example will be the basic building block of a
dielectric haloscope---a single dielectric disk as sketched in
figure~\ref{fig:scheme-single-disk}. Specifically we assume a disk of
thickness $d$, infinite in the transverse direction, with index of
refraction $n\equiv n_1>1$, surrounded by vacuum with $n_0=n_2=1$.
Here and in all following examples we assume that the external field
$B_{\rm e}$ is the same in all regions. The phase accrued by an EM
wave with frequency $\omega$ traversing this disk is
$\delta\equiv\delta_1=n\omega d$.

The transfer and axion source matrices defined in
equation~(\ref{eq:T-M-matrices}) are for a two-interface arrangement
$\TT=\GG_1 \PP_1 \GG_0$ and $\MM= \Ss_1 + \GG_1 \PP_{1} \Ss_0$,
respectively. Here, as in subsequent multi-disk systems, we are only dealing with a few types of interfaces, so
there are significant simplifications to the notation. The matrix $\GG_0$ takes us
from vacuum ($v$) to a dielectric ($\epsilon$), so we may denote it
as $\GG_{\epsilon v}$, whereas $\GG_1$ does the opposite and we call
it $\GG_{v\epsilon}$. Notice that $\GG_{v\epsilon}\GG_{\epsilon v}=
\mathbb{1}$, i.e., $\GG_{v\epsilon}=\GG_{\epsilon v}^{-1}$. Explicitly,
we find with $n=\sqrt{\epsilon}$
\begin{equation}
\GG_{\epsilon v}=\frac{1}{2n}
\(\begin{array}{cc}n{+}1 &~ n{-}1\\ n{-}1 &~ n{+}1\end{array}\)
\quad\hbox{and}\quad
\GG_{v\epsilon}=\frac{1}{2}
\(\begin{array}{cc}n{+}1 &~ 1{-}n\\ 1{-}n &~ n{+}1\end{array}\)\,,
\end{equation}
which are indeed the inverse of each other. Likewise,
the source matrix $\Ss_0$ applies to an interface with vacuum
on the left and a dielectric $\epsilon$ on the right, so we call it
$\Ss_{\epsilon v}$ and the opposite for $\Ss_1$ which we call
$\Ss_{v\epsilon}$. The two matrices are related by $\Ss_{v\epsilon}=-\Ss_{\epsilon v}$
and are explicitly
\begin{equation}
\Ss_{\epsilon v}=\frac{1-n^2}{2n^2}\,\mathbb{1}
\quad\hbox{and}\quad
\Ss_{v\epsilon}=\frac{n^2-1}{2n^2}\,\mathbb{1}\,.
\end{equation}
Finally, the matrix $\PP_1$ advances the phases through the dielectric disk and
is generically called $\PP_\epsilon={\rm diag}(e^{+i\delta},e^{-i\delta})$
with $\delta\equiv\delta_\epsilon=\delta_1=n\omega d$.

\begin{figure}[t]
\centering
\includegraphics[width=10cm]{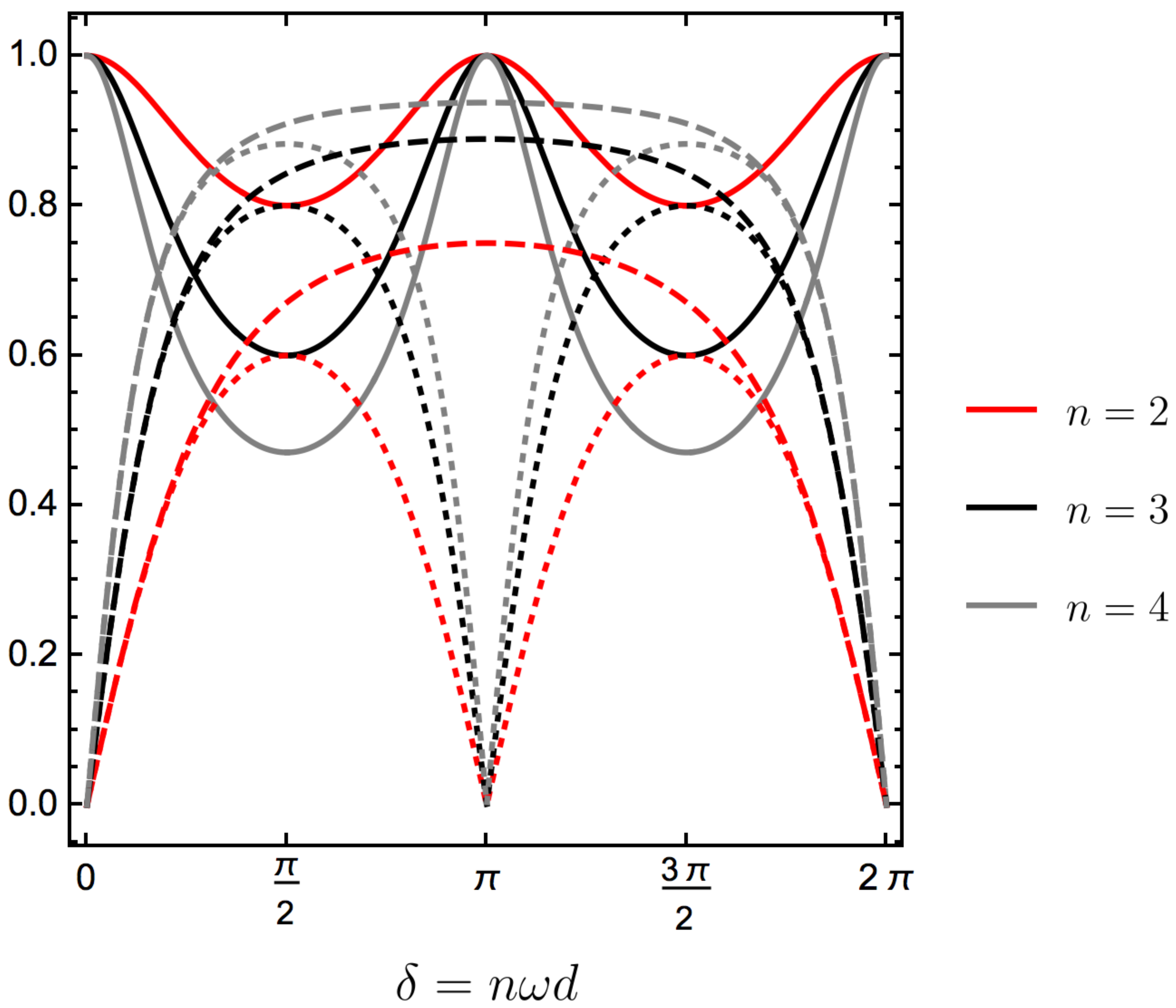}
\caption{Transmission coefficient $|{\mathcal T}_{\rm D}|$ (solid), reflection
  coefficient $|{\mathcal R}_{\rm D}|$ (dotted), and boost factor $\beta_{\rm D}=|\B_{\rm D}|$
  (dashed) for a dielectric disk. The refractive indices
  are $n=2$, 3 and 4 (red, black and gray lines).  The horizontal axis
  is the phase depth of the disk $\delta = n \omega d$, with $d$ its
  physical thickness.}
\label{owl}
\end{figure}

Notice that the regions are labelled from left to right, but the
transfer and source matrices are built up in opposite
order. The matrix in the right-most position describes the
left-most interface.  The matrices $\GG_{\epsilon v}$ or
$\Ss_{\epsilon v}$ mean that vacuum is on the left of the
dielectric. Therefore, the total transfer and axion source matrices
for a dielectric disk are
\begin{subequations}
\begin{eqnarray}
\label{1slabG}
\TT_{\rm D}&=&\GG_{v\epsilon} \PP_\epsilon \GG_{\epsilon v} =
\(
\begin{array}{cc}
\cos\delta+i\,\frac{n^2+1}{2 n}\sin\delta & ~i\,\frac{n^2-1}{2 n}\sin\delta \\[1ex]
-i\,\frac{n^2-1}{2 n}\sin\delta &~\cos\delta-i\,\frac{n^2+1}{2 n}\sin\delta
\end{array}
\)\,,
\\[3ex]
\label{1slabM}
\MM_{\rm D} &=& \Ss_{v\epsilon} + \GG_{v\epsilon} \PP_{\epsilon} \Ss_{\epsilon v}
=
\frac{n^2-1}{2n^2}\(
\begin{array}{cc}
1-\frac{n+1}{2}\,e^{i\delta} & \frac{n-1}{2}\,e^{-i\delta} \\[1ex]
\frac{n-1}{2}\,e^{i\delta} & 1-\frac{n+1}{2}\,e^{-i\delta}
\end{array}
\)\,.
\end{eqnarray}
\end{subequations}
Observable quantities are the transmission and reflection coefficients
as well as the boost amplitude defined in
equations~(\ref{eq:T-R-Coefficients}) and (\ref{eq:boost-factor}),
respectively. Because the disk is perfectly left-right symmetric, the
$L$ and $R$ quantities are the same. The explicit results are
\begin{subequations}
\begin{eqnarray}
{\cal T}_{\rm D}&=&\frac{i\,2n}{i\,2n\cos\delta+(n^2+1)\,\sin\delta}\,,\label{eq:disk-transmission}
\\[1ex]
{\cal R}_{\rm D}&=&\frac{(n^2-1)\,\sin\delta}{i\,2n\cos\delta+(n^2+1)\,\sin\delta}\,,
\\[1ex]
\B_{\rm D}   &=&\frac{(n^2-1)\sin(\delta/2)}{n^2\sin(\delta/2)+i\,n\cos(\delta/2)}\,.
\end{eqnarray}
\end{subequations}
Their moduli are shown in figure~\ref{owl} for several values of the refractive index $n$.

The disk becomes fully transparent ($\mathcal{T}_{\rm D}=1$),
and then does not reflect at all
($\mathcal{R}_{\rm D}=0$), for $\sin \delta=0$, corresponding to the frequencies
\begin{equation}\label{eq:transparency-condition}
\omega_j = j\,\frac{\pi}{n d}\,,
\quad\hbox{where}\quad
j=0,1,2,\dots
\quad\hbox{(transparent)}.
\end{equation}
On the one hand, the boost factor vanishes for $\sin(\delta/2)=0$,
i.e., for even values of $j$ in
equation~(\ref{eq:transparency-condition}). On the other hand, it is
maximal for $\sin^2(\delta/2)=1$ and then takes the value
\begin{equation}\label{eq:maximum-boost-disk}
\beta_{\rm D}^{\rm max} = 1-\frac{1}{n^2}
\quad\hbox{for}\quad
j=1,3,\ldots
\quad\hbox{(odd)}.
\end{equation}
We may compare this result with a single dielectric interface given
in equation~(\ref{eq:dielectricE}),
\begin{equation}\label{eq:boost-single-interface}
\beta_{\rm interface} = 1-\frac{1}{n}\,.
\end{equation}
The maximum EM wave emerging from our disk has an amplitude enhanced by a
factor $(1+1/n)$ relative to a single interface, caused by the
constructive interference with the EM wave emitted from the second
interface. In both cases, the largest emission is obtained by
$n\to\infty$, corresponding to a perfect mirror.

The average squared boost factor in the spirit of
equation~\eqref{eq:general-area-law} is
\begin{equation}\label{eq:sum-rule-disk}
  \bigl\langle |\B_{\rm D}|^2\bigr\rangle=\(1-\frac{1}{n}\)\(1-\frac{1}{n^2}\)\,.
\end{equation}
The average can be taken over all phase depths $\delta$ in the interval $[0,2\pi]$ or alternatively over a flat spectrum of frequencies as explained in section~\ref{sec:Boost-Amplitude}.

Thus a single dielectric disk by itself offers no advantage compared
with a mirror. It is intriguing, however, that for frequencies around
maximum boost factor, the disk is completely transparent. This observation
immediately suggests an arrangement of $N$ disks spaced
such that the EM waves from all disks interfere constructively,
enhancing the overall amplitude by a factor of $N$. We will study this
``transparent mode'' in section~\ref{sec:Transparent-Mode} below.

\subsection{Resonant effects: cavity setup}
\label{cavity}

\subsubsection{Mirror with dielectric disk}

\begin{figure}[t]
\centering
\includegraphics[width=10cm]{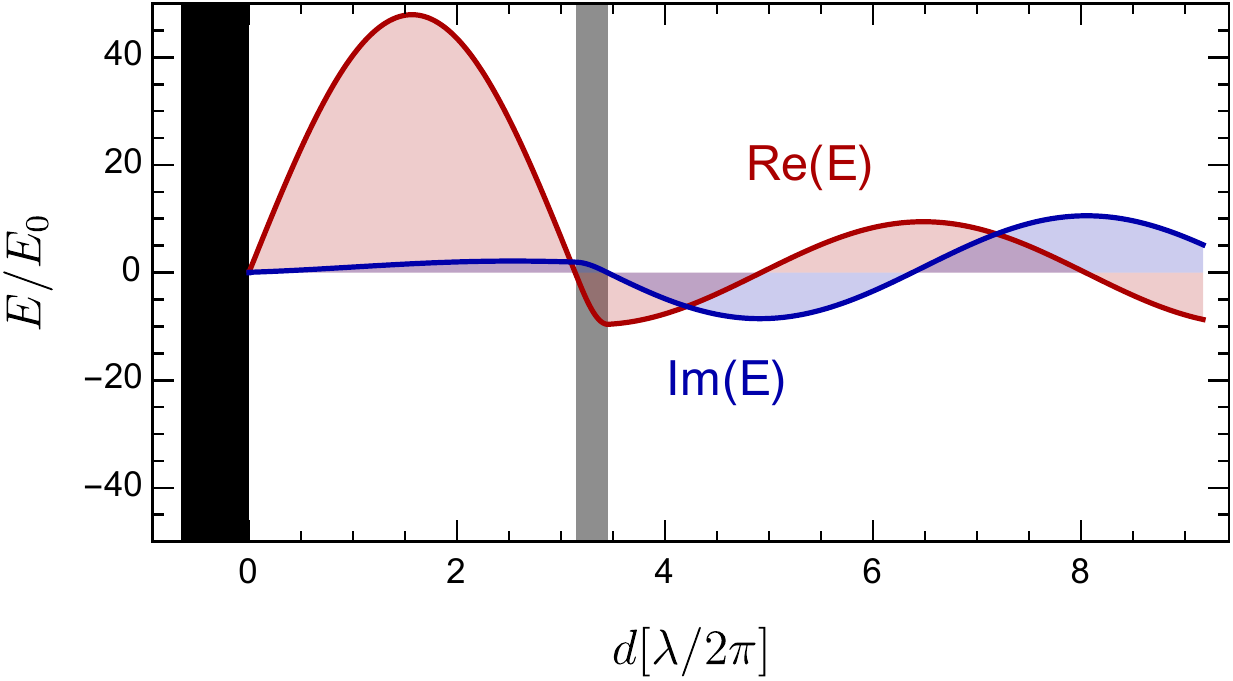}
\caption{Schematic of a resonant cavity consisting of a single
  dielectric disk (grey) and a mirror (black), showing the real and
  imaginary parts of the electric field distribution (red and blue
  lines). For scaling purposes, we choose a realistic refractive index
  $n=5$ for the disk, which is not strongly resonant (low Q). Note
  that the dielectric is arranged such that the distance between the
  mirror and dielectric is $d_v=\lambda/2$, and the thickness of the
  disk is $d_\epsilon=\lambda/4n$.  The shown electric field amplitude
  is for the sum of the produced EM wave and the homogeneous
  axion-induced field.}
   \label{fig:cavityfield}
\end{figure}

The simplest way to obtain an actual enhancement relative to
a mirror is to actually use a mirror and place
at some distance a dielectric disk with a thickness such
that it is partially reflecting. In figure~\ref{fig:cavityfield} we
show a schematic arrangement and its electric field
distribution. Assuming a realistic dielectric constant of
$\epsilon=25$ and thus a refractive index $n=\sqrt\epsilon=5$ (approximately that of LaAlO3), the
emerging EM wave has an electric field boosted by a factor of
around~10.

To derive the boost amplitude analytically, we need the same matrices
$G_{\epsilon v}$ and $S_{\epsilon v}$ as defined in the previous
section. We now write the matrix which advances the phase in the
vacuum gap between mirror and disk as
$\PP_v={\rm diag}(e^{+i\delta_v},e^{-i\delta_v})$, where $\delta_v=\omega d_v$
and $d_v$ is the physical width of the gap. For the disk we now write
$\PP_\epsilon={\rm diag}(e^{+i\delta_\epsilon},e^{-i\delta_\epsilon})$
where $\delta_\epsilon=n\omega d_\epsilon$ is the phase depth of the disk.
Moreover, we need the transfer matrix for the interface from mirror to
vacuum $G_{v\sigma}$ where $\sigma$ symbolizes the large conductivity
of the mirror, and we also need the source matrix $S_{v\sigma}$, which
all follow from our previous expressions.  Writing the compound
expression, as usual from left to right, we find
\begin{subequations}
\begin{eqnarray}
  \TT_{\rm C}&=&\GG_{v\epsilon}\PP_\epsilon\GG_{\epsilon v}\PP_v\GG_{v\sigma}\,,
\\[1ex]
  \MM_{\rm C}&=&\Ss_{v\epsilon}+\GG_{v\epsilon}\PP_\epsilon\Ss_{\epsilon v}
  +\GG_{v\epsilon}\PP_\epsilon\GG_{\epsilon v}\PP_v\Ss_{v\sigma}
\end{eqnarray}
\end{subequations}
for the full 1D cavity.

\subsubsection{Properties of the boost amplitude}

\begin{figure}[ht]
\centering
\includegraphics[width=10cm]{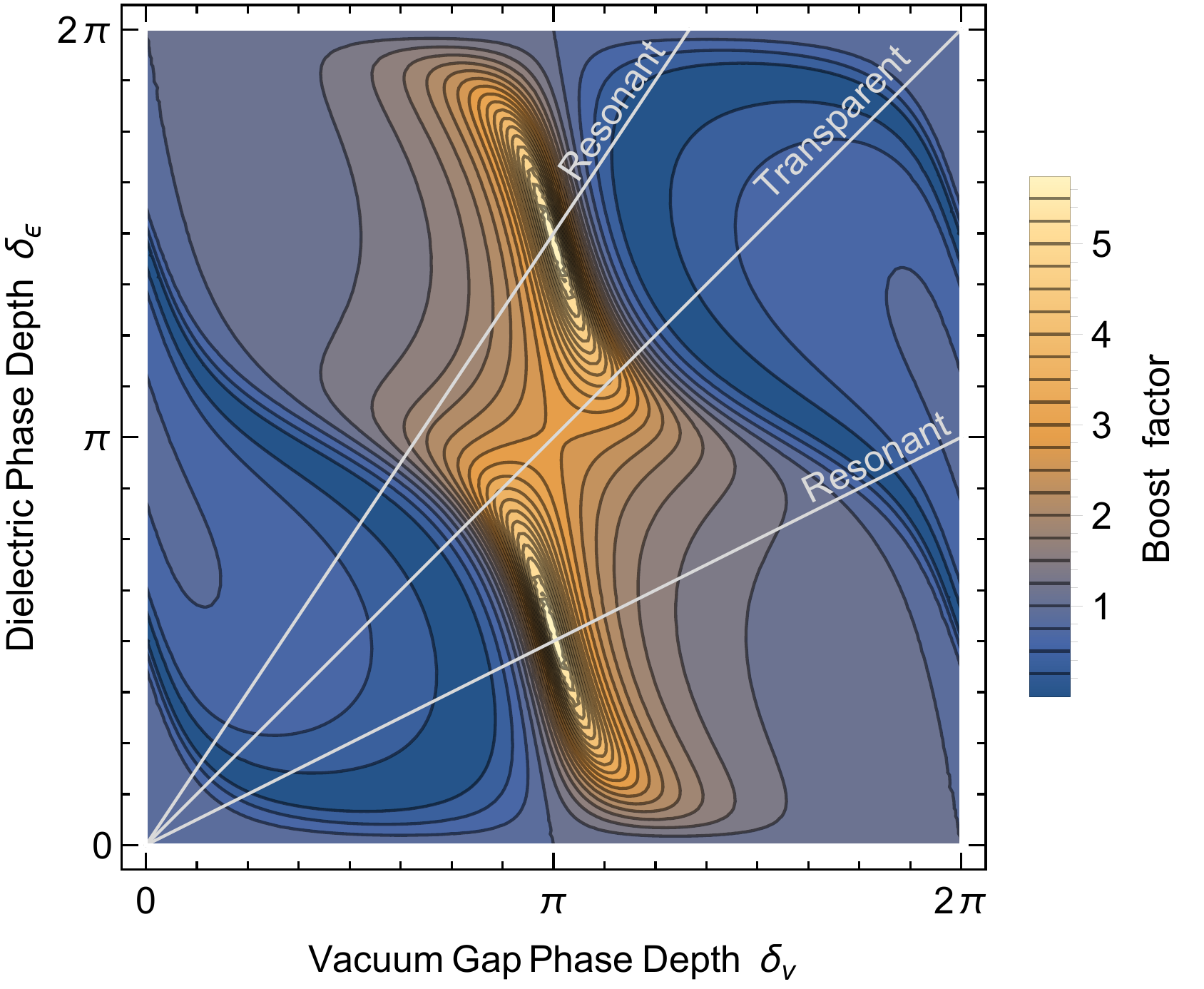}
\caption{Cavity boost factor $\beta_{\rm C}=|\B_{\rm C}|$ of a setup
  consisting of a mirror and a dielectric disk as a function of the
  phase depths of the vacuum gap, $\delta_v=\omega d_v$, and of the
  dielectric disk, $\delta_\epsilon=n\omega d_\epsilon$.  The
  refractive index is $n=3$. For a fixed distance between mirror and
  disk, varying the frequency $\omega$ takes us along the diagonal
  lines.  The ``resonant'' lines correspond to
  $\delta_\epsilon/\delta_v=n\, d_\epsilon/d_v=1/2$ or $3/2$, whereas
  the ``transparent'' line corresponds to
  $\delta_v/\delta_\epsilon=1$. These lines correspond
    to keeping the physical distances fixed, while scanning over
    frequency. Fully resonant and transparent behaviours only occur at
    a single point along each line in this plot.}
\label{fig:cavity-boost-contour}
\end{figure}

If the mirror is perfect (its refractive index is infinite), from equation~(\ref{eq:boost-factor}) we then find for the boost
amplitude
\begin{equation}
\B_{\rm C} =
\frac{1-\(1-\frac{1}{n^2}\)\bigl[\cos\delta_v(1-\cos\delta_\epsilon)
    +n\sin\delta_\epsilon \sin\delta_v\bigr]}
     {e^{-i \delta_v}\cos\delta_\epsilon -\sin\delta_\epsilon\(\frac{i}{n}\cos\delta_v+n \sin\delta_v\)}\,,
     \label{Bcavity1}
\end{equation}
which is $2\pi$ periodic in both $\delta_v$ and $\delta_\epsilon$.
In figure~\ref{fig:cavity-boost-contour} we show the cavity boost
factor $\beta_{\rm C}=|\B_{\rm C}|$ as a function of $\delta_v$ and
$\delta_\epsilon$. As expected, there are resonant structures for
certain combinations of phase depths of the disk and vacuum gap.

As discussed earlier, we may consider $\beta^2$, averaged over all possible configurations of $\delta_v$ and $\delta_\epsilon$, or alternatively, for any fixed configuration, averaged over all frequencies, and find
\begin{equation}\label{eq:sum-rule-cavity}
  \bigl\langle|\B_{\rm C}|^2\bigr\rangle
  =1+2\(1-\frac{1}{n}\)\(1-\frac{1}{n^2}\)\,.
\end{equation}
This is the same as what one gets from a mirror plus two
dielectric disks---the reflectivity of the mirror means that the dielectric disk must be counted twice. In other words, the emitted power, averaged over all
possible values of vacuum gap and disk thickness, is the average power
emitted by the mirror alone, by the outer dielectric surface alone, and
by the inner dielectric surface alone, which is reflected by
the mirror. As expected,
in the expression for the average power, all interference terms
disappear (see also appendix~\ref{sec:area-law}).

For a fixed configuration of dielectric disk and vacuum gap, we may
consider $\beta$ as a function of driving frequency, which corresponds to diagonal lines in figure~\ref{fig:cavity-boost-contour}.
The lower of these lines denoted ``resonant'' corresponds to $\delta_v/\delta_\epsilon=2$, where
the resonance occurs when the dielectric accommodates $\lambda/4$ of the EM wave,
whereas the gap accommodates $\lambda/2$. There is a second ``resonance line'' where one has $\lambda/2$ in the vacuum gap, and $3\lambda/4$ within the dielectric. There is an infinite family of such resonances, corresponding to larger odd multiples of $\lambda/4$ in the dielectric. Discussion of these cases would take us too far afield and so we will neglect them.
 The ``transparent'' line corresponds to
$\delta_v/\delta_\epsilon=1$ and in the center of the plot, both the dielectric and the gap accommodate $\lambda/2$. The Area Law implies that integrating along either line
gives us the same value for $\langle |\B|^2\rangle$ that was given in
equation~\eqref{eq:sum-rule-cavity}. For integrations along other lines, if taken only within the shown square of parameter space, the integral somewhat differs. For example,
for lines between the ``resonant'' and ``transparent'' cases and for $n=3$, the deviation is less then 2\%. For larger $n$, the deviation is even smaller.
The Area Law is only exact if
the average is taken over a full period of $|\B|^2$ as a function of $\omega$.

\subsubsection{Resonant mode}

\begin{figure}[t]
\centering
\includegraphics[width=10cm]{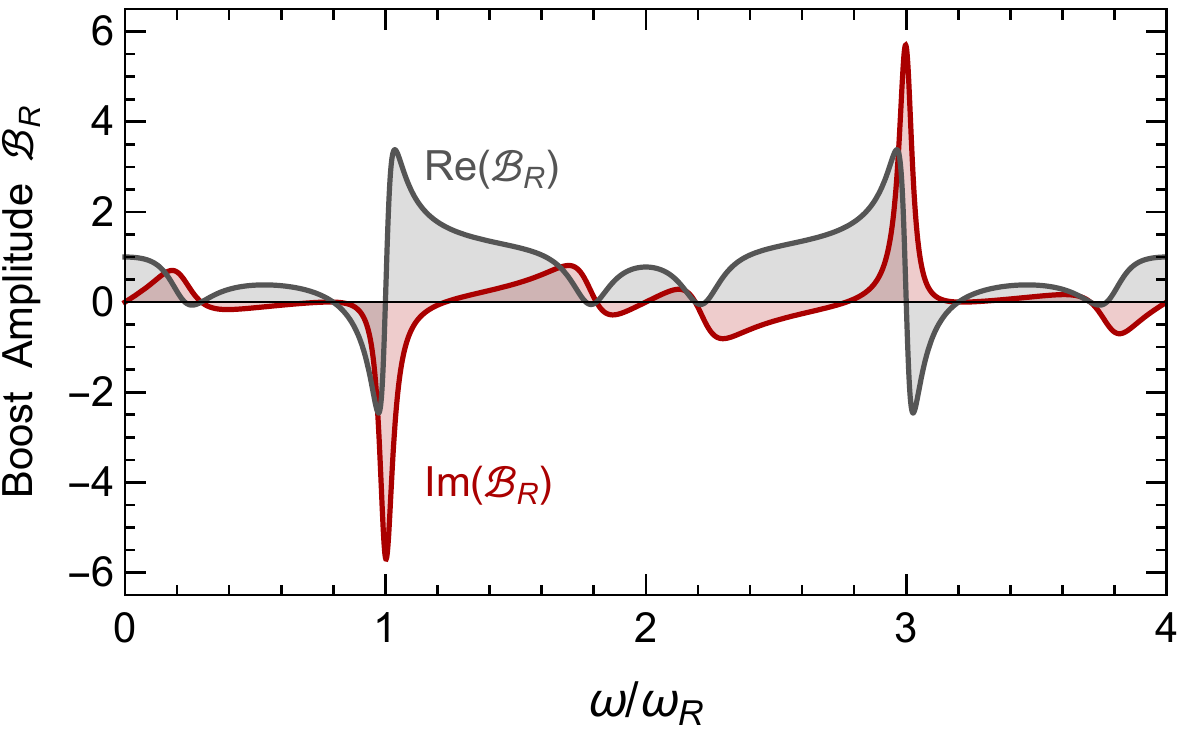}
\caption{Boost amplitude $\B_{\rm R}$ for a cavity consisting of a
  mirror and a dielectric disk where $d_v=2n d_\epsilon$ and
  $\omega_{\rm R}^{-1}=2nd_\epsilon/\pi$, which forms a resonance.
  The real part is in gray and imaginary part is in red. The
  dielectric disk was chosen to have $n=3$. For larger $n$, the
  resonance structure becomes more pronounced.}
\label{fig:boost-resonator-frequency}
\end{figure}

We briefly consider two special cases of the cavity.
The most conspicuous feature of figure~\ref{fig:cavity-boost-contour}
is the resonance which is obtained when the phase depths are
$\delta_v=\pi$ and $\delta_\epsilon=\pi/2$.
There exists a frequency
$\omega_{\rm R}=\pi/(2n\,d_\epsilon)=\omega_\epsilon/2$, where the cavity is on
resonance, corresponding to $\lambda/4$ in the dielectric and $\lambda/2$ in the vacuum gap ($\omega_\epsilon$ will correspond to the transparent mode).
The resonant enhancement of the output amplitude is
\begin{equation}
  \beta^{\rm max}_{\rm C}=2n-\frac{1}{n}\,.
\end{equation}
For the resonant setup we define the boost amplitude as a function of frequency as
\begin{equation}
\B_{\rm R}(n,\omega)=\B_{\rm C}\(n,\pi\,\frac{\omega}{\omega_{\rm R}}, \frac{\pi}{2}\,\frac{\omega}{\omega_{\rm R}}\)\,.
\end{equation}
In figure~\ref{fig:boost-resonator-frequency} we show its real and
imaginary parts as functions of $\omega$.  As anticipated,
$\B_{\rm R}$ shows a resonance structure at $\omega=\omega_{\rm R}$ and odd
multiples.  Near resonance, the boost amplitude should correspond to
the response of a driven lossy harmonic oscillator, so we expect a
structure of the form $\B_{\rm R}\propto (\omega-\omega_{\rm R}+i\Gamma/2)^{-1}$.

In our resonator the damping arises not from friction, but because
power escapes through the disk which is an imperfect mirror.  In the
good-cavity-limit, $n\gg 1$, the left and right moving waves inside
the cavity have to be of similar amplitude and related to the outgoing
wave by the transmission coefficient $|{\cal T}|\sim 2/n$ of one
dielectric disk (\ref{eq:disk-transmission}) at $\delta_\epsilon\sim
\pi/2$. The power stored in the cavity is proportional to the squared
electric-field amplitude inside, the power in the emitted wave outside
proportional to the squared outside electric field strength. The ratio
of the field amplitudes is $|{\cal T}|$ and therefore the ratio of the
energy density $|{\cal T}|^2$.  Therefore, the ratio of stored power
to escaping power is $|{\cal T}|^{-2}\sim n^2$, thus we expect the
quality factor $Q$ of the resonator to be $Q\sim n^2$.

To find the near-resonant $\B_{\rm R}$ explicitly we start from
equation~(\ref{Bcavity1}) and substitute
$\delta_v=(1+s)\,\pi$ and $\delta_\epsilon=(1+s)\,\pi/2$, where
$s=(\omega-\omega_{\rm R})/\omega_{\rm R}$. Next we expand the
numerator and denominator separately as a Taylor expansion in $s$ to
linear order and find
\begin{equation}
\B_{\rm R}=\frac{\(2n-\frac{1}{n}\)+\(1+\frac{1}{2n}-\frac{1}{n^2}-\frac{1}{2n^3}\)n^2\pi s}
               {i+\(1+\frac{1}{2n}\)n^2\pi s}
               \to\frac{2n+n^2\pi s}{i+n^2\pi s}\,,
\end{equation}
where the second expression pertains to large $n$.
This an excellent approximation
when $n^2 \pi s\lesssim 1$. The linear term in the numerator explains the
slightly skewed response around the resonance. Indeed, the maximum of $\beta_{\rm R}$
is slightly shifted relative to the nominal resonance point.
For $n\gg 1$ the boost amplitude becomes
\begin{equation}\label{eq:resonance-response}
  \B_{\rm R}(n{\gg}1)=1+2n\,\frac{\Gamma/2}{\omega-\omega_{\rm R}+i\Gamma/2}
    \quad\hbox{where}\quad
  \Gamma=\frac{2\,\omega_{\rm R}}{\pi n^2}\,.
\end{equation}
The quality factor corresponds to the power damping
rate as $\Gamma=\omega_{\rm R}/Q$, so our resonator has effectively
$Q=\pi n^2/2$.

The cavity boost amplitude as a function of frequency shown in
figure~\ref{fig:boost-resonator-frequency} has the squared average value
given in equation~\eqref{eq:sum-rule-cavity}.
Using a large value of $n$ does not enhance the power emission
averaged over all frequencies, but concentrates it near the resonance. The height of resonance peak of $\beta_{\rm R}^2$ scales with $n^2$, its width with $1/n^2$.

In principle, one can obtain an arbitrarily large boost factor. In practice,
the available dielectrics with small absorption (small loss tangent
$\tan\delta$) have moderate indices of refraction, $n\sim 3$
(Sapphire) and $n\sim 5$ (LaAlO3), so one would be forced to replace
the dielectric with another mirror and extract a small amount of power
through an antenna or a hole, tuning the frequency with the distance
between the mirrors. Such an arrangement
would be analogous to Sikivie's haloscope,
used in the ADMX, ADMX-HF and CAPP experiments, and in the Orpheus
project~\cite{Rybka:2014cya}.

\subsubsection{Transparent mode}
\label{mirrortransparent}
Another special case that will be of interest for our general approach is when
$\delta_\epsilon = \pi$ or an odd multiple. Now the disk becomes
transparent and $\B_{\rm C}$ becomes the simple sum of three waves,
one emitted from the mirror and transmitted, one emitted directly from
the disk outward, and the last one emitted from the disk toward the mirror,
reflected, and transmitted outward, leading to a boost factor
\be
\beta_{\rm C}(\delta_\epsilon=\pi)  = \left|1-\(1-\frac{1}{n^2}\)2\cos\delta_v\right|.
\ee
For a gap between disk and mirror such that $\delta_v=\pi$ or an odd
multiple, this is $1+2(1-1/n^2)$ and thus what one expects from a
coherent superposition of the emission from a perfect mirror and twice
the emission from a dielectric disk given in
equation~(\ref{eq:maximum-boost-disk}). For $n\to\infty$, the maximum
boost factor becomes $3$. Notice, however, that in this limit the disk must become infinitely thin to maintain
$\delta_\epsilon=\pi$, so this
is a somewhat unphysical limiting case.

\begin{figure}[ht]
\centering
\includegraphics[width=10cm]{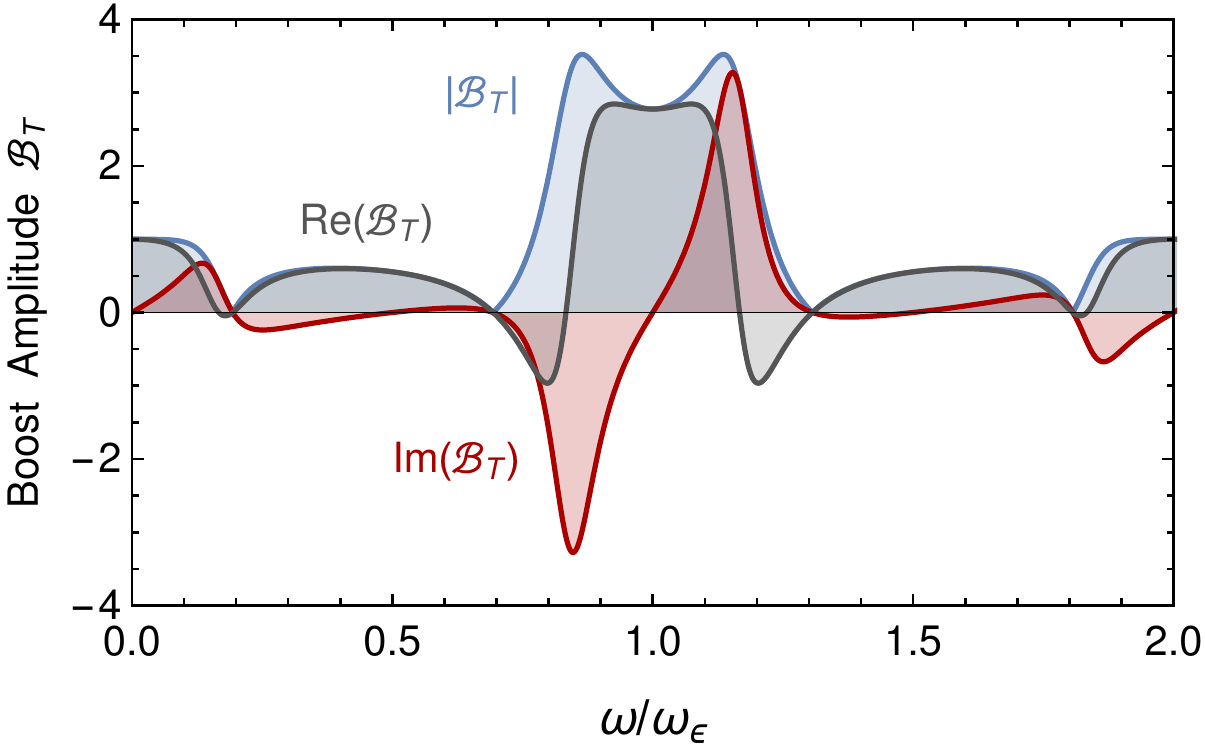}
\caption{Boost amplitude $\B_{\rm T}$ for a cavity in transparent mode
  where $d_v=2n d_\epsilon$ and $\omega_{\rm \epsilon}^{-1}=nd_\epsilon/\pi$.
  Real part in gray, imaginary part in red, and absolute value
  (boost factor) in light blue.  The dielectric disk was chosen to have
  $n=3$. For larger $n$, the two-peak structure becomes more pronounced.}
\label{fig:cavity-transparent-boost}
\end{figure}

We show the cavity boost amplitude $\B_{\rm T}(\omega)$ in transparent
mode in figure~\ref{fig:cavity-transparent-boost}. It is evident that
it has the structure of two overlapping resonances which, for moderate
$n$, produce a nearly box-shaped boost factor as a function of
frequency. This general shape can be gleaned as well from the way the
``transparent'' diagonal line cuts through the resonance in
figure~\ref{fig:cavity-boost-contour}.

\subsection{Equally-spaced disks}
\label{transparent}

\subsubsection{General homogeneous setup}

An alternative approach to boost the outgoing waves is to use multiple
dielectric layers. As a first simple case we consider a sequence of
$N$ equally sized, equally spaced dielectrics with index of refraction
$n$ and thickness $d_\epsilon$ (phase depth $\delta_\epsilon=n\omega
d_\epsilon$). In terms
of the transfer matrices used earlier, we now have a sequence of
layers (from left to right) consisting of a disk plus a vacuum gap,
i.e., each such layer has the transfer matrix
$\GG_{v\epsilon}\PP_\epsilon\GG_{\epsilon v}\PP_v$, iterated $N$
times, and un-doing one surplus vacuum gap at the end. We then have
\begin{subequations}\label{eq:transfer-homogeneous}
\begin{eqnarray}
\TT &=& \(\GG_{v\epsilon} \PP_{\epsilon} \GG_{\epsilon v} \PP_{v}\)^N \PP^{-1}_{v}=
\(\TT_{\rm D} \PP_{v}\)^N \PP^{-1}_{v},
\\[1ex]
\MM &=& \(\Ss_{v\epsilon}+\GG_{v\epsilon} \PP_{\epsilon} \Ss_{\epsilon  v}\)
+\(\GG_{v\epsilon} \PP_{\epsilon}\GG_{\epsilon v} \PP_{v}\Ss_{v\epsilon}
+\GG_{v\epsilon}  \PP_{\epsilon}\GG_{\epsilon v}\PP_{v}\GG_{v\epsilon} \PP_{\epsilon}\Ss_{\epsilon v }\)+\ldots
\nonumber\\
&=&\(\Ss_{v\epsilon}+\GG_{v\epsilon} \PP_{\epsilon} \Ss_{\epsilon  v}\)
+\GG_{v\epsilon} \PP_{\epsilon}\GG_{\epsilon v} \PP_{v}\(\Ss_{v\epsilon}
+\GG_{v\epsilon} \PP_{\epsilon}\Ss_{\epsilon v }\)+\ldots
\nonumber\\
&=&\MM_{\rm D}+\TT_{\rm D}\PP_v\MM_{\rm D}+\TT_{\rm D}\PP_v\TT_{\rm D}\PP_v\MM_{\rm D}+\ldots
\nonumber\\
&=& \(\sum_{s=0}^{N-1}(\TT_{\rm D} \PP_{v})^s\) \MM_{\rm D},
\end{eqnarray}
\end{subequations}
where from equations~\eqref{1slabG} and \eqref{1slabM} we have
recognised the transfer
($\TT_{\rm D}=\GG_{v\epsilon} \PP_{\epsilon} \GG_{\epsilon v}$)
and source matrix
($\MM_{\rm D}=\Ss_{v\epsilon}+\GG_{v\epsilon}\PP_{\epsilon}
\Ss_{\epsilon v}$)
of a single dielectric disk.

As this system is governed by two parameters $\delta_\epsilon$ and $\delta_v$, in principle one can study the full parameter space. Moreover, one can derive an analytic solution for the boost amplitude as a function of $N$, $n$, $\delta_\epsilon$ and $\delta_v$ because the matrix powers which appear in equation~\eqref{eq:transfer-homogeneous} can be solved explicitly in a basis where these matrices are diagonal. However, the resulting expressions for matrices
are too complicated to be illuminating even for small $N$.

\begin{figure}[ht]
\centering
\hbox to\textwidth{\includegraphics[height=8.8cm]{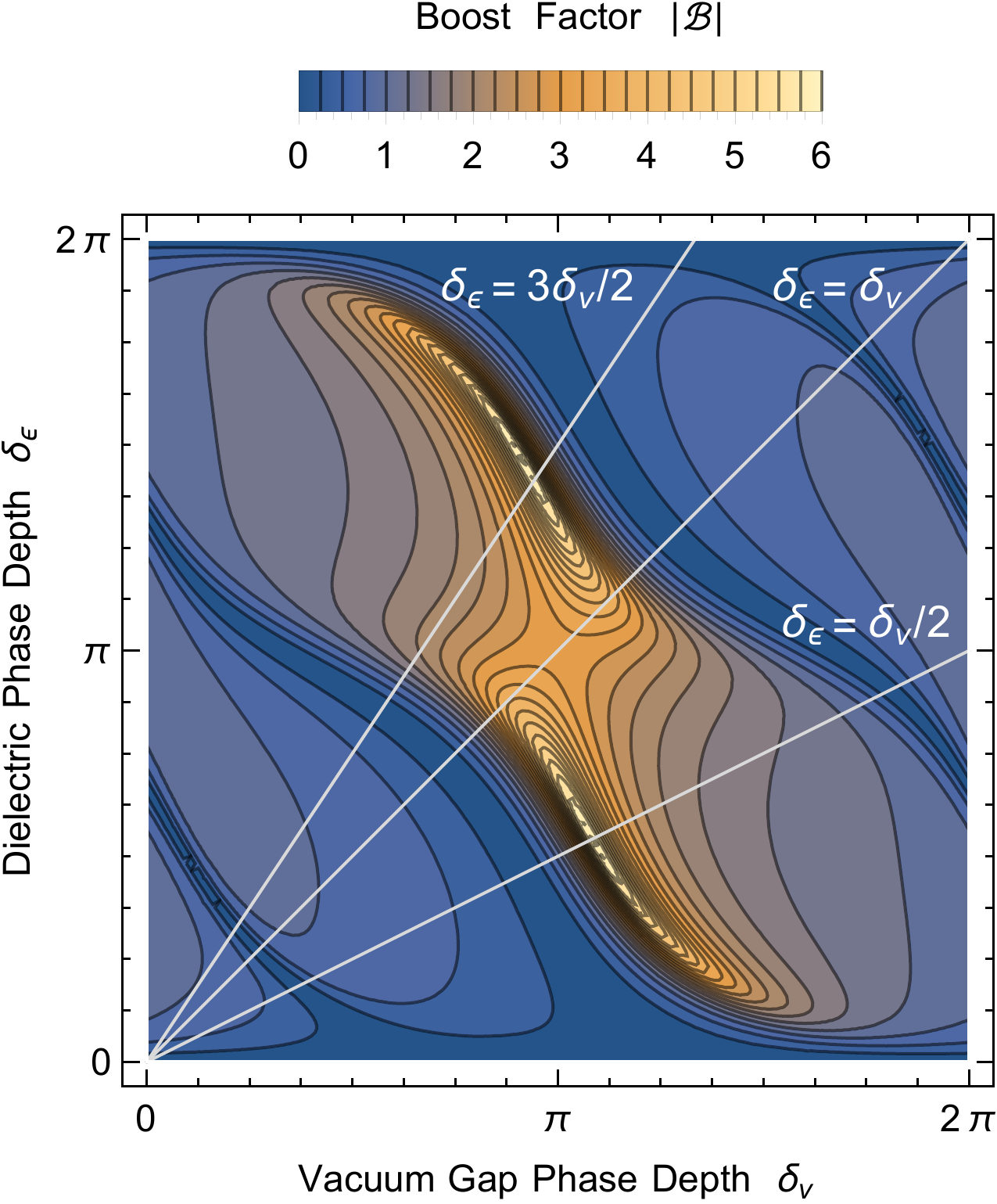}\hfill
\includegraphics[height=8.8cm]{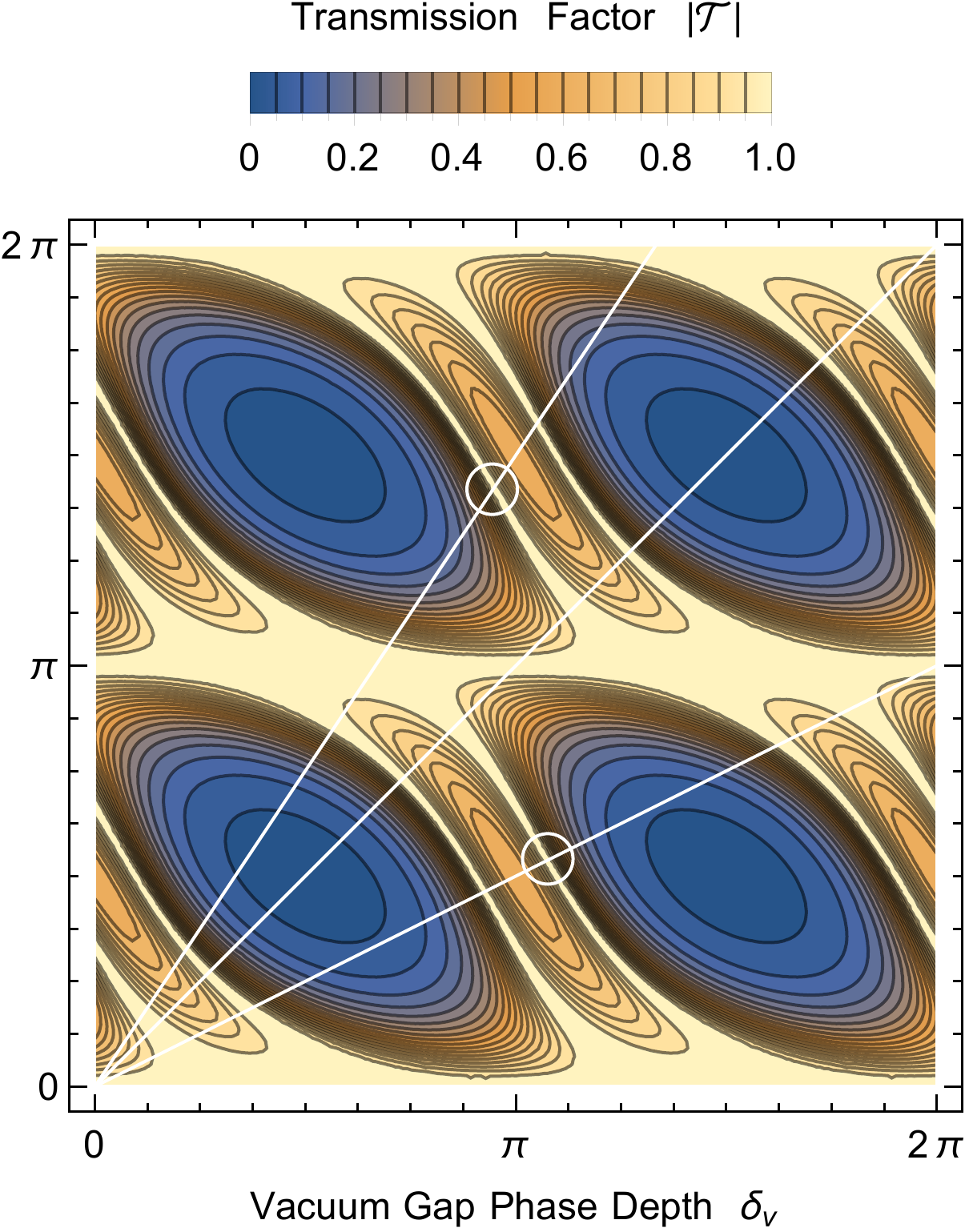}}
\caption{$N=3$ equally spaced dielectric disks with refractive index
  $n=3$.  {\em Left:\/} Boost factor $\beta=|\B|$ as a function of the
  phase depth of the dielectric layers $\delta_\epsilon = n \omega
  d_\epsilon$ and of the vacuum gaps $\delta_v = \omega d_v$, with
  $d_\epsilon$ the thickness of the disks and $d_v$ the disk spacing.
  The diagonal lines show the locus of the indicated values for
  $\delta_\epsilon/\delta_v$.  {\em Right:\/} Transmission factor
  $|\mathcal{T}|$.  The resonant peak of the boost factor in the left
  panel is marked with a white circle in the right panel.}
\label{fig:homogeneous-boost}
\vspace{0.5cm}
\centering
\hbox to\textwidth{\includegraphics[height=8.8cm]{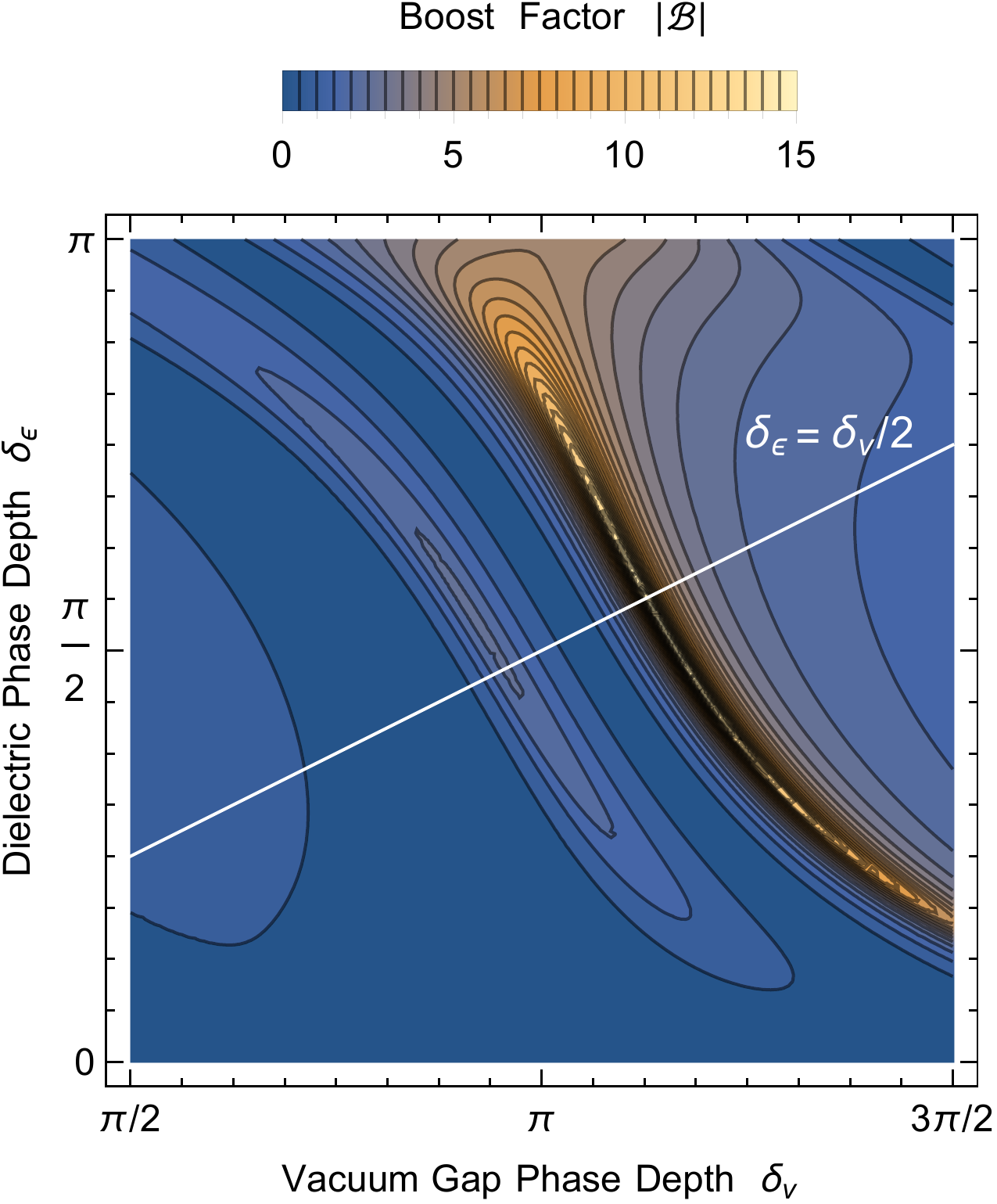}\hfill
\includegraphics[height=8.8cm]{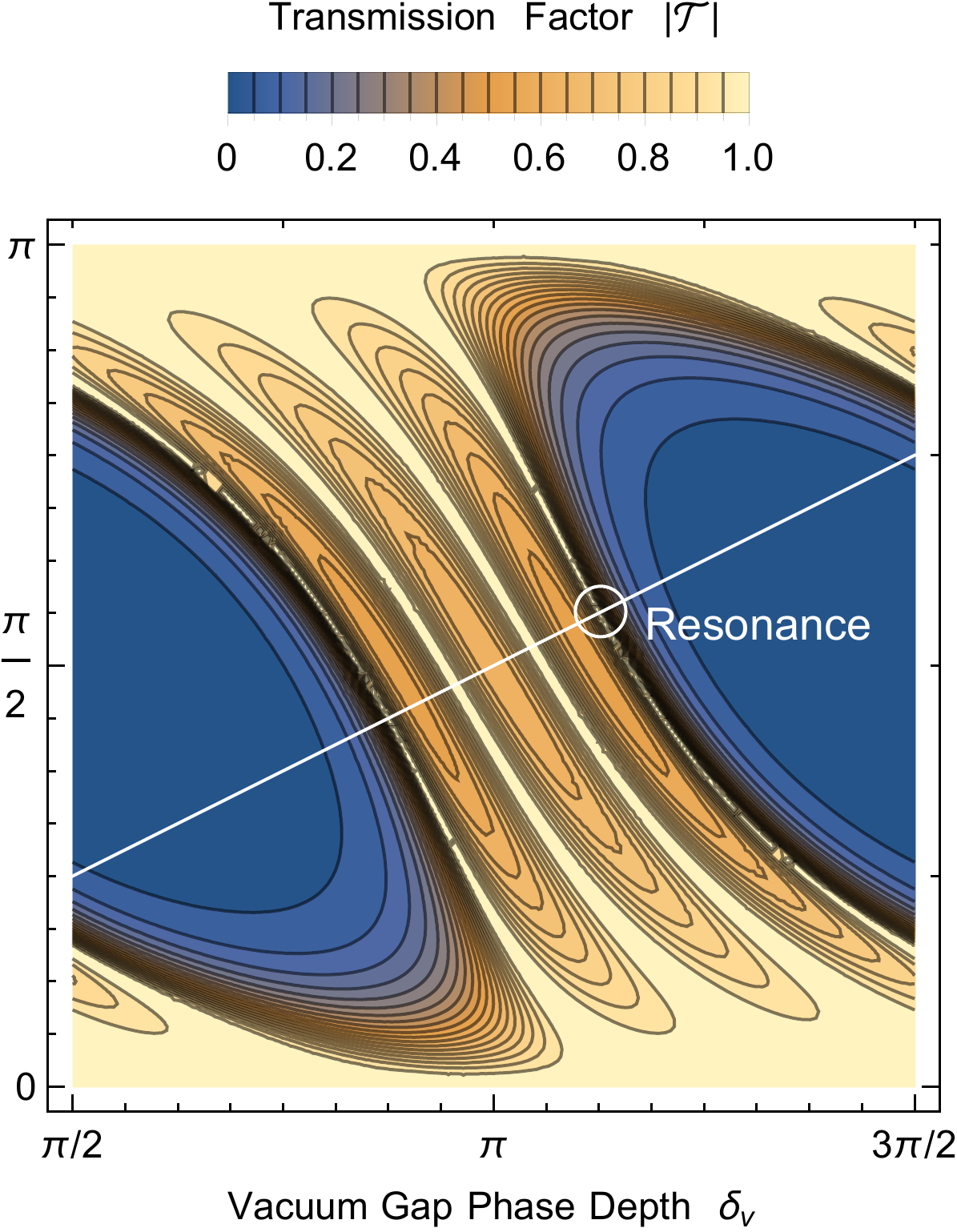}}
\caption{$N=5$ equally spaced dielectric disks with refractive index $n=3$,
  analogous to figure~\ref{fig:homogeneous-boost}.}
\label{fig:homogeneous-boost-5}
\end{figure}

For a numerical example we consider $N=3$ disks with refractive index $n=3$ where we show the boost
factor in the left panel of figure~\ref{fig:homogeneous-boost} . Near $\delta_\epsilon\sim \pi/2$ and
$\delta_v\sim \pi$ and appropriate multiples
there is a resonant enhancement of $\beta$ over the
simple sum of the emitted waves, similar to the resonator studied
earlier. 
(Another numerical example for the boost factor in the case of $N=5$ disks with refractive index $n=3$ 
is shown in the left panel of figure~\ref{fig:homogeneous-boost-5}.)
For the now-familiar squared average, we find
\begin{equation}\label{eq:sum-rule-transparent-1}
  \bigl\langle |\B_{\rm R}|^2\bigr\rangle=N\(1-\frac{1}{n}\)\(1-\frac{1}{n^2}\)\,,
\end{equation}
which is $N$ times of what we get from a single disk as given in
equation~\eqref{eq:sum-rule-disk}.

\subsubsection{Resonant mode}
\label{sec:Resonant-Mode}

For realistic refractive indices in the range $n=3$--5, the overall resonance arises from the strongly coupled ``cavities'' formed by every neighboring pair of disks. One might expect there to be a large number of resonances. However, in a plot for $|\B|$
like the left panel of figure~\ref{fig:homogeneous-boost} for any $N$ there is only one dominant resonance, except of course for repeating structures as periodic multiples of the phase depths.

\begin{figure}[ht]
\centering
\includegraphics[width=9cm]{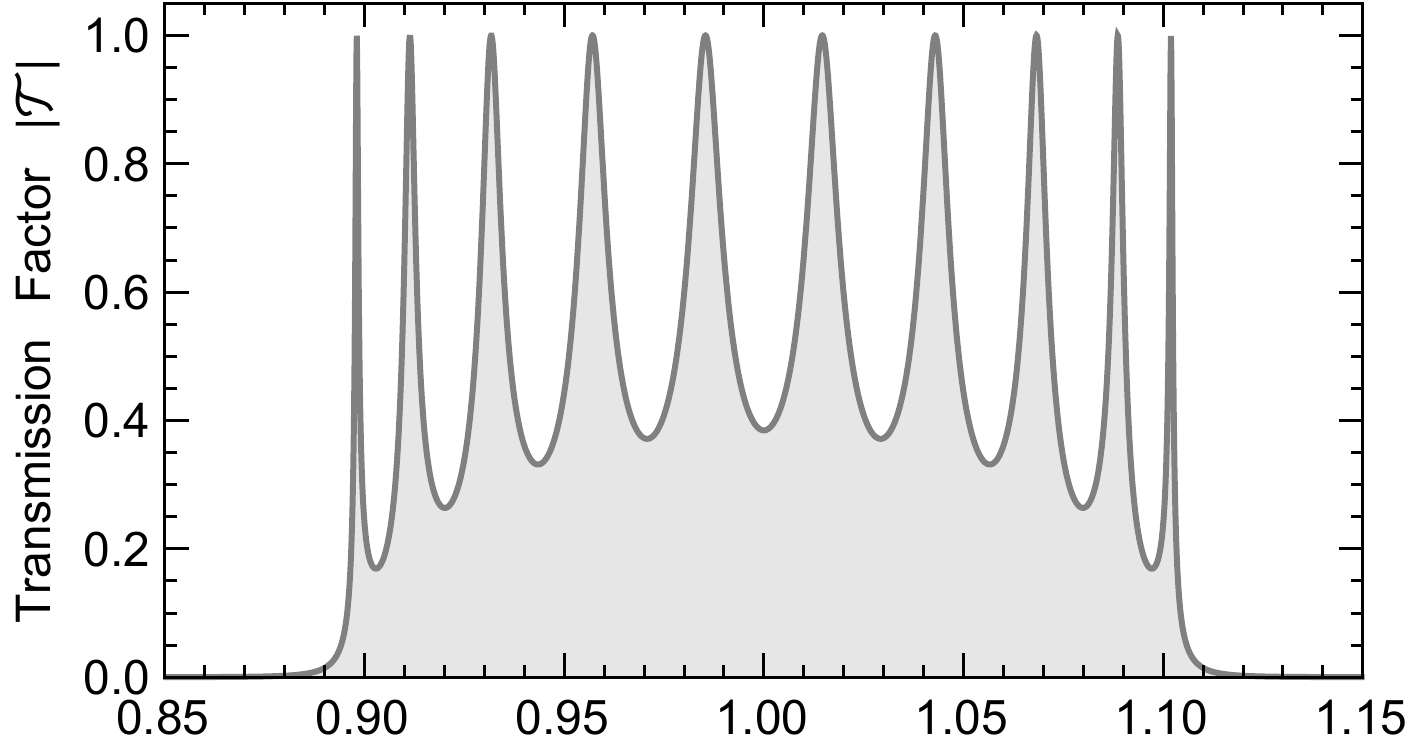}
\vskip6pt
\includegraphics[width=9cm]{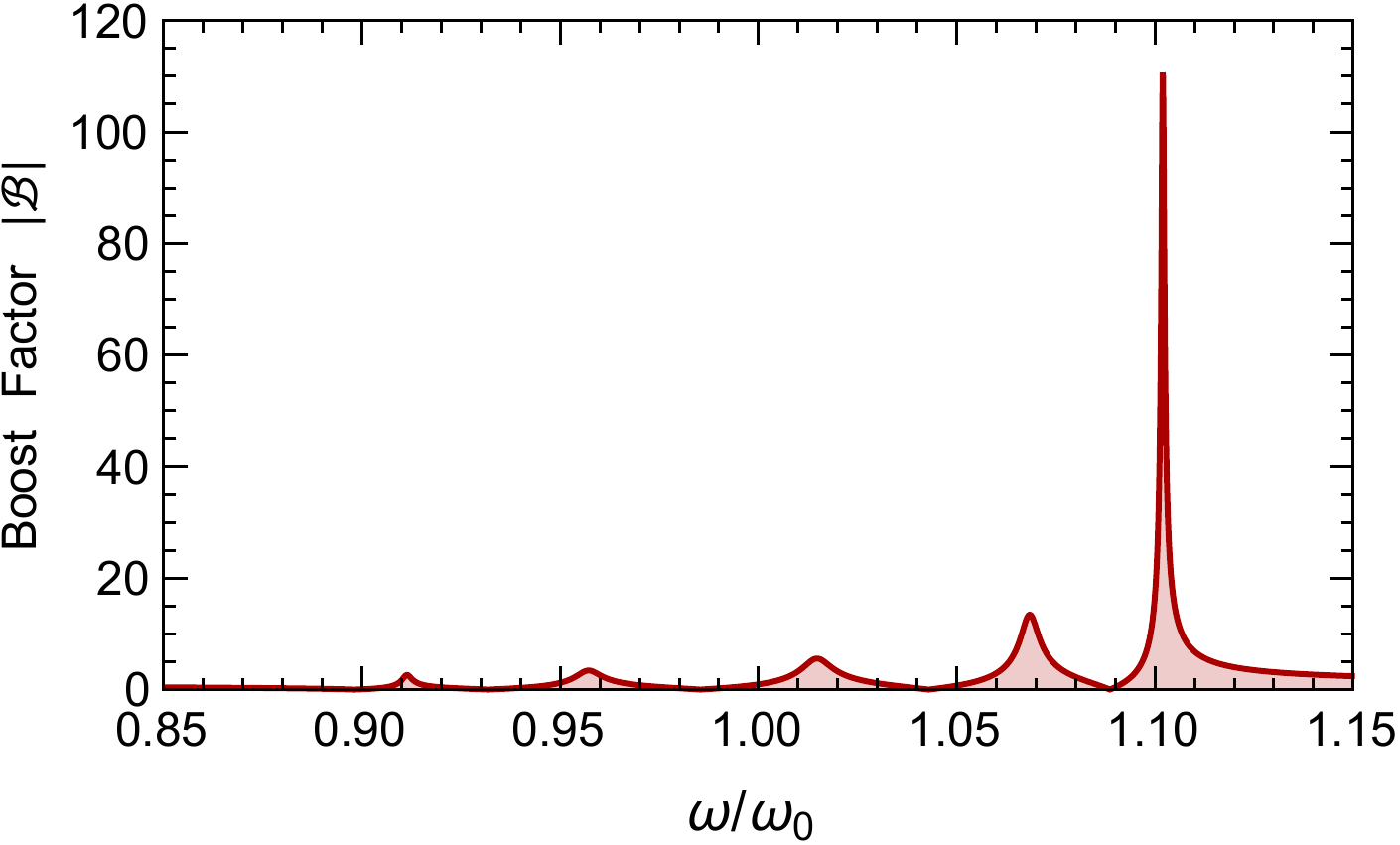}
\caption{$N=11$ equally spaced disks with refractive index $n=5$. The phase depths are arranged
as $\delta_\epsilon/\delta_v=1/2$, the base frequency $\omega_0$ corresponds to $\lambda/4$ in the
dielectric. {\em Top:\/} Transmission factor.
 {\em Bottom:\/} Boost factor.}
\label{fig:homogenous-boost-11}
\end{figure}

More insight is gained from the transparency of the system which we
show in the right panel of figure~\ref{fig:homogeneous-boost}. Here we
do see multiple regions of transparency. The lower diagonal line
corresponds to $\delta_\epsilon=\delta_v/2$ and in particular, for
$\delta_\epsilon=\pi/2$ and $\delta_v=\pi$, it corresponds to the
naive resonance condition where each dielectric layer accommodates
$\lambda/4$ of the EM wave, whereas each vacuum gap accommodates
$\lambda/2$. For a fixed arrangement of the haloscope, and scanning
over frequencies $\omega$, we scan along such lines. Near the naive
resonance point, the transparency indeed shows several ``resonances''
in the form of perfect transparency $|\mathcal{T}|=1$, whereas the
boost factor shows only one dominant peak. Notice also that the boost
resonance does not occur at the naively expected point of
$\delta_v=\pi$ and $\delta_\epsilon=\pi/2$ but somewhat
displaced. This can be understood by analogy with a series of $N-1$
identical coupled cavities. By diagonalising the Hamiltonian one sees
that the couplings induce a splitting of the resonant frequency into
$N-1$ modes, as seen in figure~\ref{fig:homogenous-boost-11}. The
axion itself couples most strongly to the mode in which the $E$-fields
in each cavity are aligned, leading to a dominant resonant peak in
$\beta$. Further, when the number of disks are odd, for half the modes
the integral of the $E$-fields vanishes leading to $\beta=0$.

Adding more disks makes the resonance of the boost factor more pronounced, but there remains only one dominant resonance as mentioned above.
We show an example with $N=5$ in figure~\ref{fig:homogeneous-boost-5}. The transmission factor, on the other hand, shows more resonances. Scanning along the white line in figure~\ref{fig:homogeneous-boost-5} as a function of frequency, the highest-$\omega$ transmission resonance corresponds to the one boost resonance. This correlation suggests a practical way to identify the resonance frequency by a transparency measurement. For different setups, there usually remains a correlation between $\mathcal R$, $\mathcal T$ and $\mathcal B$.

In figure~\ref{fig:homogenous-boost-11} we show the same effect for $N=11$ disks and a larger refractive index $n=5$. Even for these moderate parameters, the boost factor on its resonance reaches around 111 and thus exceeds our nominal benchmark value of~100. For these parameters, the full width at half maximum (FWHM) of the $|\B|^2$ resonance is $0.7\times10^{-3}$, so if $\nu_0=\omega_0/2\pi$ is 25~GHz, then the FWHM is around 18~MHz, so $\beta\sim100$ is reached for only a very narrow range. The integrated $|\B|^2$ scales with $N$, whereas the peak of $|\B|^2$ scales with $N^4$ and its width with $N^{-3}$ to
satisfy the Area Law. So increasing $N$ further, while quickly increasing the peak signal, very quickly makes it exceedingly narrow, in practice probably too narrow for a controlled and stable operation.

\subsubsection{Transparent mode}
\label{sec:Transparent-Mode}

\begin{figure}[b]
\centering
\includegraphics[width=15cm]{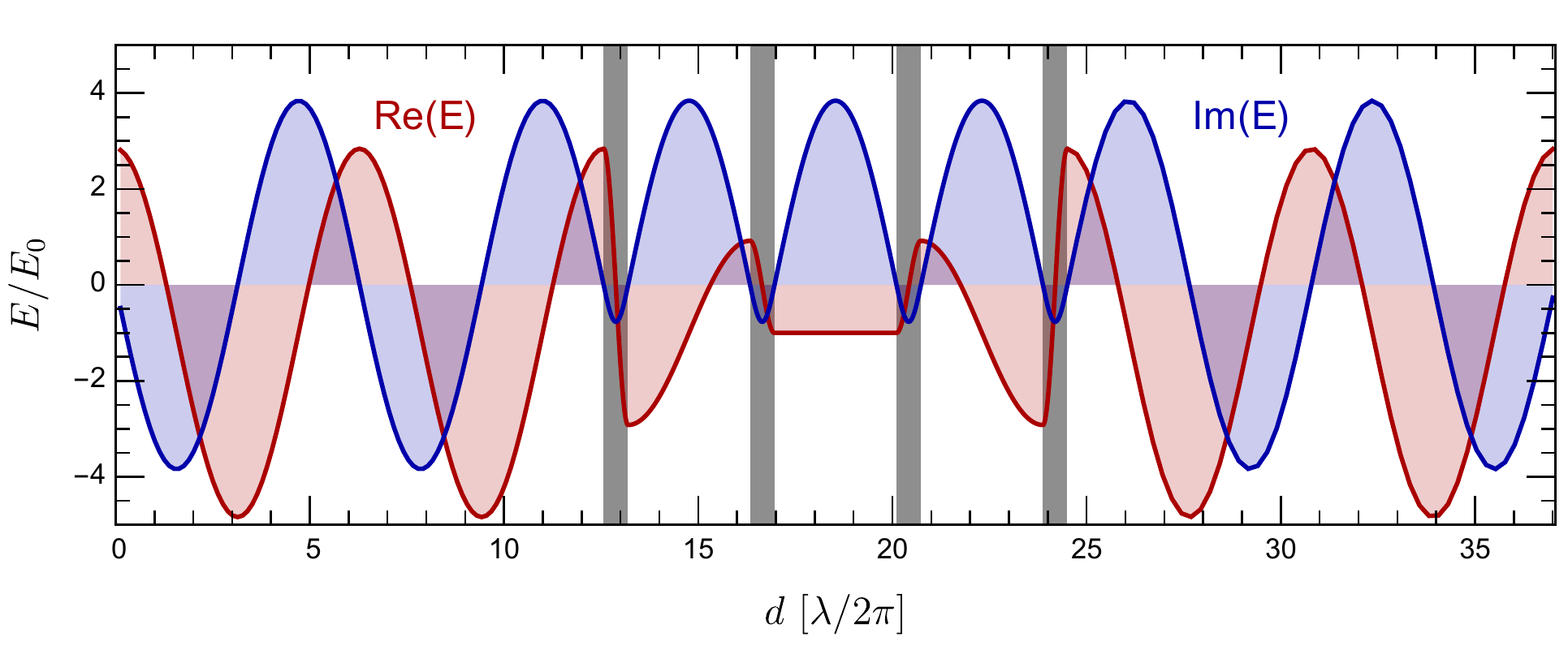}
\caption{Schematic of four dielectric layers (refractive index $n=5$)
  with thickness $d_\epsilon=\lambda/2n$ and separated by
  $d_v=\lambda/2$, i.e., in transparent mode. The red (blue) line
  shows the real (imaginary) electric field distribution, including
  left and right moving waves, as well as the axion induced electric
  field. Note the real and imaginary parts of the $E$-field outside of
  the system are not quite $\pi$ out of phase (as one would expect for
  regular EM waves) because we show the coherent sum of the EM wave
  and the axion-induced $E$-field.}
\label{fig:Transparentfield}
\end{figure}

To avoid very narrow resonances and yet get a large boost,
we now consider the opposite extreme where the response functions for a given number of plates are very broad.
In particular, we consider the homogeneous haloscope adjusted such that
the distance between disks is $d_v=nd_\epsilon$, implying that the phase depths are the same,
$\delta_v=\omega d_v=\delta_\epsilon=n\omega d_\epsilon$, the transparent setup
\cite{Jaeckel:2013eha}. For the frequency
$\omega_0 = \pi/d_v=\pi/(n d_\epsilon)$ both the dielectric layers and the vacuum gaps each accommodate $\lambda/2$ of the EM wave.
In this case, all the dielectrics emit EM waves whose amplitude is given by the boost factor $\beta_{\rm D}$ given by equation~\eqref{eq:maximum-boost-disk}. The phase depth of every gap and every disk is $\pi$, so the phase accrued by propagation through a vacuum gap plus the next disk is $2\pi$ and so all these amplitudes add up coherently, leading to a combined boost factor of
\be
\beta_{\rm T} = N \beta_{\rm D}=N \(1-\frac{1}{n^2}\)\,. \label{BN}
\ee
We show the electric field distribution for a four-disk example in
figure~\ref{fig:Transparentfield}.

For frequencies de-tuned from full transparency by a small amount
$\omega=\omega_0(1+s)$ with $|s|\ll 1$, after passing a disk plus a gap,
the EM wave accrues an extra phase $2\pi s$, and after $N$ disks and gaps this amounts to
$2\pi s N$, leading to fully destructive interference if $s=1/(2N)$. Therefore, we expect $\Delta\omega\sim \omega_0/N$ for the frequency range over which the boost factor is strongly enhanced. The emitted power scales as $\beta^2\propto N^2$, meaning that the ratio of the power to the bandwidth increases linearly with $N$. This behavior is in stark contrast to increasing the quality factor $Q$ in a resonantor, where $P\propto Q$ and the bandwidth $\propto Q^{-1}$. The difference to a resonant cavity is also underscored by the $E$-field distribution throughout the system shown in figure~\ref{fig:Transparentfield} as compared to that in a resonator that was shown in figure~\ref{fig:cavityfield}. Unlike
a cavity, the $E$-field inside and outside the system is here of similar magnitude and indeed larger outside.

\begin{figure}[t]
\centering
\includegraphics[width=9cm]{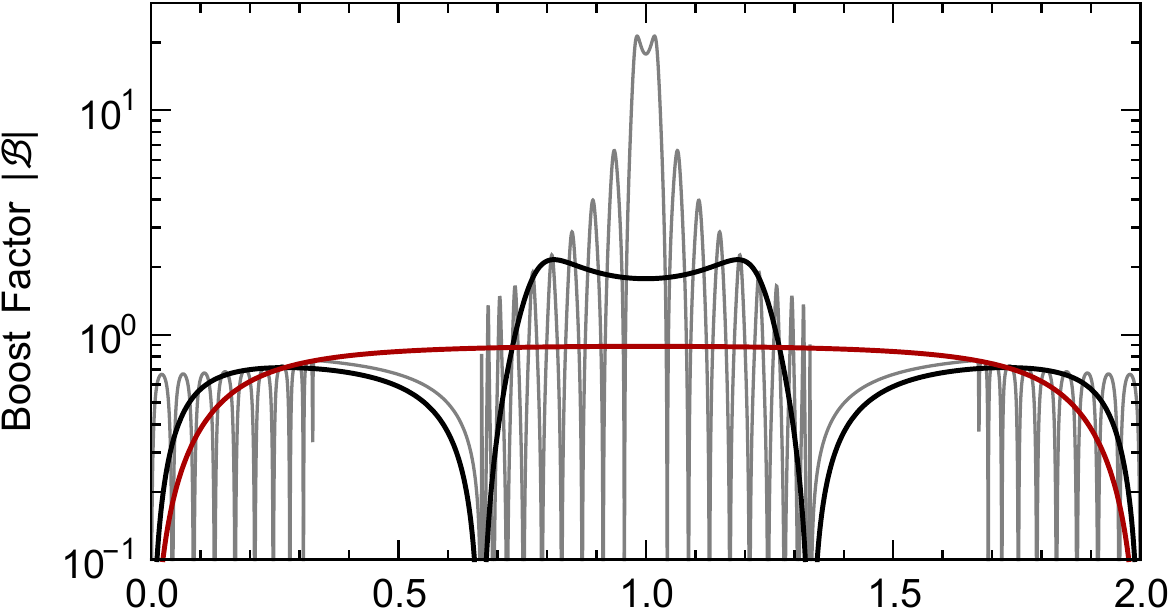}
\vskip6pt
\includegraphics[width=9cm]{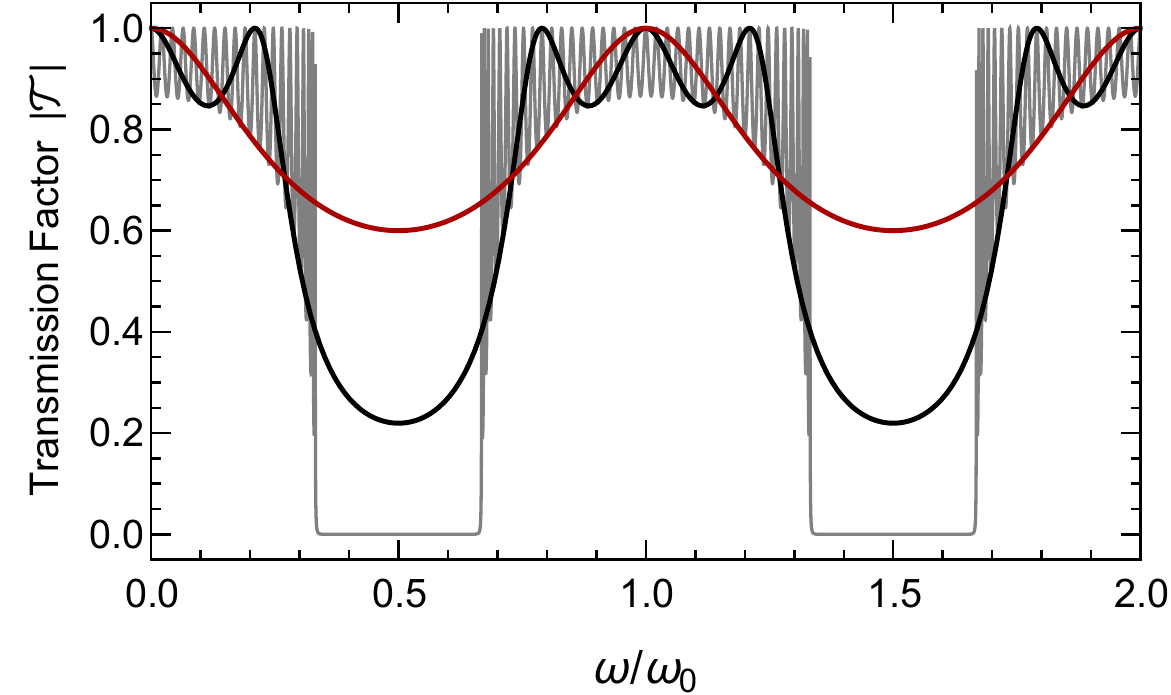}
\caption{Boost factor $|\B|$ (top)
and transmission factor $|{\mathcal T}|$ (bottom) for sets of $N=1$, 2, 20
(red, black, gray) dielectrics ($n=3$). The disks have
thickness $d_\epsilon=\pi/n\omega_0$ separated by vacuum gaps $d_v=\pi/\omega_0$, i.e.,
for $\omega=\omega_0$ we have the transparent mode.}
\label{base}
\end{figure}

For frequencies other than $\omega_0$ we can use the the full expressions for
$\TT$ and $\MM$ to obtain the response of the haloscope.
In figure~\ref{base} we show the boost factor (top) and transmission coefficient $\mathcal T$ (bottom) of a homogeneous setup with $N=1$, 2 and 20 dielectric layers ($n=3$). We observe that indeed the central peak of the boost factor becomes larger and narrower as $N$ increases.

\begin{figure}[t]
\centering
\includegraphics[width=9cm]{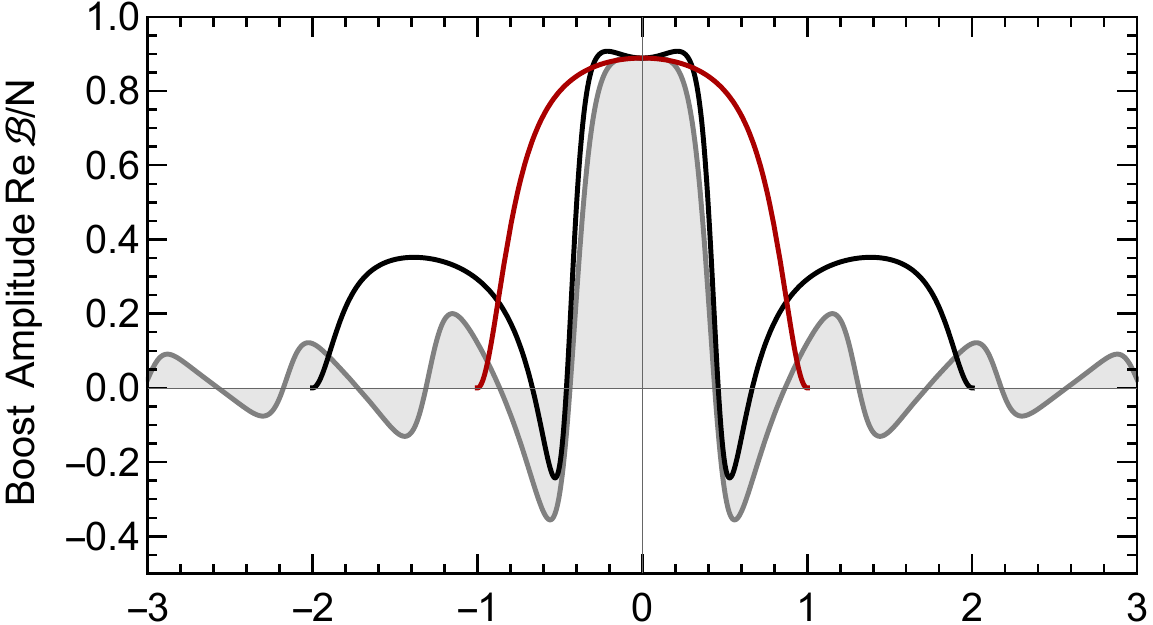}
\vskip6pt
\includegraphics[width=9cm]{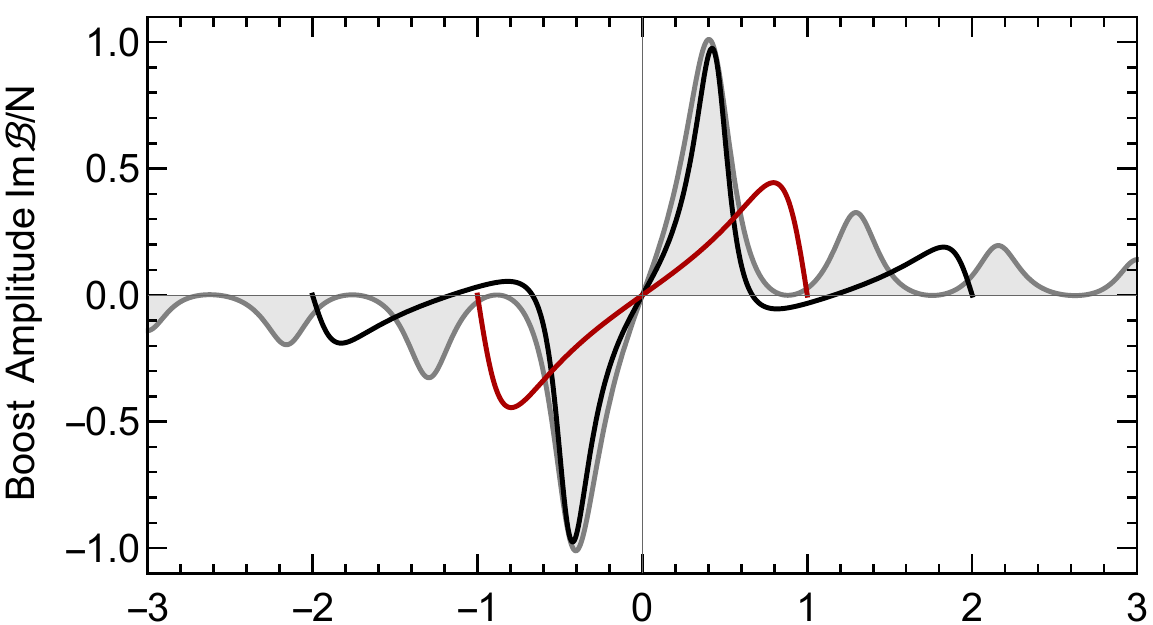}
\vskip6pt
\includegraphics[width=9cm]{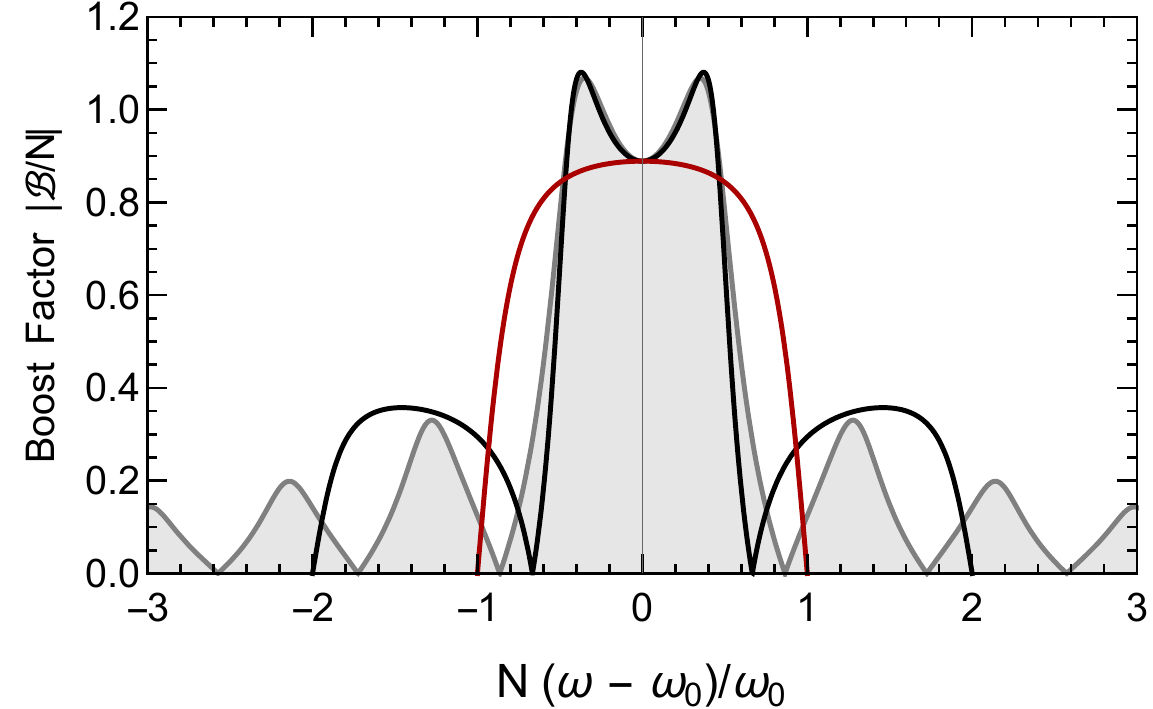}
\caption{Boost amplitude (real and imaginary part) and boost factor
for sets of $N=1$, 2, 20
(red, black, gray) dielectrics as in figure~\ref{base}.
Amplitude scaled with $1/N$, frequency with $N$, so that the central peaks
have similar size for all $N$ values.}
\label{fig:transparent-boost}
\end{figure}

To see this scaling more explicitly we show in
figure~\ref{fig:transparent-boost} the real and imaginary parts of $\B/N$
as a function of $N(\omega-\omega_0)/\omega_0$. The central peak of the scaled boost factor as a function of $\omega$ for two disks is quite similar to that for twenty disks, i.e., it is a nearly universal function. At the edge of the central peak, the boost amplitude is dominated by the imaginary part, i.e., the emerging EM wave is strongly phase shifted relative to the primary axion field. However, it is the real part of $\B$ which dominates over most of the frequency range. The boost factor is almost box-shaped as a function of frequency. Its frequency average was given in  equation~\eqref{eq:sum-rule-transparent-1} for any homogeneous haloscope.

Adding more disks concentrates the response around $\omega=\omega_0$, corresponding to our earlier discussion that the height of the central peak increases with $N$, whereas its width decreases with $1/N$. Likewise, increasing the refractive index narrows the boost profile and enhances it. At the same time, the dips in the transmission and boost factors get more pronounced. The central depression of the boost factor has a value given by equation~\eqref{BN} and does not change much with increasing $n$. However, the height of the two bumps increases, so the profile loses its box shape for larger $n$.

\begin{figure}[t]
\centering
\includegraphics[width=9cm]{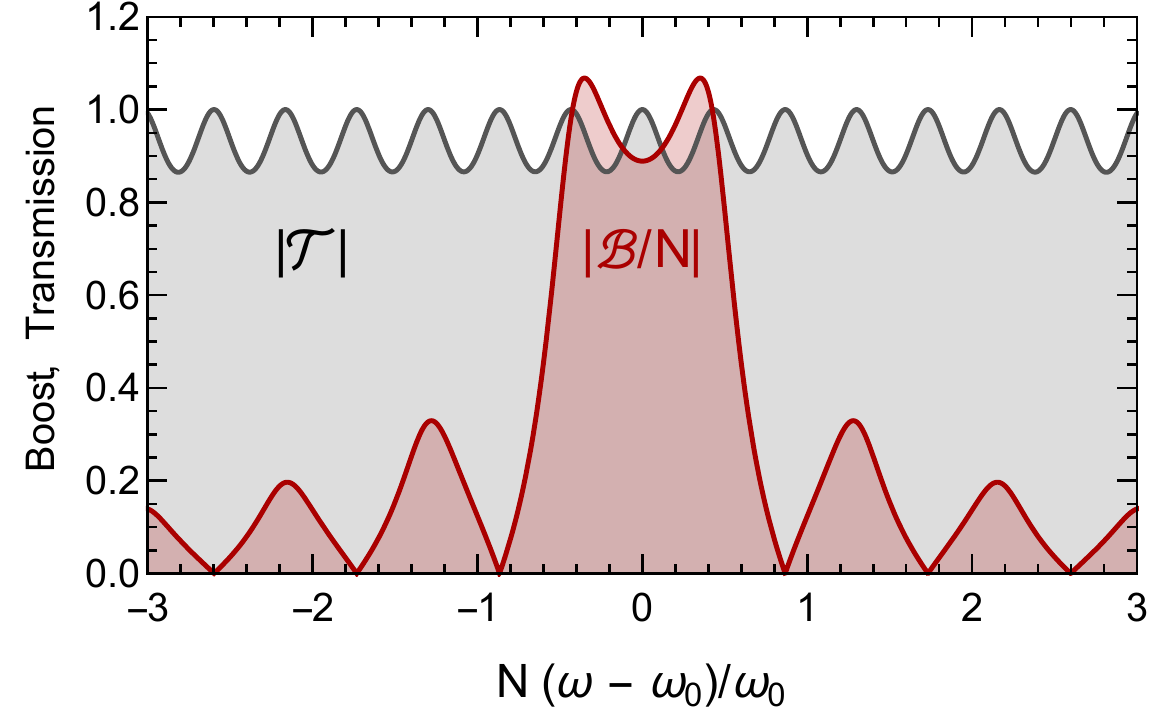}
\caption{Scaled boost factor and transmission coefficient for large $N$.}
\label{fig:transparent-boost-transmission}
\end{figure}

One may wonder if the boost factor can be empirically determined or experimentally confirmed once the haloscope has been set up. We could test the response by measuring the transmission as a function of frequency. However, for $\omega$ around $\omega_0$ the layered dielectrics are essentially transparent, with small dips as function of frequency, covering a broad $\omega$ range as seen in figure~\ref{base}. Increasing $N$ increases the number of dips, but
on the frequency range where the boost factor is large, the transmission factor also becomes a universal function as shown in figure~\ref{fig:transparent-boost-transmission}. Therefore, measuring the transmission
does not reveal the frequency $\omega_0$ to which the boost factor has been tuned. This situation is very different in the resonant case as discussed in the previous section.

\subsection{Sensitivity to inaccurate disk positioning}
\label{positioning}

In a realistic setup, the distance between dielectrics cannot be made arbitrarily precise and thus cannot be precisely adjusted to $d_r=\pi/\omega_0$ for every gap $r=1,\ldots,N-1$ (neglecting the dielectric regions). Rather, each gap distance
will have an uncertainty $\Delta d_r$. If these errors are small, the boost factor for the many disk transparent mode discussed in section~\ref{sec:Transparent-Mode} becomes roughly
\be
\beta_{\rm T} \sim \biggl|\sum_{r=1}^{N-1} e^{i\omega_0 \sum^r_{r'=1}  \Delta d_{r'} }\biggr| \, .
\ee
The average shift of the power boost factor is negative and can be estimated by assuming Gaussian positioning errors with standard deviation $\langle \Delta d^2_r\rangle = \sigma^2$ as  
\be
\left\langle \Delta \beta_{\rm T}^2 \right\rangle = -\omega_0^2\sigma^2 \frac{(N-1)(N-2)(N-3)}{6}\,.
\ee
This is to be contrasted with the simple cavity setup with large $n$ of section~\ref{cavity}, where inaccurate positioning of the singular dielectric instead gives
\be
\langle \Delta \beta_{\rm C}^2\rangle \sim  -\omega_0^2\sigma^2(4 n^6-5  n^4-n^2)\,,
\ee
where here we have assumed $n$ real for simplicity.

Rewriting these results as a function of the boost factor itself we get
$\langle \Delta \beta_{\rm T}^2\rangle\sim - \omega_0^2\sigma^2 \beta_T^3/6$,
while we get a similar result for the cavity
$\langle \Delta \beta_{\rm C}^2 \rangle\sim - \omega_0^2\sigma^2 \beta_{\rm C}^3/2$.
In order to secure the control over the boost factor we would need
\be
\frac{\langle \Delta \beta^2 \rangle}{\beta^2} \sim -\omega_0^2 \sigma^2 \beta \ll 1
\ee
which implies $\sigma \ll \lambda/(2\pi\sqrt\beta)$ with $\lambda$ the wavelength. Therefore, we need
\be
\sigma \ll 200\, \mu {\rm m} \(\frac{10^2}{\beta}\)^{1/2}\(\frac{100\,\mu\rm eV}{m_a}\) , \label{errorbound}
\ee
which is consistent with numerical simulations (i.e., to have the average variation be less than 10\% one would need $\sim 20~\mu$m precision for $\beta=100$). As $\sigma\propto \beta^{-1/2}$ the sensitivity varies slowly with $\beta$.

This condition also bounds the allowed deviations of the surface from being perfectly flat. In general, the error involved in moving the disks should be more significant than manufacturing errors. A similar bound should apply to errors in the disk alignment, but a full 3D study of finite disks has not been done. Note that as the sensitivity to errors increases with $\beta$, for a practical experiment one would wish to avoid extremely large boost factors, or equivalently, high $Q$ in the cavity case. As the sensitivity increases linearly with the axion mass, for very high axion masses one will be unable to position the dielectrics accurately enough. While this limitation depends on the concrete mechanical setup, we expect that achieving a positioning precision better than $\mu$m will be impractical. Thus we would expect that one is restricted to $\beta\lesssim 10^{3-4}$, however the exact sensitivity to error will depend on the specific setup.

Numerically, we tend to see that small errors (those satisfying equation~\ref{errorbound}) correspond to small shifts to the boost factor in frequency space without altering the shape. Note that despite the fact that a dielectric haloscope in the transparent mode might require $\mathcal O(100)$ surfaces to reach the same $\beta=100$ as a cavity with a single $n=50$ dielectric, the sensitivity to misalignments is very much similar. This similarity is because many of the positioning errors will cancel their effects on average. However, not all setups with the same $\beta$ are equally sensitive to errors: setups arranged so that the boost factor is a local maximum with respect to changing the positions of the dielectrics can be more susceptible to error as all errors necessarily reduce the boost factor. Further, as we saw in section~\ref{transparent}, a setup with many dielectrics gives the possibility of a much larger bandwidth than a cavity.

As we expect to have similar sensitivity to errors in the positioning if the resonator is working in the transparent or cavity case (for equal boost factors), we expect that one can use setups which contain features of both the transparent and resonant cases with similar robustness with respect to positioning errors.

\section{Broadband response}
\label{optimum}

For a realistic haloscope, a resonant response function is not necessarily optimal, depending on various practical considerations. For a fixed set of dielectric disks with given uniform thickness, adjusting each vacuum spacing independently offers a large number of degrees of freedom to tailor a frequency-dependent response function within the limitations of the Area Law. We here study a few examples of constructing a nearly top-hat shaped response function.

\subsection{Motivation and setup}

So far we have been concerned with simple and demonstrative setups with high levels of symmetry, which however are not necessarily optimal for a practical experiment. A necessary trait is the ability to search a wide $m_a$ range via a scanning procedure. The transparent and resonant setups discussed in section~\ref{Genericexamples}, as well as those considered in references~\cite{Morris:1984nu,Jaeckel:2013eha}, require specifically chosen thicknesses for the dielectrics and would need a new set of disks for each measurement, which is clearly impractical.

For a realistic experiment it is desirable to have both
broad rectangular (top-hat) responses and narrow resonant ones \cite{TheMADMAXWorkingGroup:2016hpc}. Broadband setups reduce the sensitivity to positioning errors and compensate for the potentially non-negligible time $t_{\rm R}$ required to readjust the disk positions between measurements (see appendix~\ref{scanoptimisation}). One could scan a large frequency range in one go, and then use narrow resonances to confirm or reject a potential detection. While the homogeneous setup discussed in section~\ref{transparent} can be adjusted between a wide range of frequencies, one lacks the ability to move from broadband to narrow-band responses at the same central frequency.

To show that such versatile setups are possible, we here consider dielectric disks of fixed uniform thickness, with each vacuum gap as a separate degree of freedom. Adjusting these spacings, we will see that one can control the height, width, shape, and position of the boost factor as a function of frequency.

Henceforth we will consider one side of the device to be closed by a perfect mirror as in reference~\cite{TheMADMAXWorkingGroup:2016hpc}. This setup ensures that the full power emerges in a single direction and can be measured by a single detector. Otherwise two detectors would be needed, introducing new issues about adding the signals coherently or incoherently and introducing two sources of detector noise. Our specific examples use a set of 20~aligned dielectric disks (1~mm thick, refractive index $n=5$), providing enough degrees of freedom to make our point, yet few enough to handle them with relative ease. Our benchmark frequency is $\nu=25$~GHz ($m_a=103.1~\mu$eV), corresponding to a vacuum wavelength of $\lambda_0=1.20$~cm and a wavelength in the dielectric of
$\lambda_\epsilon=\lambda_0/n=0.24$~cm, i.e., there is no special relation between the disk thickness and chosen central frequency.

\begin{figure}[t]
\centering
\includegraphics[width=10cm]{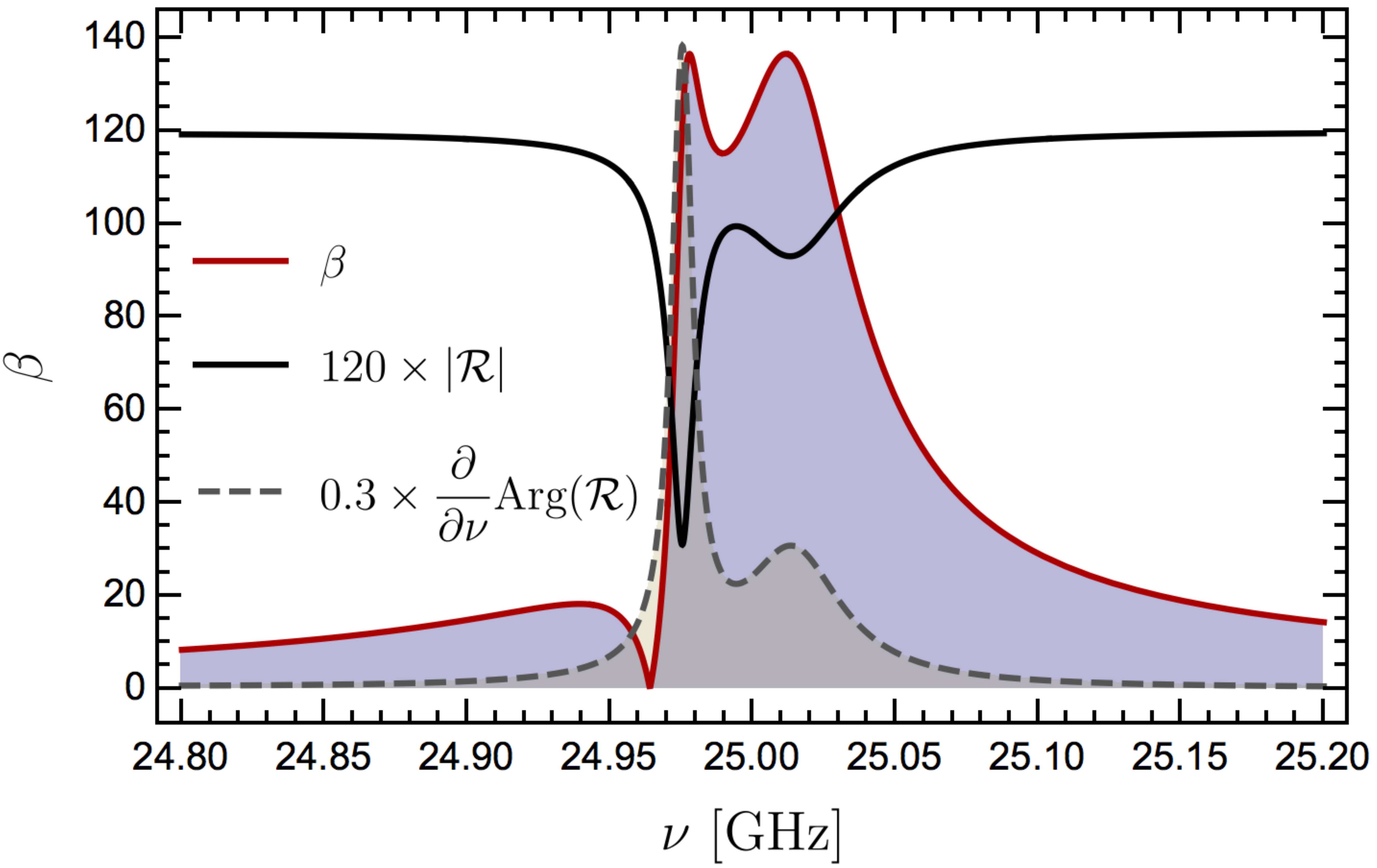}
\caption{Boost factor $\beta$ (red), reflectivity $|\mathcal R|$ (black) and group delay $\frac{\partial}{\partial\nu}\text {Arg}(\mathcal R)$ (dashed gray) as a function of frequency $\nu$ for our configuration B50 (20~disks, 1~mm thick, refractive index $n=5$, mirror on one side, bandwidth 50~MHz centred on 25~GHz).
The reflectivity is illustrated for exaggerated dielectric losses ($\tan\delta=5\times10^{-3}$) to show the non-trivial structure. The reflectivity and group delay have been scaled by the factors 120 and 0.3, respectively.}
\label{fig:50mhz}
\end{figure}

\begin{figure}[t]
\centering
\includegraphics[width=14cm]{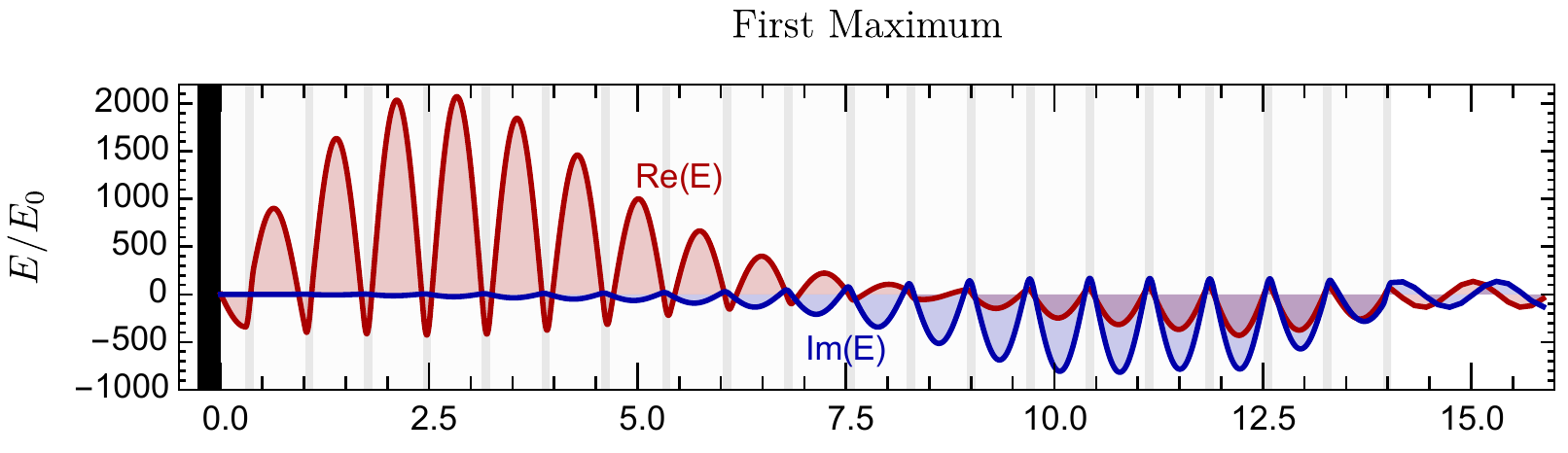}
\includegraphics[width=14cm]{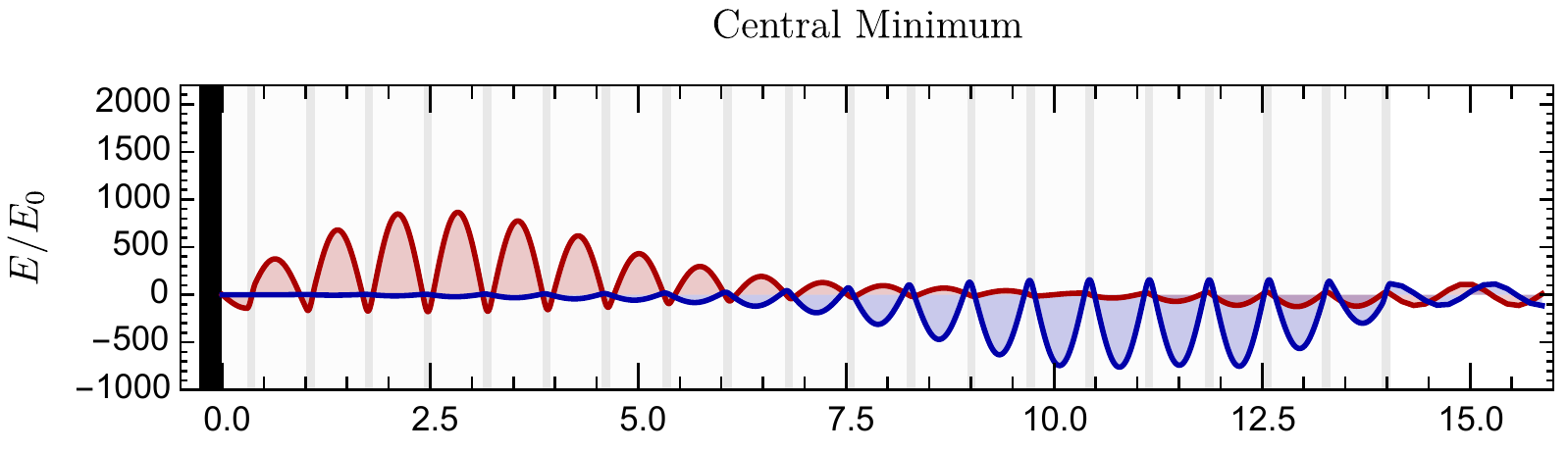}
\includegraphics[width=14cm]{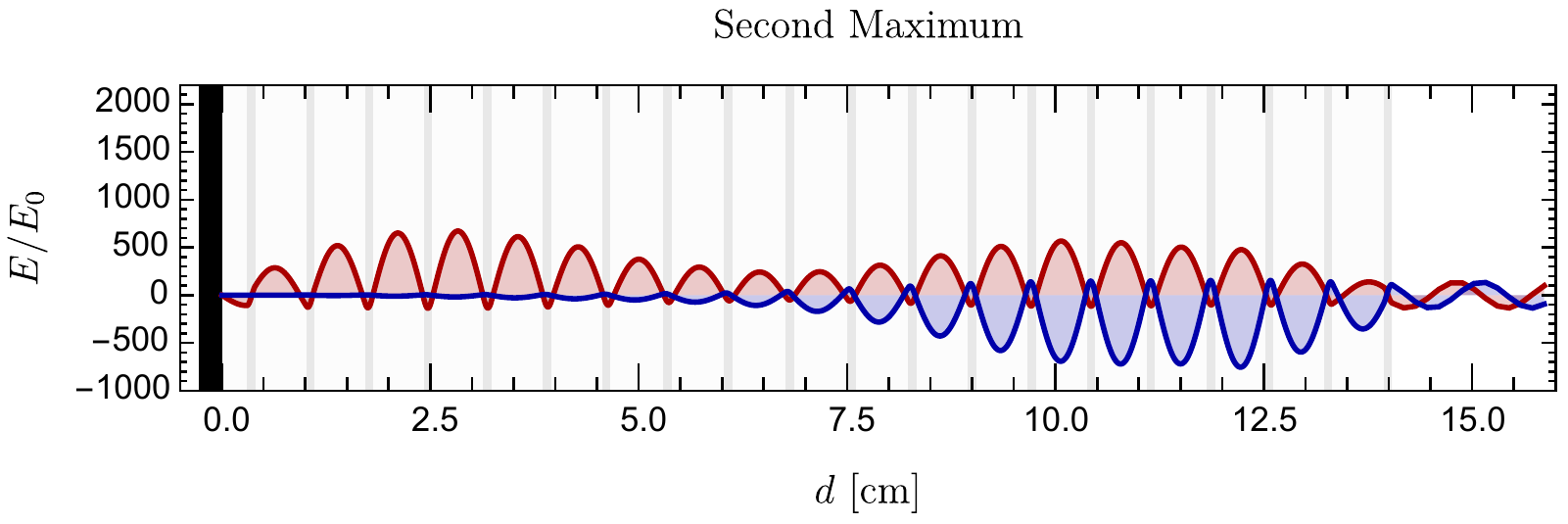}
\caption{$E$-field distribution as a function of distance $d$ from the mirror
in our B50 configuration of a dielectric haloscope. Real part of $E$ in red, imaginary part in blue. As labeled, the panels refer to the frequencies of the first maximum, the minimum, and the second maximum of the boost factor curve shown in figure~\ref{fig:50mhz}.
In each panel the locations of the mirror and the dielectric disks are indicated respectively by the black and the light-gray vertical bars.}
\label{fig:E-fields2}
\end{figure}

\begin{figure}[t]
\centering
\includegraphics[width=14cm]{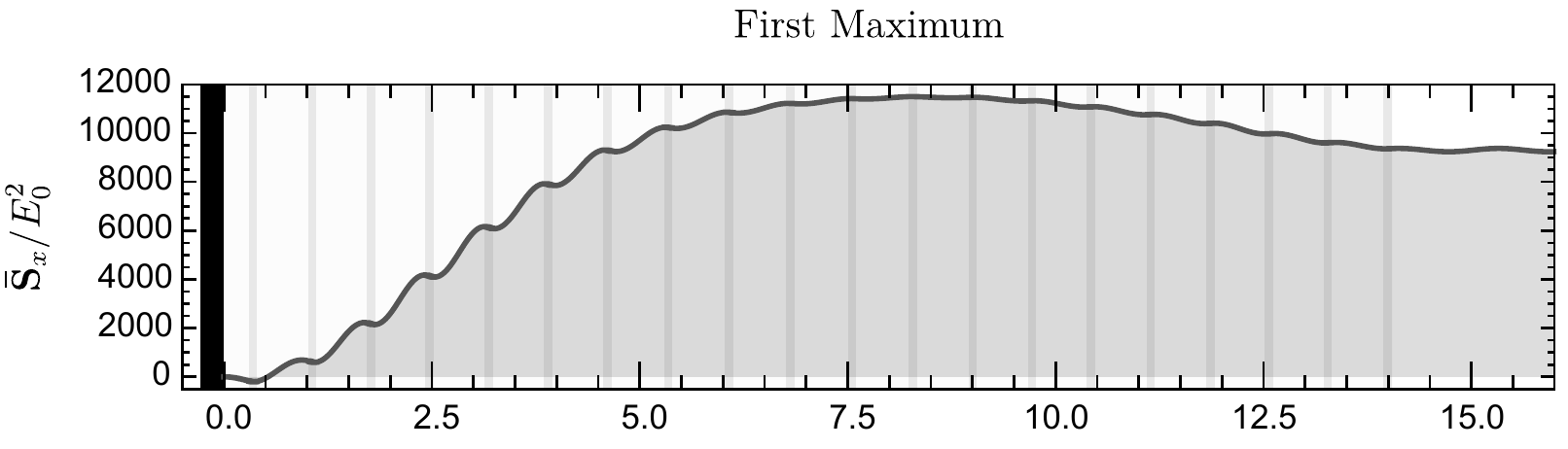}
\includegraphics[width=14cm]{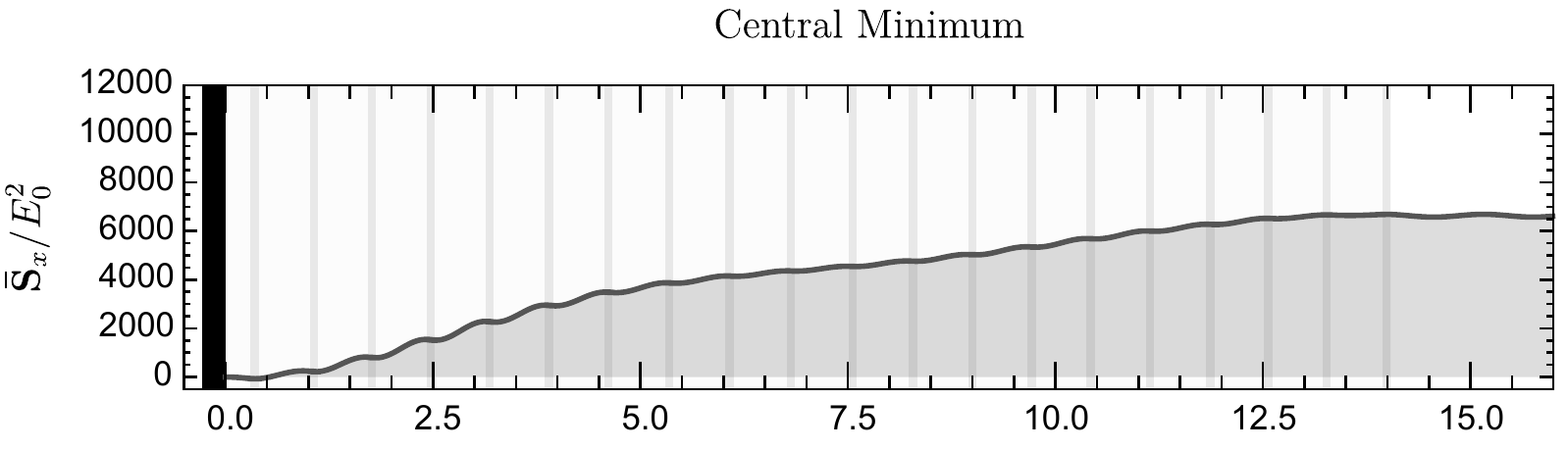}
\includegraphics[width=14cm]{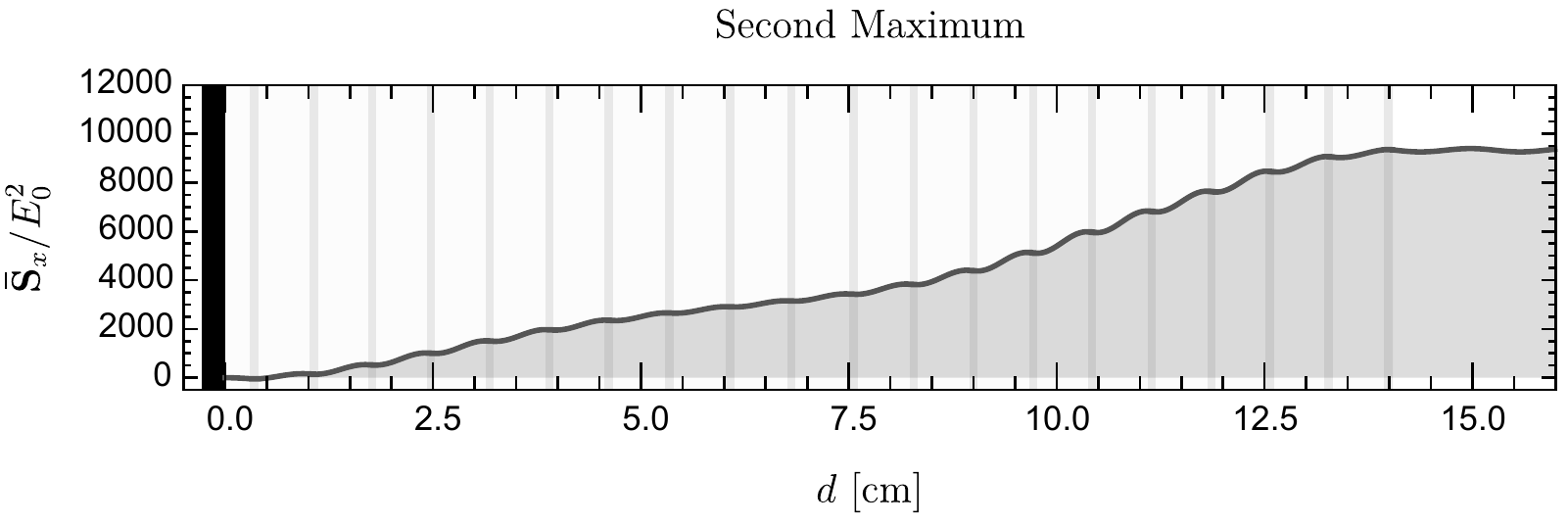}
\caption{Cycle averaged Poynting flow $\bar{\bf S}_x$ in our B50 haloscope, corresponding to the $E$-field configurations shown in
figure~\ref{fig:E-fields2}. Note that the sign of $\bar{\bf S}_x$ indicates the direction of energy flow. The mirror and dielectric disk locations are indicated as in figure~\ref{fig:E-fields2}.}
\label{fig:P-fields2}
\end{figure}

\subsection{Configuration with 50 MHz bandwidth}
\label{50mhzop}

For a first example, termed configuration B50 (B for broadband), we choose a bandwidth of 50~MHz, which is quite broad in comparison to the axion line width $\Delta\nu_a={\cal O} (10~{\rm kHz})$. We seek a response function which covers
this bandwidth as uniformly as possible, i.e., which comes close to a top-hat shape. To achieve a desired signal-to-noise ratio across this range, the lowest boost factor in this interval is the critical figure of merit. To find the optimal
disk positions, we sample a set of frequencies $\nu_i$ in the chosen interval using a random walk in the 20~dimensional parameter space to find the configuration that maximises the minimal value
$\text{Min}\[\beta(\nu_i)\]$. Our best configuration reaches
$\text{Min}\[\beta(\nu_i)\]\sim 115$. The frequency-dependent boost factor is shown in figure~\ref{fig:50mhz}. It has approximately rectangular shape, but in detail shows two distinct peaks.

For a heuristic understanding of this shape, we look to the EM response functions. Because our haloscope is closed by a perfect mirror on one end, the transmission coefficient vanishes, whereas the modulus of the reflection coefficient $\mathcal R$ is unity. However, the phase of $\mathcal R$ carries non-trivial information. While the microwaves are always reflected, the path-length depends on frequency: near frequencies that experience a large number of internal reflections the phase of the reflected radiation must change rapidly. Thus the derivative of this phase, the group delay $\frac{\partial}{\partial\nu}\text {Arg}(\mathcal R)$, maps out resonances in the system. Of course, $\mathcal R$, $\frac{\partial}{\partial\nu} \mathcal R$, and $\text{Arg}(\mathcal R$) carry the same information, but it is brought out most clearly in the group delay. When losses are included, regions with more internal reflections will experience more damping, so $|\mathcal R|$ also maps out the resonant structure, although it does not seem to provide new information over the group delay. We can see this correlation in figure~\ref{fig:50mhz}. While the peaks coming from $\mathcal R$ are shifted slightly relative to those in $\beta$ (and the relative peak heights change), it seems plausible that one could use the group delay to verify the boost factor and correct for errors.

The double peak structure in both $\beta$ and $\cal R$ suggests a combination of two resonances. We confirm this interpretation in figure~\ref{fig:E-fields2} by looking at the $E$-field distribution in the B50 haloscope at the maxima and minimum of the boost-factor curve. Indeed the sharper peak on the lhs seems to correspond to resonant enhancement by the dielectrics close to the mirror, and the rhs peak corresponds to a less pronounced resonance on the side opposite the mirror. This
behavior is echoed in $\mathcal R$, with the higher peak corresponding to the stronger resonance. The $E$-field is not simply a standing wave as one would expect in a true resonator, rather both standing wave and significant traveling waves exist. The latter are gleaned from the phase changing spatially throughout the haloscope. Note that the real and imaginary parts of $E$ are not exactly $\pi/2$ out of phase as the axion-induced electric field $E_a$ is included.

Another way to gain more intuition about these coupled resonators is to look at the cycle-averaged Poynting flow that was defined in equation~\eqref{eq:Poynting}.  Figure~\ref{fig:P-fields2} reveals that most of the emergent microwave power is generated in the haloscope region near the mirror for the first resonance, although the entire haloscope contributes. The central minimum and second maximum show less localised power generation.

\subsection{Shifting the response in frequency space}

\begin{figure}[t]
\centering
\includegraphics[width=12cm]{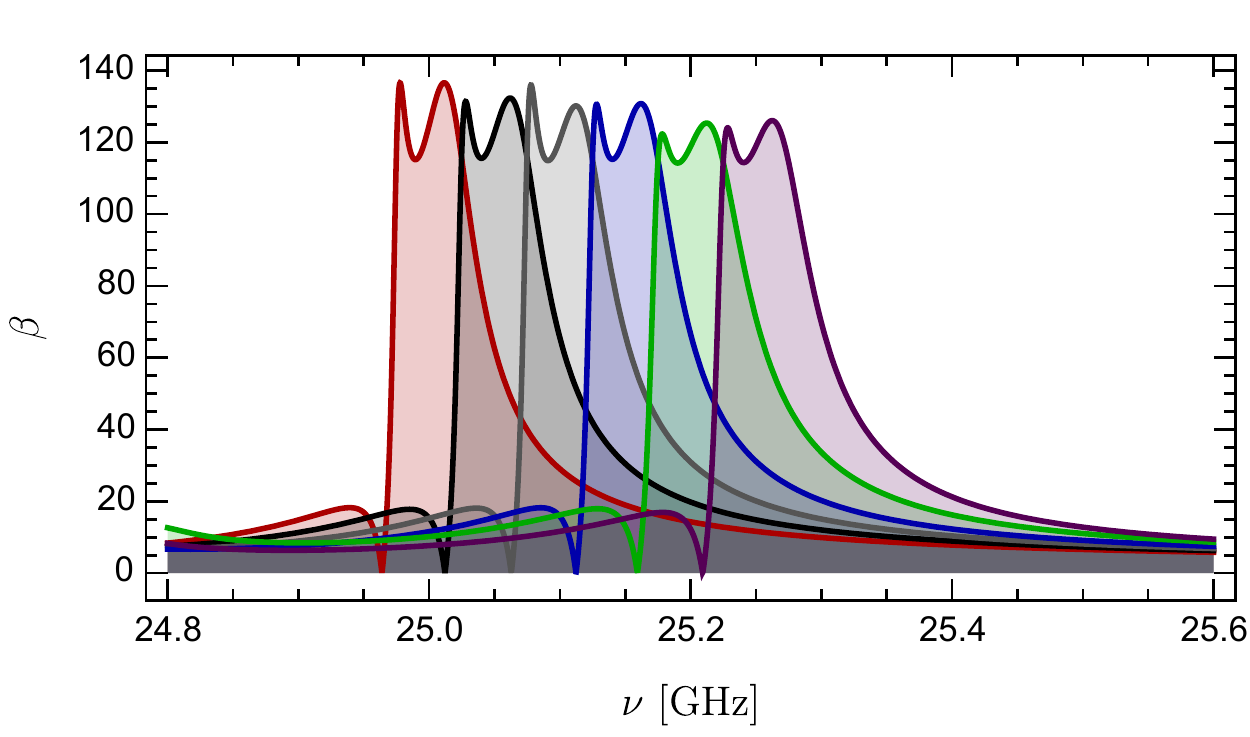}
\caption{Example of a scan across a frequency range of 300~MHz using six
B50 configurations with shifted central frequency. The left-most curve is identical with the boost factor of figure~\ref{fig:50mhz}, corresponding to the original case centred on 25~GHz.}
\label{fig:scan}
\end{figure}

In a realistic experiment one needs to scan over a much broader frequency range than given by our B50 arrangement. It is one of the main attractions of our approach that one can easily shift the B50 response function from its original center at 25~GHz, for example in steps of 50~MHz to achieve contiguous coverage of a broader search range. In figure~\ref{fig:scan} we show an example of six B50 response functions that seamlessly cover a 300~MHz interval. The disk positions for a shifted configuration are very similar to the previous ones and indeed we always used the previous configuration as a starting point for finding the next shifted one. Notice that the ability to shift the response function in this way depends on the dielectrics not having a special thickness relative to the chosen frequencies, i.e., they are neither transparent nor fully reflective.

Note that we do not quite reach the same boost factor for all configurations, i.e., the central frequency relative to the chosen dielectric thickness is not irrelevant. Indeed with a single set of dielectrics it is not possible scan all frequencies. When the disks are transparent, as considered in sections \ref{mirrortransparent} and \ref{sec:Transparent-Mode}, $\text{Min}\[\beta(\nu_i)\]\leq 2N+1$. Further, as can be seen in section~\ref{slab}, the disks do not emit any radiation at $\nu=2,4,6,...\times\pi/\sqrt{\epsilon}d$. Therefore, a realistic dielectric haloscope will require at least two sets of disks of different thicknesses to avoid such issues.

\subsection{Varying the number of disks}

\begin{figure}[t]
\centering
\includegraphics[width=13cm]{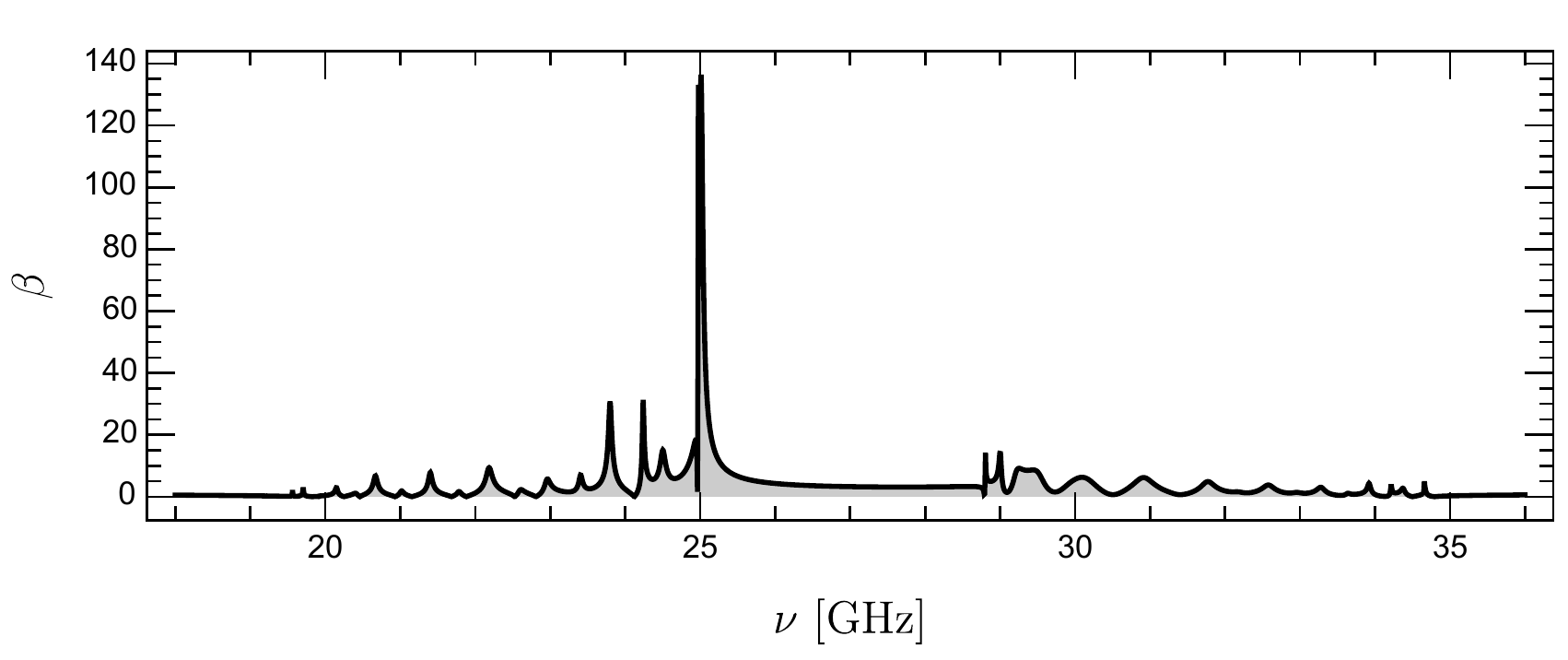}
\caption{Boost factor of the B50 configuration as in figure~\ref{fig:50mhz} for a broader
range of frequencies. The dominant peak appears in the 50~MHz region around the central frequency of 25~GHz.}
\label{fig:largeband}
\end{figure}

Starting with the B50 configuration, which has 20 disks, we can use the Area Law to extrapolate these results to settings with a different number $N$ of disks. From the Area Law~\eqref{eq:arealawa2} we know that the area under $\beta^2(\nu)$ is due to a sum over the interfaces and so should increase linearly when one adds disks. In figure~\ref{fig:largeband} we see that for
a wide frequency region surrounding our frequency of interest, the boost factor is dominated by the $\beta$ peak we optimised for, implying that the Area Law applied to this peak alone gives us a good estimate of the height$^2{\times}$width of our top-hat boost factor curve.

\begin{figure}[ht]
\centering
\includegraphics[width=9cm]{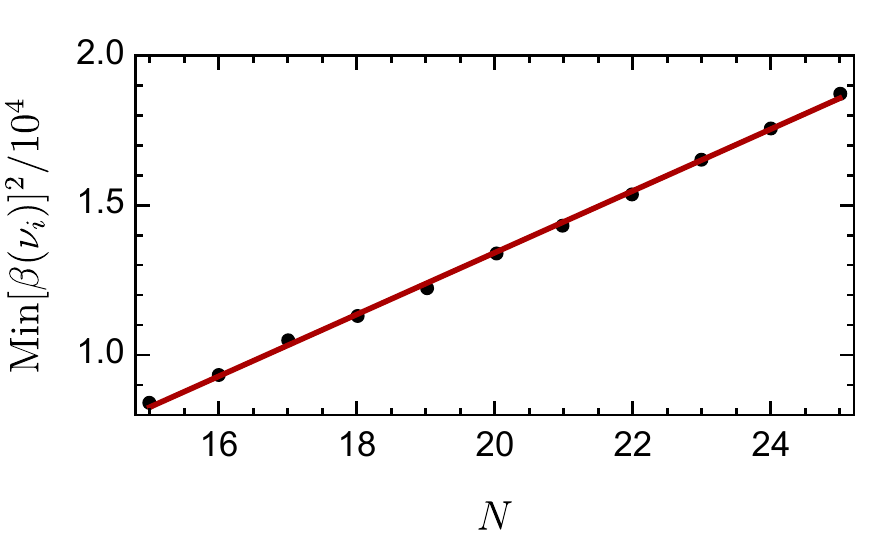}
\caption{Squared boost factor $\text{Min}\[\beta(\nu_i)\]^2$ achieved by our B50 configuration with
fixed 50~MHz width, but varying the number $N$ of disks. The variation is indeed approximately
linear.}
\label{fig:B50-variableN}
\end{figure}

We consider our B50 configuration, keeping the 50~MHz width fixed, but varying the number $N$ of disks. We perform the same optimization procedure and determine the boost factor
$\text{Min}\[\beta(\nu_i)\]$ within the chosen bandwidth. In figure~\ref{fig:B50-variableN} we show the result  for $N=15$--25 and find excellent agreement with the predicted
linear variation of the squared boost factor with $N$. In other words, we can use the modified number of disks to modify the boost factor in the given frequency band, or we can modify the band while keeping the boost factor fixed.

\subsection{Configuration with 200 MHz bandwidth}
\label{difband}

\begin{figure}[ht]
\centering
\includegraphics[width=10cm]{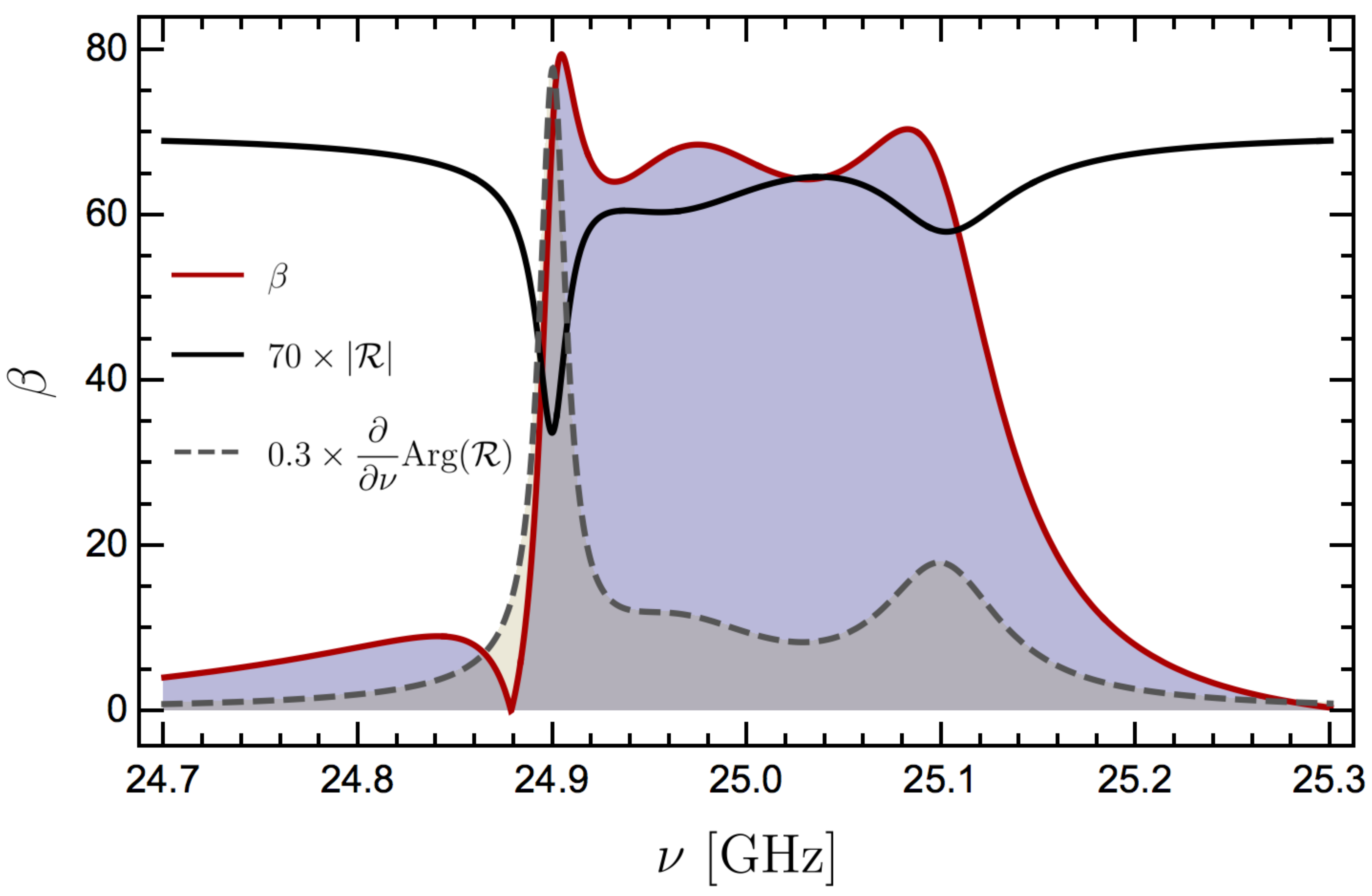}
\caption{Boost factor $\beta$ (red), reflectivity $|\mathcal R|$ (black) and group delay $\frac{\partial}{\partial\nu}\text {Arg}(\mathcal R)$ (dashed gray) as a function of frequency $\nu$ for our configuration B200 (20 disks, 1~mm thick, refractive index $n=5$, mirror on one side, bandwidth 200~MHz centred on 25~GHz) in analogy to figure~\ref{fig:50mhz}.
The reflectivity is illustrated for exaggerated dielectric losses ($\tan\delta=5\times10^{-3}$) to show the non-trivial structure. The reflectivity and group delay have been scaled by the factors 70 and 0.3, respectively.}
\label{fig:200mhz}
\end{figure}

We return to our configuration with a fixed number of $N=20$ disks, but now consider a broader frequency band of 200~MHz while keeping all else fixed: the B200 configuration. We perform an analogous procedure and show in figure~\ref{fig:200mhz} the corresponding response functions. We recognise that one can indeed trade between bandwidth and boost factor,
now reaching $\text{Min}\[\beta(\nu_i)\]\sim 65$. Actually this value is slightly larger than expected from a naive application of the Area Law. The response function is here more rectangular, i.e., the smallest boost factor within the chosen frequency interval does not dip down as far.

\begin{figure}[ht]
\centering
\includegraphics[width=10cm]{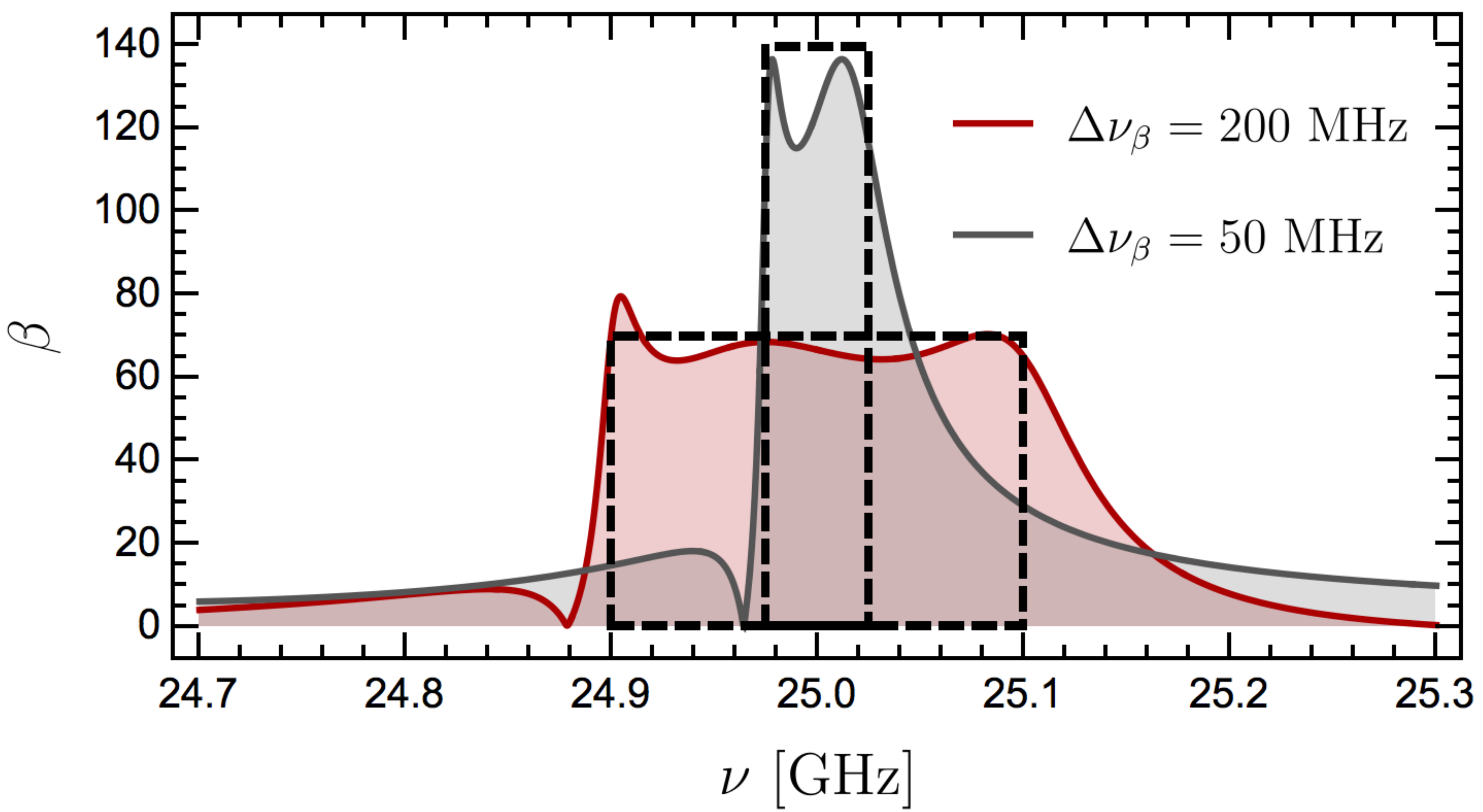}
\caption{Comparison of the B50 (gray) and B200 (red) response functions. These are contrasted with hypothetical ideal top-hat configurations (black dashed) based on what one might expect from the Area Law as applied to the main peak of the B200 configuration.}
\label{fig:bandwidths}
\end{figure}

The same point can be made by comparing the response functions of B50 and B200 directly in figure~\ref{fig:bandwidths}. We also compare the numerical responses with idealised top-hat
profiles which have the chosen width of 50 and 200~MHz and heights derived from the Area Law applied to the main peak of the B200 configuration. So the broader bandwidth case, being more rectangular, uses the available area more efficiently.

\begin{figure}[ht]
\centering
\includegraphics[width=10cm]{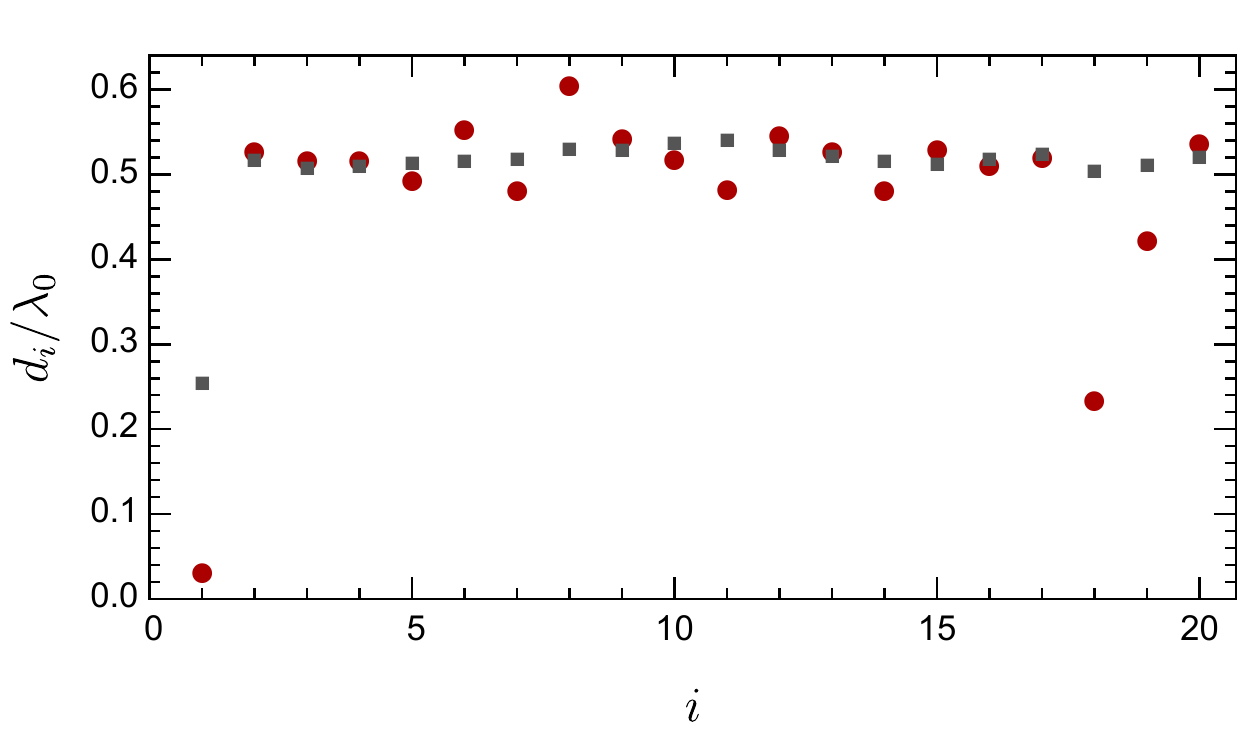}
\caption{Positions of the 20 disks for the B50 (gray) and B200 (red) configurations. We show the
distances $d_i$ of disk $i$ relative to disk $i{-}1$, where for $i=1$ it is the distance to the mirror which is counted as $i=0$. We normalise the distances to the wavelength of the central frequency $\lambda_0=1.20$~cm. }
\label{fig:diskpositions}
\end{figure}

While an analytic understanding of the positions required for these broadband setups is not available at present, we can gain some insight from the numerical results of our B50 and B200 configurations. In figure~\ref{fig:diskpositions} we show the disk separations in these two configurations. Most of the separations are around $\lambda_0/2$, although the dielectric next to the mirror is much closer to it. The configuration with a broader bandwidth (B200, red) shows a greater dispersion of the disk separations, corresponding to this configuration being less resonant.

\begin{figure}[ht]
\centering
\includegraphics[width=14cm]{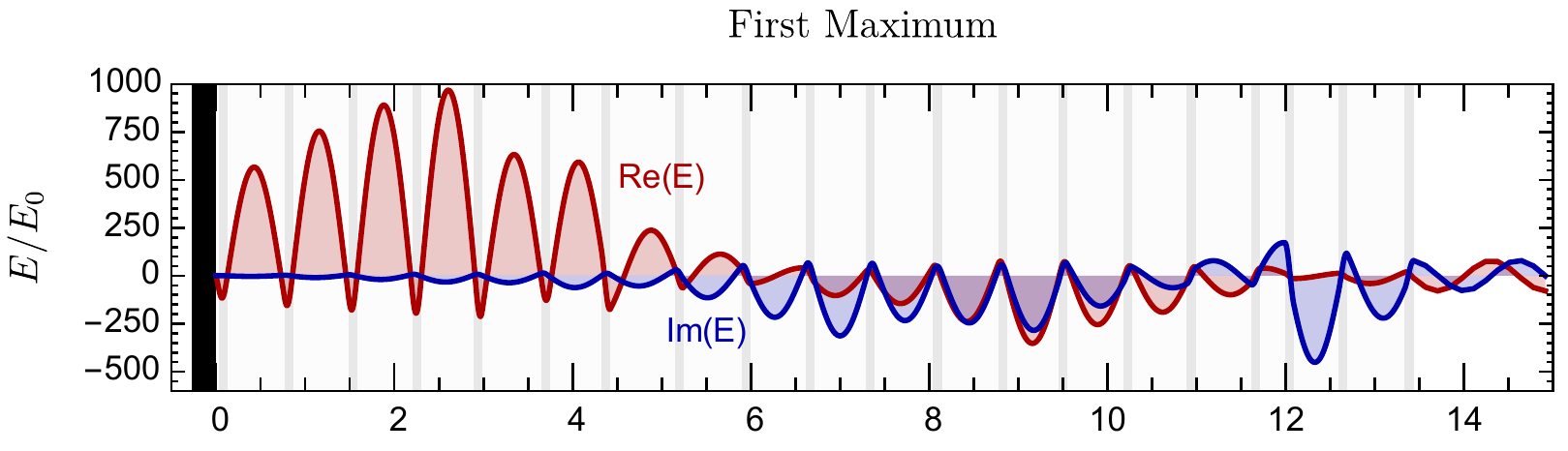}
\includegraphics[width=14cm]{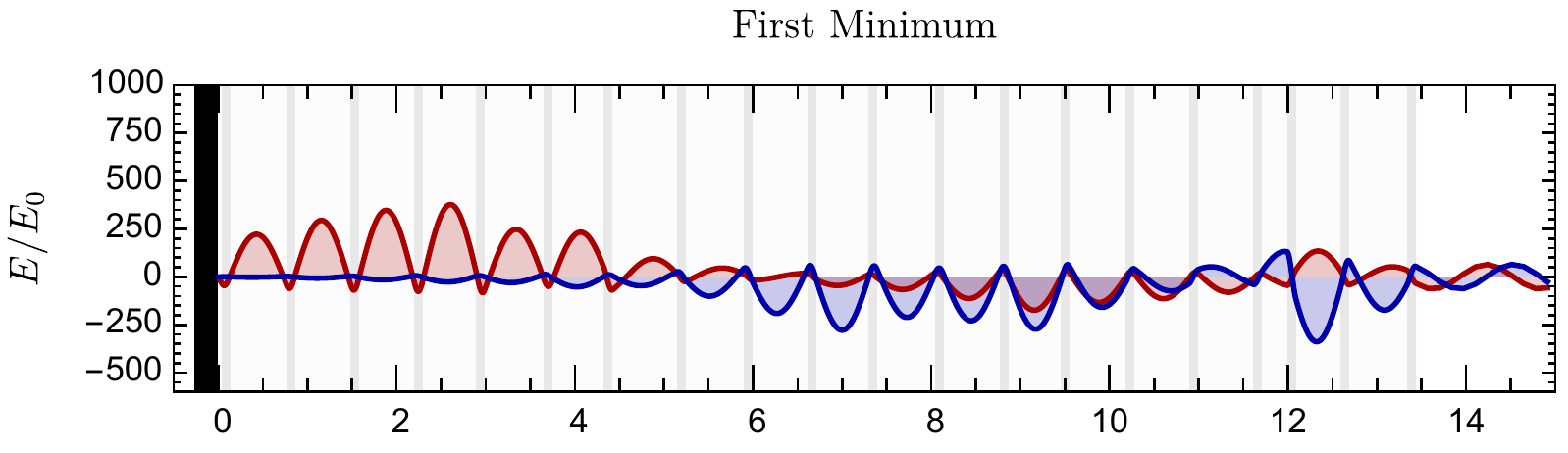}
\includegraphics[width=14cm]{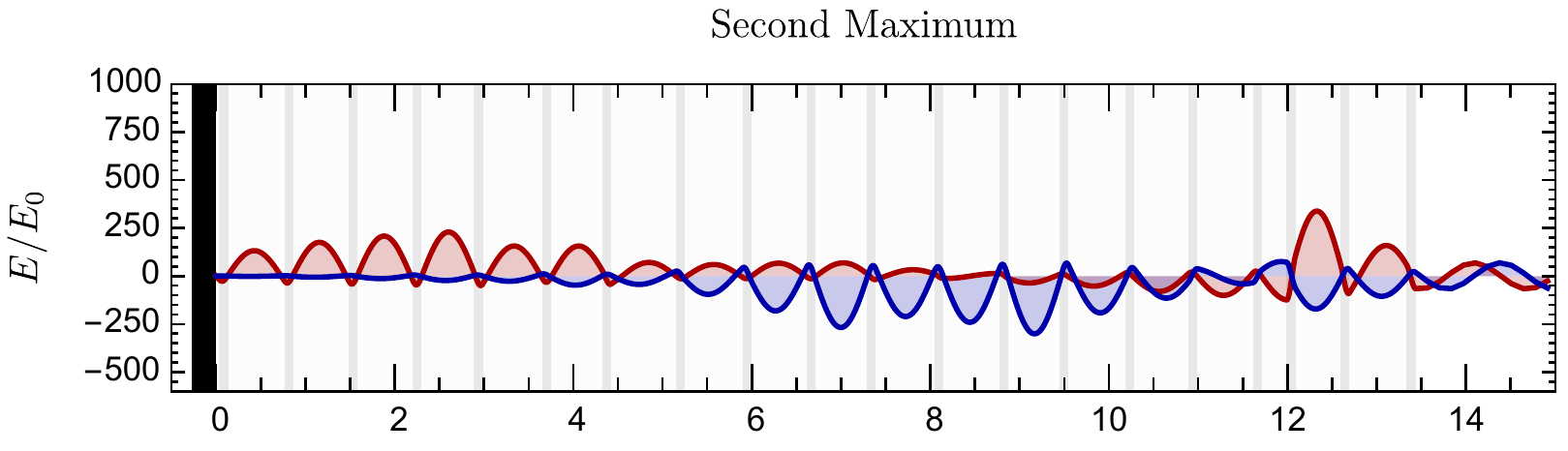}
\includegraphics[width=14cm]{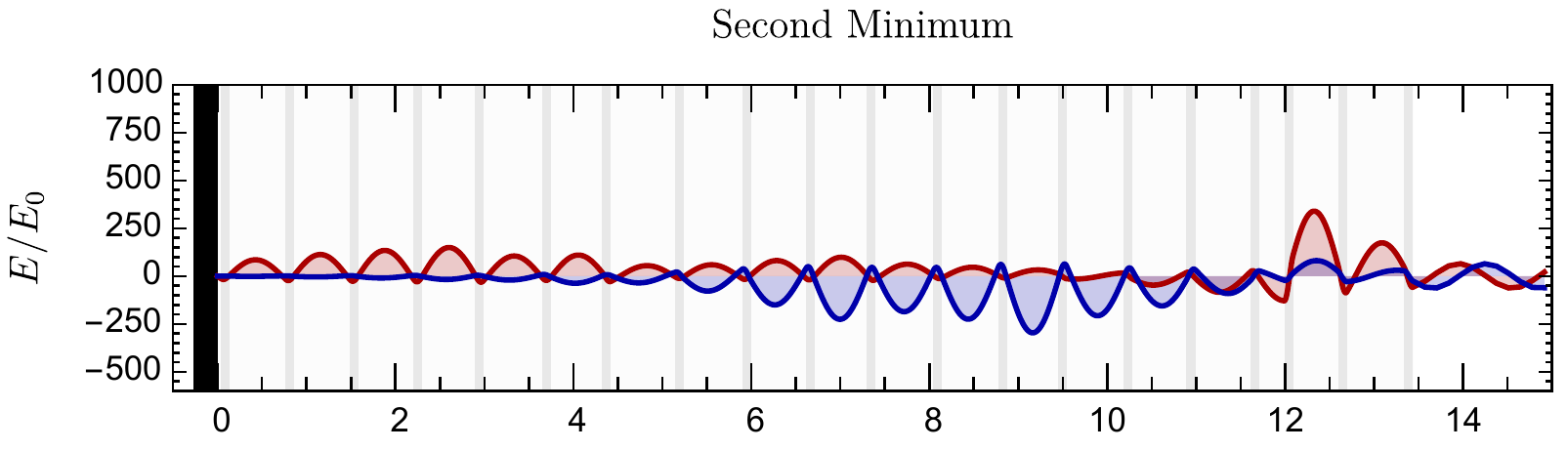}
\includegraphics[width=14cm]{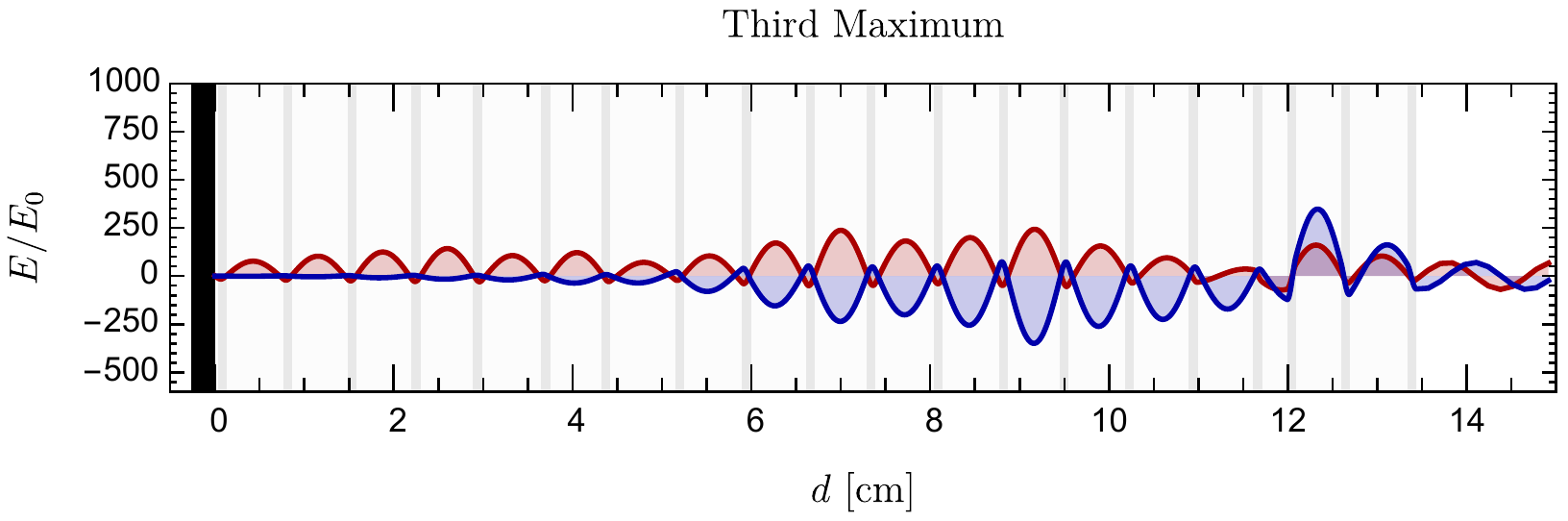}
\caption{$E$-field distribution as a function of distance $d$ from the mirror
in our B200 configuration. Real part of $E$ in red, imaginary part in blue. As labeled, the panels refer to the frequencies of the maxima and minima of the boost-factor curve shown in figure~\ref{fig:200mhz}. The mirror and dielectric disk locations are indicated as in figure~\ref{fig:E-fields2}.}
\label{fig:E-fields1}
\end{figure}

\begin{figure}[ht]
\centering
\includegraphics[width=14cm]{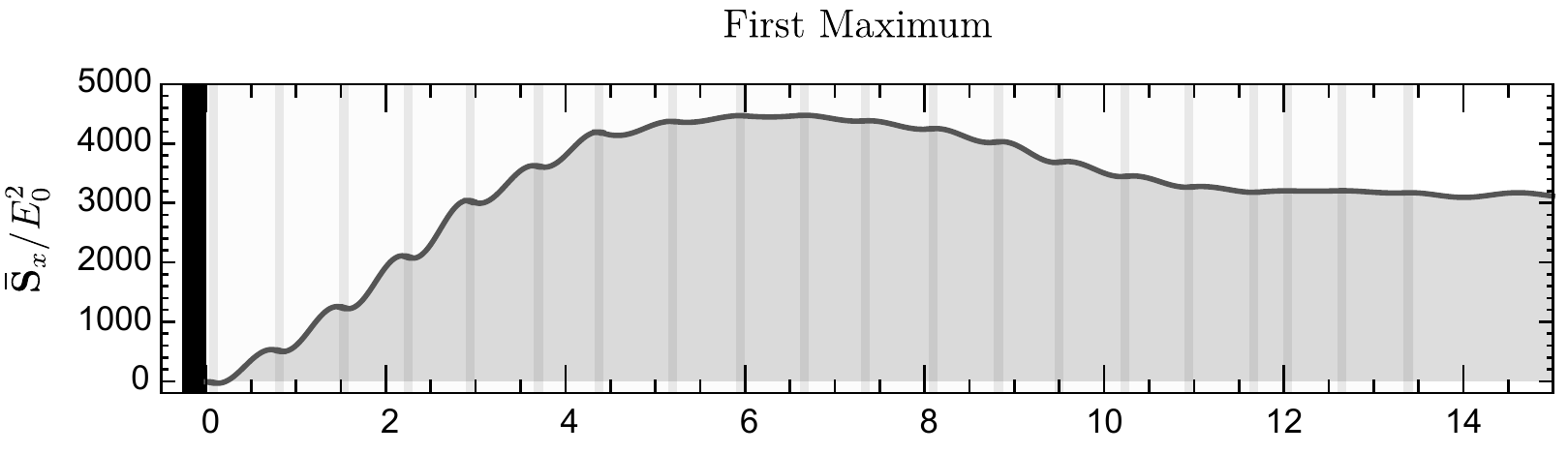}
\includegraphics[width=14cm]{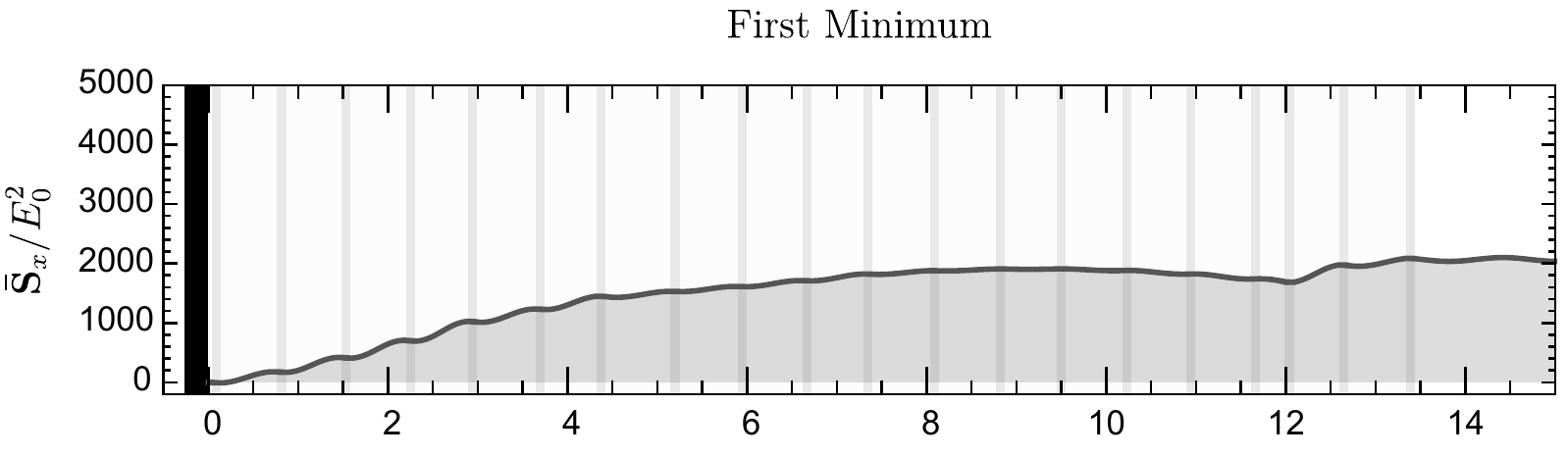}
\includegraphics[width=14cm]{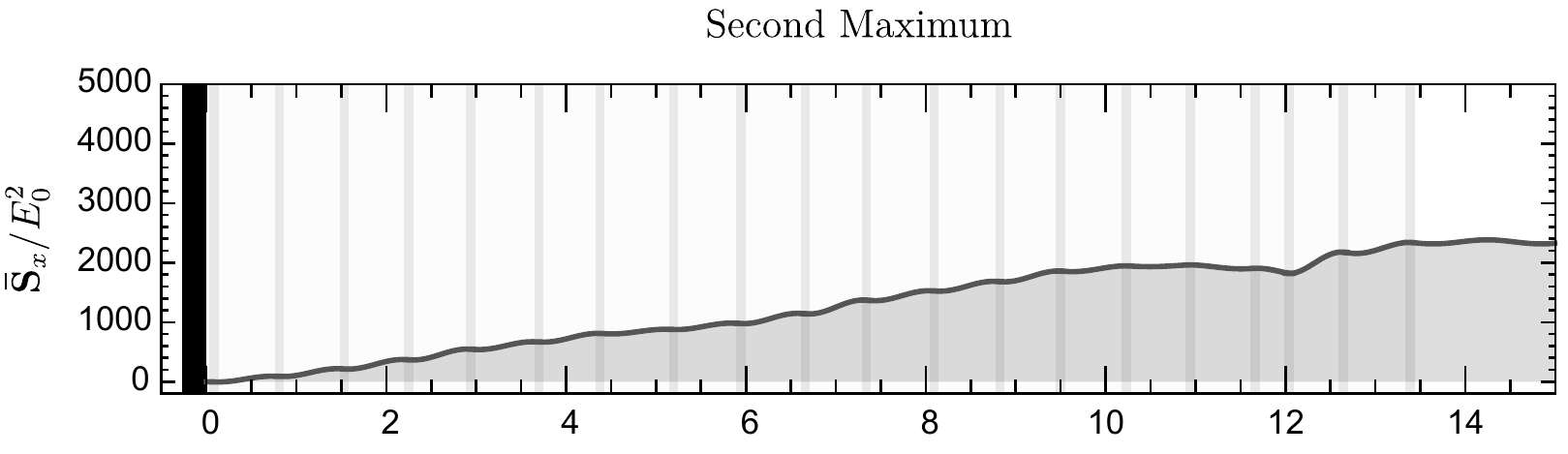}
\includegraphics[width=14cm]{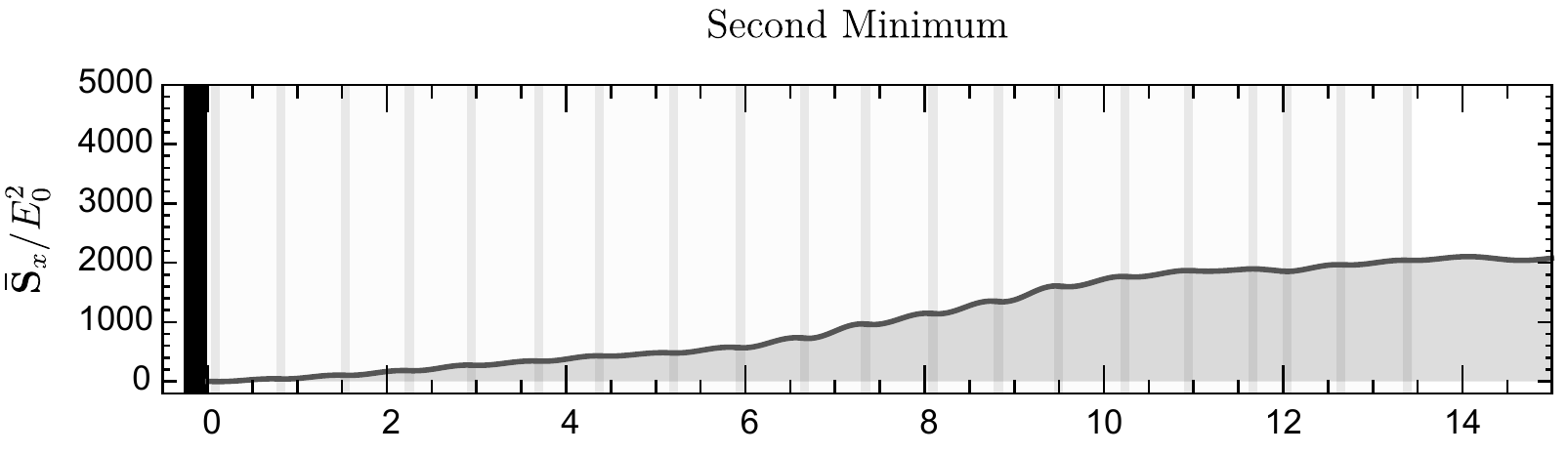}
\includegraphics[width=14cm]{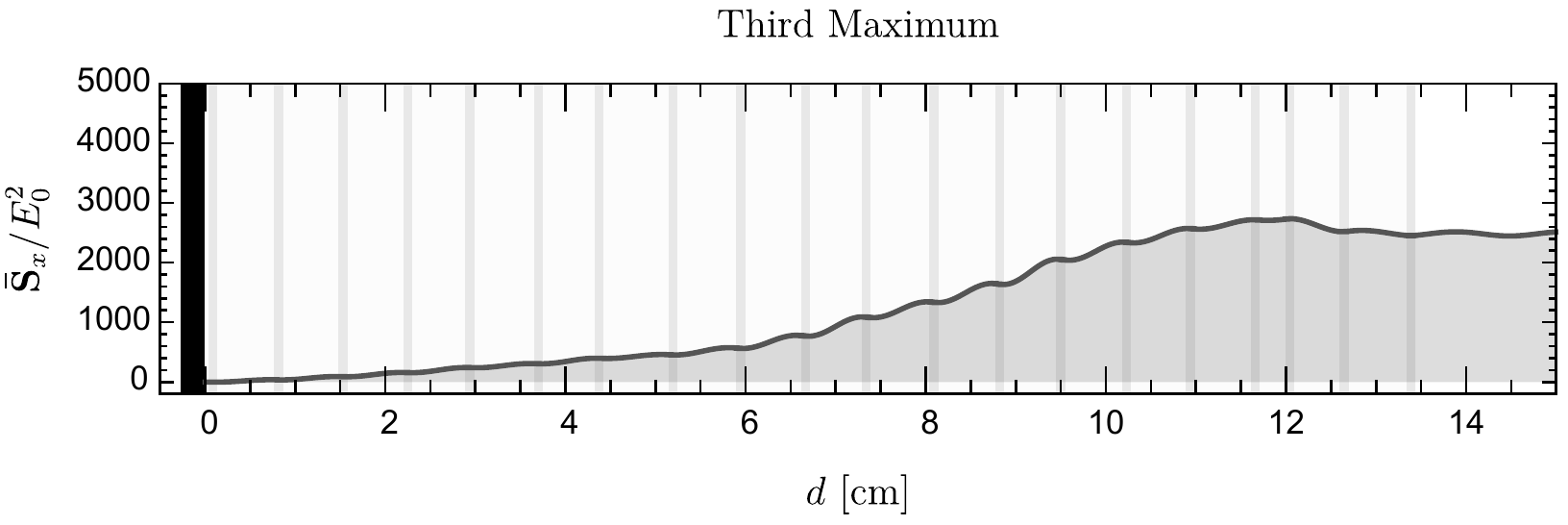}
\caption{Cycle averaged Poynting flow $\bar{\bf S}_x$ in our B200 configuration, corresponding to the $E$-field configurations shown in
figure~\ref{fig:E-fields1}. The mirror and dielectric disk locations are indicated as in figure~\ref{fig:E-fields2}.}
\label{fig:P-fields1}
\end{figure}

That the behaviour of the B200 configuration is less resonant is indicated in $\mathcal R$: the middle peak is not pronounced, suggesting that it does not correspond to strongly resonant behaviour. Figure~\ref{fig:E-fields1} shows the E-fields at the maxima and minima of the boost factor curve in figure \ref{fig:200mhz}. The sharp cut-off on the lhs of the $\beta$ curve seems to be due a resonant mode near the mirror, as in the B50 configuration. The interpretation we had of two resonant regions for our B50 configuration is not quite as clear for the B200 configuration: we do not see three obviously distinct resonant modes creating the three-peak structure, although the left and right peaks do seem to correspond to resonances. Again, as can be seen in the significant spatial variation of the phase, traveling waves are present. By looking at the cycle averaged Poynting flow in figure~\ref{fig:P-fields1}, one can see that the power is generated more or less evenly throughout the haloscope for the central maximum and is not localised to any one region, unlike the left and right peaks.

\clearpage

\begin{figure}[b!]
\centering
\includegraphics[width=10cm]{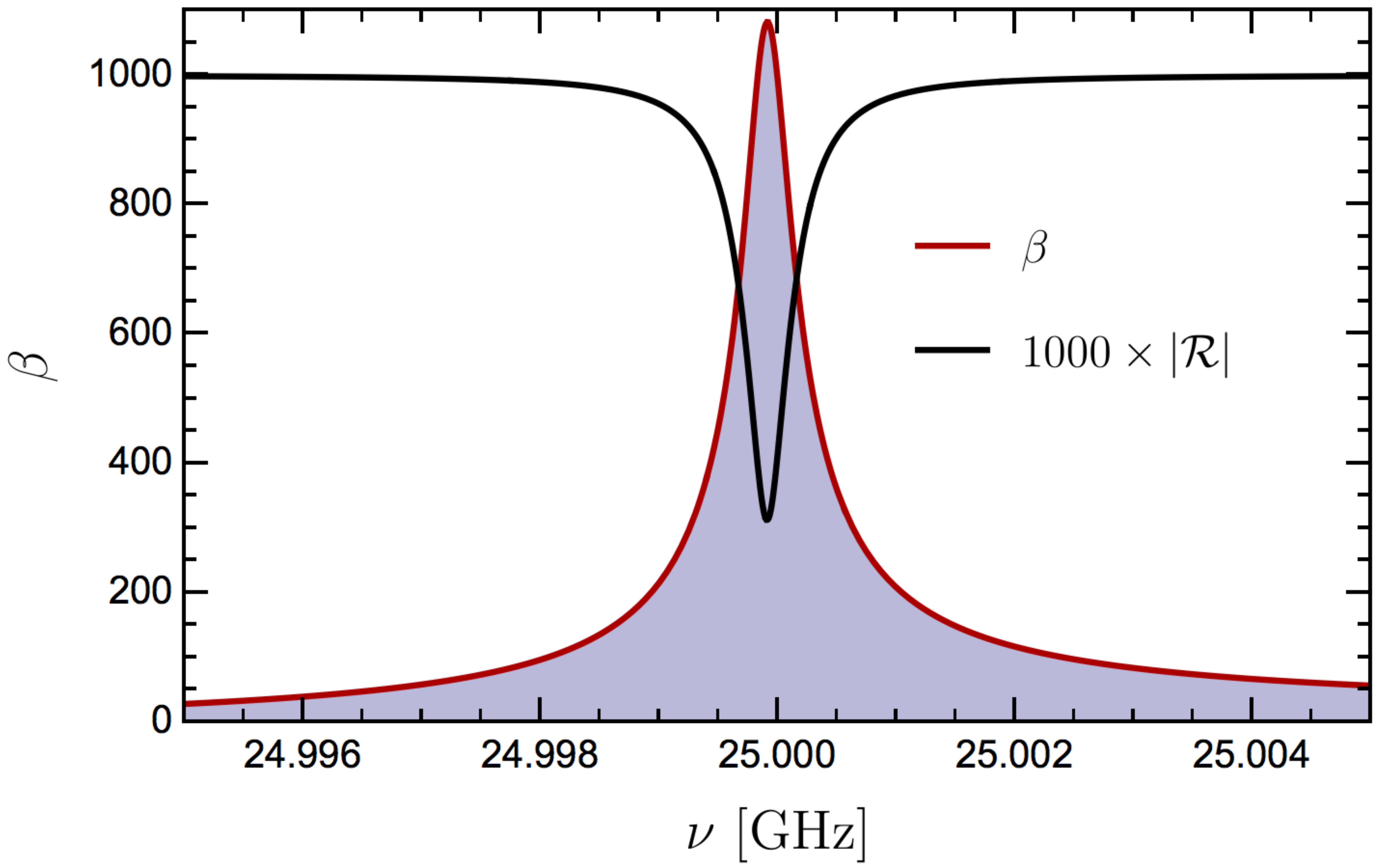}
\caption{Boost factor $\beta$ (red) and reflectivity $|\mathcal R|$ (black) as a function of frequency for
our B1 configuration (20 disks, 1~mm thick, refractive index $n=5$, mirror on one side, bandwidth 1~MHz centred on 25~GHz). The reflectivity is illustrated with exaggerated dielectric losses ($\tan\delta=2\times10^{-4}$) to show non-trivial structure. The reflectivity has been scaled
by a factor 1000.}
\label{fig:1mhz}
\vskip20pt
\centering
\includegraphics[width=13cm]{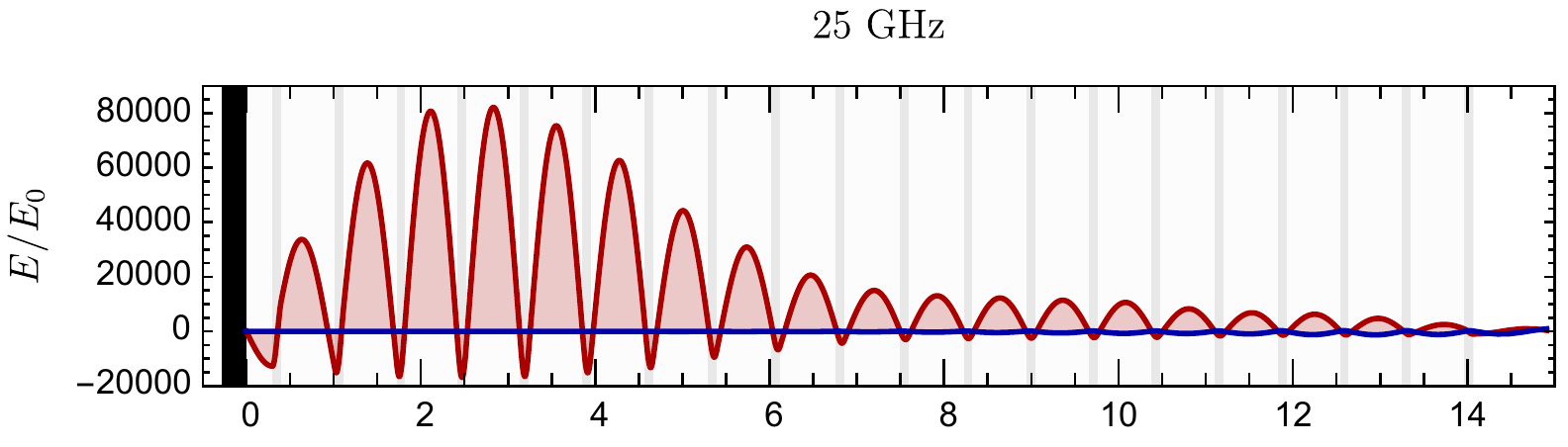}
\includegraphics[width=13cm]{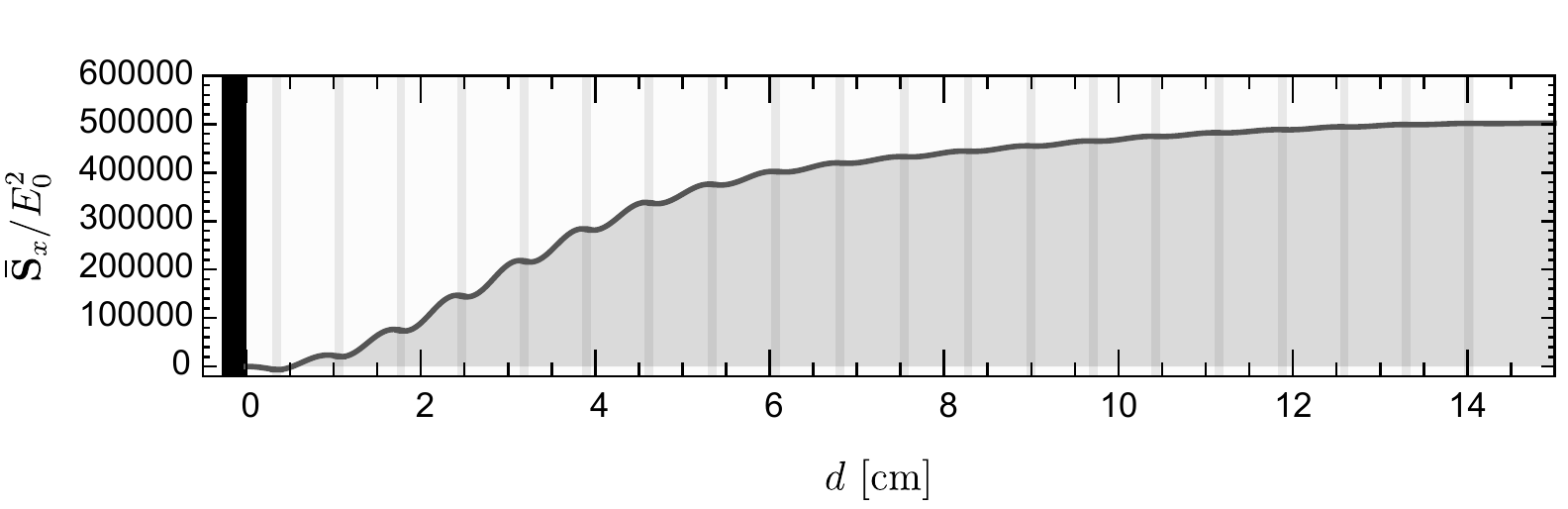}
\caption{E-field distribution (top panel, red: real part, blue: imaginary part) and cycle-averaged Poynting flow $\bar{\bf S}_x$ (bottom panel)
for the B1 configuration at the central frequency of 25~GHz. The mirror and dielectric disk locations are indicated as in figure~\ref{fig:E-fields2}.}
\label{fig:P-fields3}
\end{figure}

\subsection{Configuration with 1 MHz bandwidth}

As a final example we consider a configuration with a 1~MHz bandwidth, all else being equal to the previous examples, our B1 configuration. Proceeding in the same way as earlier we find the boost factor curve shown in figure~\ref{fig:1mhz} which reaches $\text{Min}\[\beta(\nu_i)\]\sim 600$. The shape is now far from rectangular---qualitatively it looks more like Lorentzian. If it were more rectangular, the Area Law would suggest $\text{Min}\[\beta(\nu_i)\]\sim 800$. While the boost factor at the central frequency is much larger than for B200 or B50, an even larger value of
$\beta\gtrsim 10^4$ at 25~GHz can be achieved by judicious disk positioning. In this sense, the B1 configuration still qualifies as ``broadband,'' although it really consists of a single resonance as one can also see from $\cal R$. From the $E$-field distribution and cycle-averaged Poynting flow shown
in figure~\ref{fig:P-fields3} we can see that that this single resonance is much stronger than any in B50 and B200. Unlike in the previous two examples there are no significant traveling waves in the haloscope.

\subsection{Summary}

We have studied specific examples for broadband configurations of a dielectric haloscope. We have seen that even a relatively small number of disks enables significant versatility to tailor the frequency-dependent haloscope response to the axion field. But there also limitations. We have seen that achieving a large boost factor as in our 1~MHz example meant we had to give up on the rectangular shape of the ideal response function. Exploring the options and devising an optimal strategy in terms of the required sets and numbers of disks and their positioning will require considerable effort.

However, as a first estimate it appears justified to use the Area Law to extrapolate to other configurations, those with different bandwidth or with different numbers of disks.
In the following section we will take advantage of this insight to estimate the conceivable reach of a dielectric haloscope to detect axion dark matter.

\section{Discovery potential}
\label{discoverypotential}

We now turn to the discovery potential of dielectric haloscopes for axion dark matter (DM), expanding quantitatively on the estimates provided in reference~\cite{TheMADMAXWorkingGroup:2016hpc}. Unlike what is generally assumed for cavity haloscopes, we operate under the assumption that the limiting factor to the scanning rate is not how large a boost factor one can achieve, but rather by the time required to adjust the spacings between dielectrics from one configuration to the next. The broadband nature of the dielectric-haloscope response can compensate for this limitation, allowing one to scan across most of the unexplored high-mass axion parameter space.

\subsection{Target parameter space}

To motivate and guide a search for axions, we turn to cosmology. The relic abundance of DM axions depends on the axion field initial conditions and the subsequent cosmological evolution, but it also retains a dependence on the axion mass. For a given cosmological history, we can then compute the relic abundance and find for which axion mass we would saturate the observed DM abundance. There are two main classes of cosmological histories of relevance to this question, depending on the epoch where cosmological inflation took place. We start by assuming that, at some early time, the axion field $a$ is uncorrelated at distances greater than the causal horizon. This is the case immediately after the Peccei--Quinn symmetry, $a\to a+2\pi f_a$, becomes spontaneously broken in a phase transition. We call {\bf Scenario~A} when inflation occurs after this ``Peccei--Quinn epoch''
and {\bf Scenario~B} when inflation occurred before it.

In Scenario~A, a sub-horizon sized patch of the universe, where the axion field is essentially homogeneous, is blown up during inflation to a size much larger than our observable universe. We can then take homogeneous initial conditions for the axion field $a(x)=a_{\rm I}=\theta_{\rm I} f_a$, where $0\leq\theta_{\rm I}<2\pi$ is the initial misalignment angle, for the purpose of comparing with the overall DM abundance measured by Planck: $\Omega_{\rm DM}h^2 = 0.120\pm0.003$~\cite{Ade:2015xua}. The cosmic axion abundance depends on both $\theta_{\rm I}$ and $m_a$, so the observed DM density can be matched for essentially any value of $m_a$ allowed by astrophysical bounds~\cite{Raffelt:2006cw} for a suitable $\theta_{\rm I}$. The required value, assuming standard radiation domination during the onset of the axion mass, has been recently calculated~\cite{Borsanyi:2016ksw}. It spans from $\theta_{\rm I}=10^{-4}$ to $\theta_{\rm I}\simeq \pi$ in the $f_a$ range \hbox{$10^9$--$10^{19}$~GeV}, i.e., in the $m_a$ range $10^{-6}$--$10^3~\mu$eV. The largest values of $f_a$ require very small $\theta_{\rm I}$, while for the smallest $f_a$ values, $\theta_{\rm I}$ has to be exquisitely tuned to~$\pi$.

In Scenario~B, inflation plays no role in setting the initial conditions and thus the axion abundance is given by the average over random initial conditions and the decay of accompanying cosmic strings and domain walls. In this case the abundance depends on the domain-wall number ${\cal N}$ that determines how many degenerate CP conserving vacua are realised in the Peccei--Quinn mechanism. In the simplest ${\cal N}>1$ cases (such as the DFSZ axion), the network of domain walls and cosmic strings is stable and leads to a Universe very different from the one we observe~\cite{Sikivie:1982qv}. Various extensions have been discussed that avoid this conclusion and make these scenarios viable. One possibility is that the degeneracy of the different vacua is broken by higher-dimensional operators generated by dynamics at energy scales larger than $f_a$~\cite{Gelmini:1988sf,Larsson:1996sp}. Generically, this breaking shifts the position of the minimum of the axion potential away from the CP conserving value, so some mechanism is required to keep this effect below the requirements of a natural solution of the strong CP problem. Models where the Peccei--Quinn symmetry is an accidental symmetry of a theory containing a discrete symmetry~\cite{Georgi:1981pu} provide a well motivated realisation of this idea. A recent phenomenological study of this scenario~\cite{Ringwald:2015dsf} concludes, however, that the axion mass required to saturate the observed DM abundance exceeds the meV range, somewhat above the sensitivity range of dielectric haloscopes. (These authors
do not mention the possibility of DM dilution after the axion mass turns on, which will reduce this figure.) We note that another possibility to avoid a domain-wall dominated cosmology consists of introducing other axion-like fields that connect the different vacua and lead effectively to a ${\cal N}=1$ cosmology~\cite{Lazarides:1982tw,Choi:1985iv}.

Models with ${\cal N}=1$ such as KSVZ~\cite{Kim:1979if,Shifman:1979if} and KSVZ-like variants like SMASH~\cite{Ballesteros:2016euj,Ballesteros:2016xej} are remarkably predictive. The most updated calculations lead to a unique prediction for the axion mass that saturates the observed DM abundance, $m_a\sim 100\,\mu$eV~\cite{Hiramatsu:2012gg,Kawasaki:2014sqa}, although with some still unclear theoretical uncertainty from the efficiency of axion radiation from strings and domain walls~\cite{Fleury:2015aca}. In SMASH one estimates $50\, \mu{\rm eV}\lesssim m_a\lesssim 200\, \mu{\rm eV}$~\cite{Ballesteros:2016euj,Ballesteros:2016xej}. Dielectric haloscopes are designed to target this singular class of models. Note however that the heavy quarks required in this model can shift this
mass range downwards if they decay/annihilate after the axion mass onset. (See reference~\cite{DiLuzio:2016sbl} for a recent comprehensive study.)

The axion-photon coupling is well studied in these scenarios too. The simplest KSVZ model features a new heavy quark with zero hypercharge that implies ${\cal E}=0$ and so $C_{a\gamma}=-1.92(4)$~\cite{diCortona:2015ldu}. The heavy quark is cosmologically stable~\cite{Nardi:1990ku} and in trouble with observations~\cite{DiLuzio:2016sbl} so one needs to consider hypercharged versions of KSVZ. With hypercharge $-1/3$ or $2/3$ (as in SMASH), the heavy quark can mix with down or up quarks and decay, leading to an uncomplicated cosmology. Here one finds ${\cal E}=2/3$ or $8/3$ and thus $C_{a\gamma}=-1.25(4)$ or $0.74(4)$ respectively. These are the only ${\cal N}=1$ models with one heavy quark and thus the simplest natural targets for dielectric haloscopes.
Models with extra heavy quarks can have effectively ${\cal N}=1$ and thus point to the same 100~$\mu$eV mass range, but have different values of $C_{a\gamma}$~\cite{DiLuzio:2016sbl}.

\subsection{Example 80 disk experiment}
\label{80disk}

For a realistic setup, we assume a device consisting of a mirror and 80 disks of refractive index $n=5$ and area $A=1$ ${\rm m}^2$, contained inside a 10~T magnetic field. We also assume a conservative full day of downtime between measurements (readjustment time $t_{\rm R}=1~\rm{day}$). As there are $\mathcal O(10^6)$ channels in our search region, we require $S/N\sim 5$ to reduce the number of false positives to an acceptable amount. Depending on the ease of rescanning false positives, in particular depending on $t_{\rm R}$, it may be more efficient to allow for more false positives and rescan more frequently.

To find the time taken for a single measurement, one can use equation~\eqref{eq:scan}, searching $\Delta\nu_\beta/\Delta\nu_a$ channels in each measurement. Naively, one might expect that a narrow resonance is most efficient: the Area Law implies that $P\Delta \nu_\beta\sim {\rm const}$ and from equation~\eqref{eq:scan} we see that the scanning rate scales as $P^2\Delta \nu_\beta$, so one gains by a factor $P$. As previously noted~\cite{TheMADMAXWorkingGroup:2016hpc}, this argument ignores the time $t_{\rm R}$ necessary to reposition the disks---as shown in appendix~\ref{scanoptimisation} an optimal scanning rate requires $t_{\rm R}=\Delta t$. To compensate for a relatively long $t_{\rm R}$ one can use a broadband response to measure a larger frequency range in each measurement.

With some simplifications, the time taken to optimally measure some frequency range
$\nu_2$--$\nu_1$ can be found analytically, as in appendix~\ref{scanoptimisation}. If the boost factor required to obtain $\Delta t=t_{\rm R}$ at $\nu_2$ is $\beta_2$, then the optimal overall
scanning time is
\be
t_{\rm o}\sim\frac{2t_{\rm R}}{2p+1}\,\frac{\nu_2}{\Delta\nu|_{\nu_2}} \, ,\label{eq:optimumtime}
\ee
where $\Delta\nu|_{\nu_2}$ is the bandwidth of $\beta_2$ at $\nu_2$. As will be discussed below,
the idealised HEMT (high-electron-mobility-transistor) amplifier with constant $T_{\rm sys}$ has $p=1/4$, whereas quantum-limited amplifiers ($T_{\rm sys}=\omega$) have $3/4$.  Interestingly, to first approximation the measurement time depends only on $\nu_2$. As $\Delta\nu|_{\nu_2}$ should be a linear function of the number of disks (as seen from the Area Law), we see that the frequency range measured in a given time interval is also linear. Thus to double the frequency range scanned, one must roughly double the number of disks. Further, $\beta_2\propto C_{a\gamma}$, so the measurement time decreases with the square of $C_{a\gamma}$, rather than the forth power as one might expect from Dicke's formula~\eqref{eq:scan}.

Note that equation~\eqref{eq:optimumtime} is somewhat less conservative than the method used in reference~\cite{TheMADMAXWorkingGroup:2016hpc}, which assumed that one was optimising on a half yearly basis, rather than optimising the bandwidth for every measurement. As $\beta$ varies very slowly with frequency, both procedures give similar results, especially as we are only concerned with simple estimates.

To estimate the power generated for a given bandwidth, we extrapolate our 25~GHz configurations using the Area Law. Explicitly, we use the trend line from the bottom panel of figure~\ref{fig:largeband} to get an idea of the boost factor for an 80~disk haloscope, i.e.,  $\beta\sim 275$ across $\Delta\nu_\beta= 50$~MHz. While the exact $\beta$ one can achieve depends on the frequency (even when using a few different sets of dielectrics one will not achieve a uniform result), this should give a reasonable estimate. For consistency with other DM detection forecasts, we assume $f_{{\rm DM}} =1$, i.e., $\rho_a=0.3~{\rm GeV}/{\rm cm^3}$.

\begin{figure}[bt!]
\centering
\includegraphics[width=9cm]{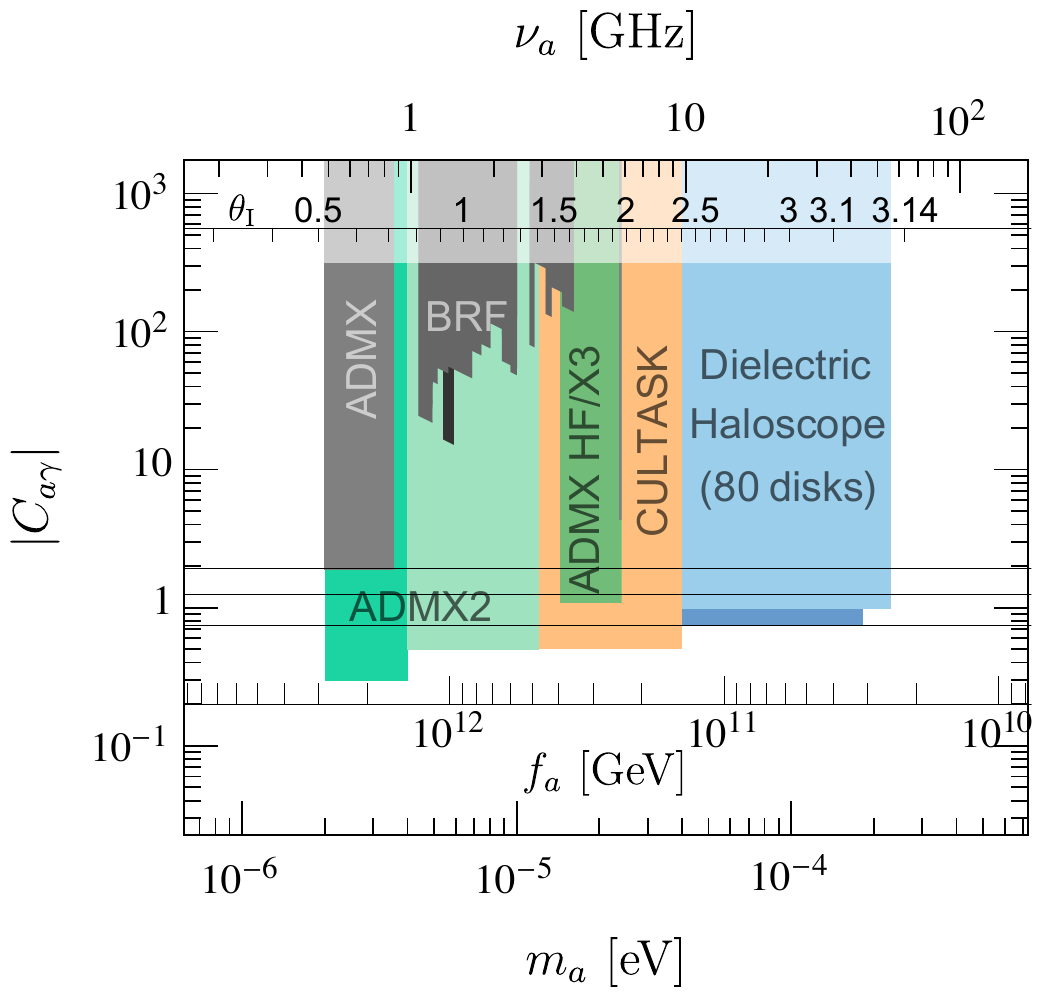}
\caption{Two examples of the discovery potential (light and dark blue) of our dielectric haloscope using 80 disks ($n=5$, $A=1\,{\rm m}^2$, $B_{\rm e}=10\,{\rm T}$, $\eta=0.8$, $t_{\rm R}=1\,{\rm day}$) with quantum-limited detection in a 3-year campaign. We also show exclusion limits (gray) and sensitivities (coloured) of current and planned cavity haloscopes~\cite{Brubaker:2016ktl,Asztalos:2009yp,Hagmann:1990tj,Wuensch:1989sa,Carosi2016Jeju,vanBibber2015Zaragoza,CULTASK}.
The upper inset shows the initial misalignment angle $\theta_{\rm I}$ required in Scenario~A \cite{Borsanyi:2016ksw}. The lower inset depicts the $f_a$ value corresponding to a given $m_a$, and the three black lines denote $|C_{a\gamma}|=1.92$, 1.25 and 0.746. Note that Scenario~B predicts $50~\mu{\rm eV}\lesssim m_a\lesssim 200~\mu{\rm eV}$ \cite{Kawasaki:2014sqa,Ballesteros:2016euj}.
(Figure taken from reference~\cite{TheMADMAXWorkingGroup:2016hpc} with permission.)}
\label{fig:reach}
\end{figure}

In figure~\ref{fig:reach} we show the discovery potential of an 80~disk experiment with a run time of three years (figure taken from reference~\cite{TheMADMAXWorkingGroup:2016hpc}). This estimate assumes 80\% power efficiency and quantum-limited detection ($T_{\rm sys}\sim m_a$).
One would be able to search a large fraction (\hbox{40--$230~\mu$eV}) of the high mass parameter space for $|C_{a\gamma}|=1$, and 40--$180~\mu$eV for $|C_{a\gamma}|=0.746$. This range would essentially cover the entire mass range predicted by Scenario~B. If the Peccei--Quinn symmetry
is not restored after inflation (Scenario~A), this range corresponds to initial misalignment angles $2.4\lesssim\theta_{\rm I}\lesssim 3.12$. Figure~\ref{fig:reach} compares this forecast with existing limits and forecasts from planned cavity haloscopes~\cite{Brubaker:2016ktl,Asztalos:2009yp,Hagmann:1990tj,Wuensch:1989sa,Carosi2016Jeju,vanBibber2015Zaragoza,CULTASK}.

It is conceivable to use a two-stage search strategy, using a mixture of commercially available HEMT amplifiers and quantum-limited detection. Using HEMT amplifiers with $T_{\rm sys}=8$~K, in a five year campaign one could scan $m_a\lesssim 120~\mu{\rm eV}$ reaching $|C_{a\gamma}|=1$ or $m_a\lesssim 100~\mu{\rm eV}$ for $|C_{a\gamma}|=0.746$. One could then upgrade to quantum-limited detection for an additional two-year scan to cover the high-mass range, $m_a\lesssim 230~{\rm or}~180~\mu{\rm eV}$.

While the mass range can be extended by increasing the run time of the experiment or by adding disks, our analysis cannot be extended beyond $\sim 250~\mu$eV. Throughout this paper we have assumed the axion field to be homogeneous on the scales of the apparatus and to oscillate coherently with monochromatic frequency throughout the helioscope. When $m_a$ gets too large, both assumptions begin to break down, although the drop off in power is not instantaneous.

Dielectric haloscopes seem uniquely suited to searching for high mass axions. In this paper we have made relatively minimal assumptions about the required technologies: like cavity haloscopes it should be possible to extend the natural range of a dielectric haloscope ever further with future technological advances. The ability to use a large transverse area as well as a high boost factor with a broadband search strategy would allow one to search a large fraction of the high-mass parameter space.

\section{Conclusions}
\label{conclusion}

A dielectric haloscope, consisting of a mirror and many dielectric disks with adjustable vacuum gaps between them placed in a strong external $B$-field, is a unique approach to search for galactic dark matter axions in the high-mass region of 40--$400~\mu{\rm eV}$. The oscillating axion field drives microwave radiation emitted orthogonally to the dielectric disks that can be picked up by a suitable detector. We have performed a systematic study of the EM response of such an apparatus, providing a firm theoretical foundation for this approach.

To this end we have calculated the $E$-fields induced by the axion field at interfaces between different dielectric media, using a transfer matrix formalism that allows one to study multilayer systems. By controlling the spacings between the interfaces it is possible to manipulate the frequency-dependent efficiency of microwave production which can be greatly enhanced. Unlike the traditional resonant cavity, it is possible to use a broadband search strategy, which is a key feature of dielectric haloscopes. We have shown that the area under the power boost as a function of frequency is independent of the chosen spacings, but grows linearly with the number of disks. For cavity haloscopes the power output is proportional to the overlap integral between the external $B$-field and the cavity-mode $E$-field. We have generalised this formalism to our case, i.e., the transfer matrix approach is equivalent to an overlap integral approach for any setup which produces a
strong amplification of the axion-induced $E$-field, regardless of resonant conditions. This equivalence allows a more direct comparison between dielectric and cavity haloscopes.

In addition to setting up this theoretical formalism, we have considered some of the practicalities involved in realising such an experiment.
We expect the required precision in placing the dielectric disks to be of the order of a few $\mu$m. Further, we have seen that the frequency-dependent phase shift
of a reflected microwave signal could be used to corroborate the boost factor curve, which potentially allows one to correct for positioning errors.

Dielectric haloscopes are an exciting new way to search for high mass (40--400 $\mu$eV) axion dark matter. The formalisms introduced here allow one to perform
detailed and systematic studies of these devices.

\section*{Acknowledgments}

We thank the MADMAX Working Group at the Max Planck Institute for Physics
for support and helpful discussions, in particular Allen
Caldwell, Gia Dvali, B\'ela Majorovits and  Olaf Reimann. We also enjoyed conversations with H.~Bart, K.~van Bibber, J.~Cari\~nena, P.~Goulart, I.~Irastorza, J.~Jaeckel,
G.~Moore, P.~Sikivie and E.~Vitagliano. We acknowledge partial support by the
Deutsche Forschungsgemeinschaft through Grant No.\ EXC 153 (Excellence
Cluster ``Universe'') and by the European Union through the Innovative 
Training Network ``Elusives'' Grant No.\ H2020-MSCA-ITN-2015/674896. J.R.\ is supported by the Ramon y Cajal Fellowship 2012-10597 and FPA2015-65745-P (MINECO/FEDER).  Part of this work was performed at the Bethe Forum ``Axions'' (7--18 March 2016), Bethe Center for Theoretical Physics, University of Bonn, Germany.

\clearpage
\appendix
\section{Proof of the Area Law}
\label{sec:area-law}

In section~\ref{sec:Boost-Amplitude} we stated the Area Law in
the form of equation~\eqref{eq:general-area-law} which we here prove.
We assume the haloscope is described by $N$ independent phase depths
forming a vector ${\bm\delta}=(\delta_1,\ldots,\delta_N)$ where $N$
could be smaller than the total number of $m{-}1$ dielectric regions
between interfaces. In practice we will often use many identical
dielectric disks, each of them having the same phase depth.
Each phase depth is given as $\delta_j=n_j d_j$ with
${\bf d}=(d_1,\ldots,d_N)$ forming the configuration vector.
We further assume that ${\bm\delta}$ consists only of real numbers, meaning that
all dielectrics are taken to be lossless.

The boost amplitude $\B$ is constructed from polynomials and powers of
all $e^{\pm i\delta_j}$. Overall, $\B$ is $2\pi$ periodic in any of the
$\delta_j$, so we may write it as an infinite Fourier series
\begin{equation}
\B=\sum_{\bf k}\,a_{\bf k}\,e^{i{\bf k}\cdot{\bm\delta}}\,,
\end{equation}
where ${\bf k}=(k_1,\ldots,k_N)$ is a $N$-dim vector of positive and
negative integers. The sum is over all integers, i.e., each individual
$k_j$ runs over integers from $-\infty$ to $+\infty$. There is a
complex Fourier amplitude $a_{\bf k}$ for every ${\bf k}$. Notice that
even for a finite series of interfaces, the sum runs to infinite $k$
values because, in the expression for $\B$, we divide by a finite
polynomial of $e^{i k_j\delta_j}$ expressions, usually leading to an
infinite series.

We now consider the Area Law in the sense of $|\B|^2$ averaged over
the $N$-dim space of all phase depths ${\bm\delta}$,
\begin{equation}
  \left\langle|\B|^2\right\rangle_{\bm\delta}
  =\left\langle\B \B^*\right\rangle_{\bm\delta}
  =\(\prod_{j=1}^N\int_{-\pi}^{+\pi}\frac{d\delta_j}{2\pi}\)
  \sum_{~{\bf k},{\bf k}'}\,a_{\bf k}a_{{\bf k}'}^*\,e^{i({\bf k}-{\bf k}')\cdot{\bm\delta}}
=\sum_{{\bf k}}\,\left|a_{\bf k}\right|^2\,.\label{eq:arealawa2}
\end{equation}
After integration, only those terms have survived which have ${\bf k}={\bf k}'$ assuming $n$ does not strongly depend on $\omega$.
Therefore, $\left\langle|\B|^2\right\rangle_{\bm\delta}$ no longer
depends either on $\omega$ or the configuration vector ${\bf d}$.Note that this sum is over interfaces---each ${\bf k}$ is associated with a $\delta$ and so is given by a corresponding interface. This is seen in, for example, \eqref{eq:sum-rule-transparent-1}.

We may achieve the same result if instead of averaging over
${\bm\delta}$ space we average over configuration space ${\bf d}$
except that we need to choose the integration volume such that each
$d_j$ covers a full period. Alternatively, we can average over an
infinite volume.

We also get the same answer if we we integrate over frequency alone,
replacing the $N$-dim integral over all phases by a 1-dim
$d\omega$-integral, keeping in mind that each $\delta_j=n_j\omega
d_j$. Explicitly,
\begin{eqnarray}
\left\langle|\B|^2\right\rangle&=&
\frac{1}{s_2-s_1}
\int_{s_1}^{s_2} ds\,\sum_{~{\bm k},{\bm k}'}\,a_{\bm k}a_{{\bm k}'}^*\,e^{i({\bm k}-{\bm k}')\cdot{\bm\delta}\,s}
\nonumber\\[1ex]
&=&\sum_{{\bm k}}\,\left|a_{\bm k}\right|^2
+\frac{1}{s_2-s_1}\int_{s_1}^{s_2} ds\,
\sum_{~{\bm k}\not={\bm k}'}\,a_{\bm k}a_{{\bm k}'}^*\,e^{i({\bm k}-{\bm k}')\cdot{\bm\delta}\,s}
\,.
\end{eqnarray}
The second integral over phases vanishes if the integration range is large enough---all relative phases average to zero.  In particular, if
the relative phase depths $\delta_j$ are commensurate (rational
fractions of each other), then $\B$ is periodic in $\omega$ and we may
integrate over one such period, which however could be very large.  In
general, $\B$ is not periodic and the Area Law strictly applies only
for a $d\omega$ integration over an infinite range.  In practice, it
is useful as an approximation if we integrate over a finite
region that contains the main resonance.

\section{Comparison to the overlap integral formalism}
\label{Sikiviecomp}

Throughout this paper, we have used a transfer matrix formalism to calculate the $E$-fields produced by axions. However many readers are probably more familiar with the overlap integral formalism
introduced by Sikivie to study cavity haloscopes~\cite{Sikivie:1985yu}. The boost factor calculation in the transfer matrix formalism involves the sum of electric field waves emitted from different interfaces. In contrast, in the overlap integral formalism the power generated is calculated with a volume integral of the form
$\int dV\,{\bf E}_\alpha\cdot {\bf B}_{\rm e}$, where ${\bf E}_\alpha$ is the electric field of a given cavity mode $\alpha$ which is excited by the axion field. While these formalisms are associated with very different physical intuitions, they should agree for cases when both of them apply. As we will show in the following, it is possible to generalise the overlap integral approach to open and non-resonant systems. In the generalised from, it can thus be applied to describe dielectric haloscopes.

The overlap integral formalism was originally developed for a closed resonator in 3D, slightly perturbed by losses. On the other hand, our transfer matrix formalism applies to general 1D open systems. A comparison requires to establish a common ground. Thus, after reviewing the overlap integral formalism for a 3D cavity, we will reduce it to a 1D resonator that may contain layers of dielectric media. In this case, the overlap integral reduces to a sum over regions with electric fields that are related by regular EM transfer matrices. To develop a generalised overlap integral formalism, we must find a similar representation in our treatment of an open and non-resonant 1D system. While the transfer matrices in section~\ref{transfermatrixformalism} are written as a sum over interfaces, we will thus use matrix identities to reorganise the sum by region rather than by interfaces. This form is then equivalent to a generalised overlap integral for large boost factors and holds regardless of resonant conditions. For a 1D resonant cavity this integral formalism agrees with that of Sikivie. However, we will find that the electric field playing the role of ${\bf E}_\alpha$ (the field configuration to be integrated over) is actually not identical with the radiation mode excited by the axion field.

\subsection{The overlap integral formalism}

The physical picture of the original overlap integral formalism is that of the axion field acting as a source inside the cavity, which feeds the resonance so that a steady state is reached where power exiting the cavity is balanced with the power injected by the axion field. If the cavity quality factor $Q$ is very large, the field strength ${\bf E}_\alpha$ of the relevant resonant mode far exceeds the source fields which are neglected.

\subsubsection{3D cavity}
\label{sec:3Dcavity}

With good approximation, the resonant mode is a standing wave with a time-dependent electric field configuration of the form
${\bf E}_\alpha({\bf x})\,\sin(\omega t)$, where the eigenfrequency $\omega$ on resonance corresponds to the axion mass $m_a$. For a homogeneous external
field ${\bf B}_{\rm e}$, the power extracted from a 3D cavity is~\cite{Sikivie:1985yu}
\be
P_{\rm cav} = \kappa {\cal G} V \frac{Q}{m_a} \rho_{a} g_{a\gamma}^2B_{\rm e}^2, \label{eq:sikivieoverlap}
\ee
where $V$ is the cavity volume and
\be
{\cal G} = \frac{\(\int dV\,{\bf E}_\alpha\cdot {\bf B}_{\rm e}\)^2}{V B_{\rm{e}}^2\int dV\,
{\bf E}_\alpha^2}
\ee
is the geometry factor. Moreover,
\be
Q=-\omega\,U/\dot U_l
\ee
is the loaded quality factor, where $U$ is the energy stored in the cavity in mode $\alpha$ and $\dot U_l$ is the rate with which energy is lost. Some of this loss is simply dissipation, whereas some part is the extracted signal
$\dot U_s$, defining the cavity coupling factor
\be
\kappa =\dot U_s/\dot U_l\,.
\ee
On resonance, electric and magnetic fields contribute equally to the energy content $U$ so that
\be
U = \frac{1}{4} \int dV\({\bf E}_\alpha^2+{\bf B}_\alpha^2\)
  =\frac{1}{2} \int dV\, {\bf E}_\alpha^2\,,
\ee
where ${\bf B}_\alpha$ is the $B$-field of mode $\alpha$. Notice that one factor $1/2$ in these expressions derives from the cycle average.
When computing $P_{\rm cav}$ given by equation~\eqref{eq:sikivieoverlap},
this integral will cancel with that in the denominator of $\cal G$.
We use equation~\eqref{eq:axionDMdensity} with $m_a=\omega$ to introduce
\be
E_0^2 = g_{a\gamma}^2 B_{\rm{e}}^2 |a_0|^2 =
\frac{g_{a\gamma}^2 B_{\rm{e}}^2}{\omega^2} \omega^2 |a_0|^2 =
\frac{g_{a\gamma}^2 B_{\rm{e}}^2}{\omega^2} 2 \rho_{a}\,.
\ee
We can now recast equation~\eqref{eq:sikivieoverlap} in the form
\be
P_{\rm cav}=-\frac{\dot U_s}{2(\dot U_l)^2} \frac{m_a^2\(\int dV\,{\bf E}_\alpha\cdot {\bf B}_{\rm{e}}\)^2}{B_{\rm e}^2} \frac{E_0^2}{2}\,.
\ee
In this form, it is the starting point to compare with the
transfer-matrix formalism.

\subsubsection{1D cavity}

\begin{figure}[t]
\centering
\includegraphics[width=14cm]{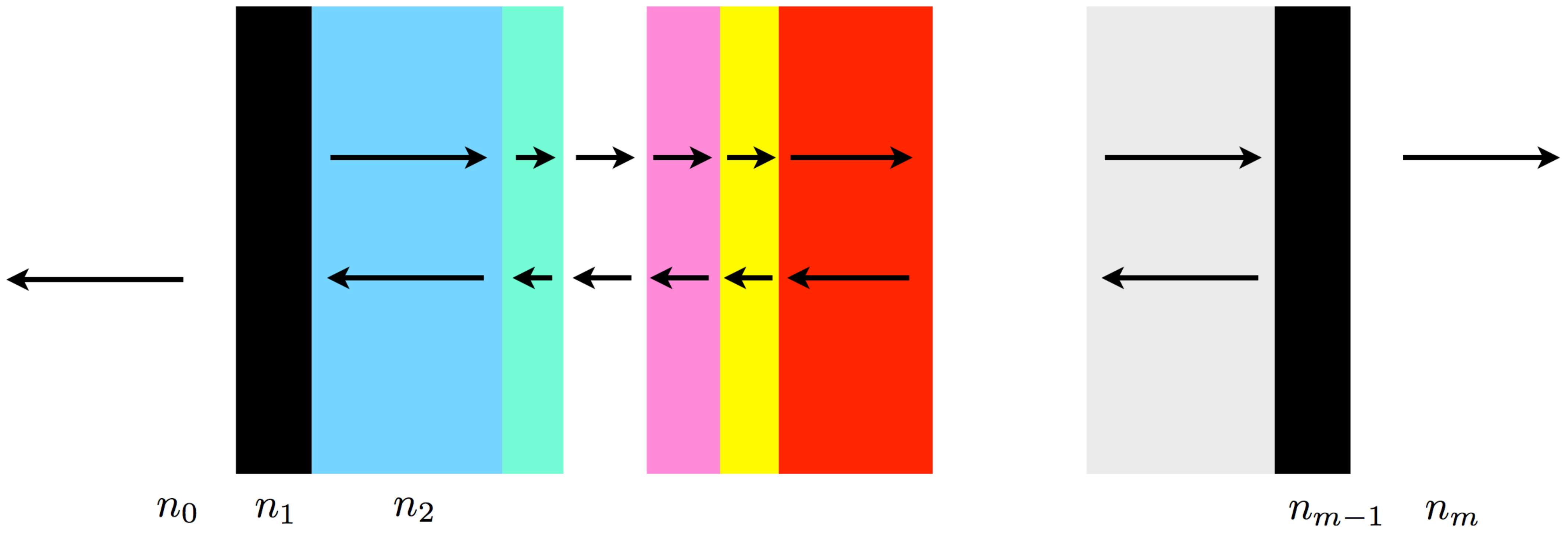}
\caption{A 1D cavity consisting of several dielectric regions $r$ with
different refractive indices $n_r$ represented by different colors.
Regions~1 and~$m-1$ are the cavity walls,
made of extremely reflective materials with a high refractive index, and
regions~$0$ and~$m$ the vacuum space outside of the cavity.
Left and right moving EM waves are represented by the corresponding arrows.}
\label{fig:cavitysetup}
\end{figure}

Now let us consider a 1D cavity made of reflecting walls and several regions $r$ with different refractive indices $n_r$ based on the same arrangements as in section~\ref{transfermatrixformalism}. The setup is shown in figure~\ref{fig:cavitysetup}
and is a special case of the one in figure~\ref{scheme}: Regions~$0$ and~$m$ are the vacuum space outside of the cavity ($n_0=n_m=1$), regions~$1$ and~$m-1$ the cavity walls made of reflective materials with a large refractive indices $n_{1}$ and $n_{m-1}$, and regions~$2$ to~$m-2$ the ones inside the cavity. The optical thickness of each region $r$ is $\delta_r=\omega n_r (x_{r+1}-x_r)$, cf.~equation~\eqref{eq:deltar}. The arrows illustrate left and right moving EM waves in each homogeneous region.

In the following we will consider a leaky cavity as well as the limit of a closed one. In the leaky case, some EM waves travel away from the mirrors in regions $0$ and $m$ as illustrated in
figure~\ref{scheme}. The refractive indices of the mirrors $n_1$ and $n_{m-1}$ are large but finite and we assume phase depths $\sin\delta_1=\sin\delta_{m-1}=1$, i.e., maximal reflectivity,
although other values of $\sin\delta_1$ and $\sin\delta_{m-1}$ can be chosen. For increasing
$n_1$ and $n_{m-1}$, the physical thickness of the mirrors would have to be chosen ever smaller.
Moreover, $n_2/n_1\ll 1$ and $n_{m-2}/n_{m-1}\ll 1$ is assumed which implicitly avoids the possibility of transparent cavity walls via a simple combination of two adjacent $\lambda/4$ layers of material with the same refractive index $n_r$. For a closed (perfect) cavity, we assume
$n_1 \sin\delta_1\to\infty$ and $n_{m-1} \sin\delta_{m-1}\to\infty$.

As in section~\ref{transfermatrixformalism}, the external magnetic field is taken to be
oriented along the $z$-direction, ${\bf B}_{{\rm e}}(x)=B_{{\rm e}}(x) {\bf\hat z}$,
and to be piecewise homogeneous. Each region $r$ will thus be characterised by a given
$B_{{\rm e},r}$ and $n_r$. To normalise $B_{\rm{e}}$ we use
\be
b(x)\equiv \frac{B_{\rm{e}}(x)}{B_{\rm{e,max}}}
\quad\hbox{and}\quad
b_r \equiv \frac{B_{\rm{e},r}}{B_{\rm{e,max}}}.
\label{eq:brdef}
\ee
We will consider modes with the electric field oriented in the $z$-direction,
${\bf E}_{\alpha}=E(x) \,{\bf \hat z}$, where mode subscripts will be suppressed to simplify our notation from here on. Then we express the overlap integral in the form
\be
\frac{\(\int dV\,{\bf E}_\alpha\cdot {\bf B}_{\rm e}\)^2}{B_{\rm e,max}^2}
= A^2 \( \int dx\, E b \)^2 ,
\ee
where $A$ is the transverse area considered.

Next we proceed as in section~\ref{transfermatrixformalism} and in each region $r$ express
the complex electric-field amplitude of a given mode in the form
\be
E^{(r)}(x)=R_r e^{+i \omega n_r \Delta x}+L_r e^{-i \omega n_r \Delta x}\,, \label{eq:eregion}
\ee
where $\Delta x=x-x_r$ with $x_r$ denoting the coordinate at which region $r$ begin (cf.~figure~\ref{scheme}). Moreover, the complex electric-field amplitudes $L_r$ and $R_r$ of the left and right moving EM waves are again defined at the left boundary of every region, except for $L_0$ which denotes the amplitude in region $r=0$ on the left of the interface at $x_1$.
We assume that there are no incoming waves outside of the cavity walls, i.e., $R_0=L_m=0$.
For a perfect cavity, the outgoing waves vanish as well ($L_0=R_m=0$), whereas they take
small nonvanishing values in the leaky cavity case.

In the leaky cavity case, the energy loss per unit area results from the waves exiting the cavity into the external media (regions~$0$ and~$m$ in figure~\ref{fig:cavitysetup}) and from internal losses, which we define by unit area $\dot U_{\rm int}\equiv-\varGamma A/2$
in terms of a damping rate $\varGamma$,
\be
\dot U_l=-A \(\frac{L_0^2}{2}+\frac{R_m^2}{2}+\frac{\varGamma}{2}\).
\ee
We take the signal from only one side of the device, $\dot U_{s,m}= A R_m^2/2$ or
$\dot U_{s,0}= A L_0^2/2$, in analogy with
our boost factor considerations, e.g., in equations~\eqref{eq:boost-factor} and~\eqref{eq:boost-factor-2}. The corresponding power per unit area is
\begin{subequations}\label{eq:powersik}
\bea
\frac{P_m}{A}
&=&  \frac{R_m^2}{(R_m^2+L_0^2+\varGamma)^2}\[\omega^2\(\int dx\, E b\)^2\]\frac{E_0^2}{2}
\label{eq:powersikpm}\,,\\
\frac{P_0}{A}
&=& \frac{L_0^2}{(R_m^2+L_0^2+\varGamma)^2}\[\omega^2\(\int dx\, E b\)^2\]\frac{E_0^2}{2}\,.
\label{eq:powersikpn}
\eea
\end{subequations}
These expressions allow us to identify the respective boost factors with
\begin{subequations}\label{eq:boostsikivie}
\bea
\beta_{m}&=&\frac{R_m}{R_m^2+L_0^2+\varGamma}\,\, \omega \int dx\, E b\,,
\\
\beta_{0} &=&\frac{L_0}{R_m^2+L_0^2+\varGamma}\,\, \omega \int dx\, E b\,.
\eea
\end{subequations}
Notice that in this derivation we have taken the electric field $E$ and the amplitudes
$R_m$ and $L_0$ to be real, which is approximately possible because the resonance
configuration forms an approximate standing wave in the cavity and the emerging waves traveling away are in phase with the cavity at the surface.

\subsubsection{Field configurations inside a closed 1D cavity}

To evaluate the overlap integral in equation~\eqref{eq:boostsikivie}, the electric field configurations of the cavity modes are required. We will now determine these field configurations in the limit of a perfect cavity. When returning to the case of a leaky cavity in section~\ref{Leakycavity} below, we will assume that the small losses to regions $0$ and $m$ will modify these field configurations in a negligible way.

The boundary conditions for a perfect cavity are $E^{(2)}(x_2)=0$ and $E^{(m-2)}(x_{m-1})=0$ because the fields vanish inside perfect mirrors, $E^{(1)}=E^{(m-1)}=0$.
These conditions require the corresponding $R_r$ and $L_r$ components to be equal in magnitude and opposite in phase,
\be
\label{limfields}
\vv{R_{2}}{L_{2}} = R_{2} \vv{1}{-1}
 \quad \mathrm{and} \quad
\vv{R_{m-2}^D}{L_{m-2}^D} = R^D_{m-2} \vv{1}{-1}\,,
\ee
where the superscript $D$  denotes fields at the rhs of the region $r$, i.e., at $x=x_{r+1}$. To specify the electric field that we are integrating over, one can simply specify the field at any point and then use the regular EM transfer matrices~\cite{Sanchez-Soto} to calculate the field everywhere inside the cavity.
(We will neglect settings with disconnected configurations, i.e., completely independent cavities.)
For example, using the transfer matrices  $\TT^a_b$ given in equation~\eqref{eq:emtransfer},
all $E^{(r)}(x)$ inside the cavity can be built from the boundary  condition at $x_2$ via
\be
\vv{R_r}{L_r} = \TT^r_2\vv{R_2}{L_2}= \TT^r_2 \vv{1}{-1}R_2\,.\label{eq:fieldbreak}
\ee

\subsubsection{Overlap integral evaluation}

Let us now evaluate the overlap integral $\int dx\, E b$ for a perfect cavity. By assumption each region is defined by a given $b_r$ and $n_r$, providing a similarly piecewise $E^{(r)}(x)$ given by equation~\eqref{eq:eregion} with, e.g., equation~\eqref{eq:fieldbreak} once $R_2$ is specified, we can split the integral by region. Each of the right and left moving waves in these regions is a simple plane wave, so evaluating the integrals simply gives us the boundary terms
\be
\omega \int dx\, E b = \sum_{r=2}^{m-2}\(R_r \frac{e^{i\delta_r}-1}{i n_r}+L_r \frac{e^{-i\delta_r}-1}{-i n_r}\)b_r\,. \label{eq:splitint}
\ee
Using equation~\eqref{eq:fieldbreak}, we obtain
\be
\label{ff}
\omega \int dx\, Eb = \vv{1}{-1}^T\[\sum_{r=2}^{m-2}\frac{b_r}{in_r} (\PP_r-\mathbb{1})\TT^r_2 \] \vv{1}{-1} R_2\,.
\ee
More generally, we see that any expression of the form
\be
\omega \int dx\, Eb = \vv{1}{-1}^T\[\sum_{r=1}^{m-1}\frac{b_r}{in_r} (\PP_r-\mathbb{1})\TT^r_2 \] \vv{a}{b} \label{eq:edotb}
\ee
can be written as the overlap integral of an electric field in the absence of axions, regardless of whether the setup is a perfect cavity or not. Expression~\eqref{eq:edotb} will be central to our comparison with the transfer matrix formalism below.

\subsubsection{Leaky cavity}
\label{Leakycavity}

We now return to the case of the leaky cavity. As long as the fields leaking to the vacuum regions are much smaller than those inside the cavity the distortion of the perfect-cavity modes will be small. Indeed, we use now the approximation that this distortion is negligible so that the leaking fields can be related to fields in regions $r=2,\, m-2$ in the form
\begin{subequations}\label{eq:insideout}
\bea
\vv{R_2}{L_2}
&=&
\TT_0^2 \vv{0}{L_0} =  \frac{i n_1}{2 n_2}\vv{1-\varepsilon_1}{-1-\varepsilon_1} L_0\,,\\[2ex]
\vv{R_{m-2}^D}{L_{m-2}^D} &=& \TT_m^{m-2}\vv{R_m}{0}=
-\frac{1}{n_{m-2}}\frac{i n_{m-1}}{2 }\vv{1+\varepsilon_{m-1}}{-1+\epsilon_{m-1}}R_m\,,
\eea
\end{subequations}
with the transfer matrices (for vacuum outside the cavity, i.e., $n_{0,m} =1$)
\begin{subequations}
\bea
\TT_0^2&=&\frac{i n_1}{2 n_2}\ma{1+\varepsilon_1}{1-\varepsilon_1}{-1+\varepsilon_1}{-1-\varepsilon_1}\,,
\\[2ex]
\TT_{m-2}^m&=&\frac{i n_{m-1}}{2}\ma{1+\varepsilon_{m-1}}{1-\varepsilon_{m-1}}{-1+\varepsilon_{m-1}}{-1-\varepsilon_{m-1}}\,.
\eea
\end{subequations}
We have introduced the small parameters $\varepsilon_{m-1}\equiv n_{m-2}/n_{m-1}$ and $\varepsilon_1\equiv n_{2}/n_1$, not to be confused with the dielectric constants. Neglecting terms $\propto \varepsilon_{1,m-1}$, the boundary conditions~\eqref{limfields} hold for
\be
\label{outa}
R_2 = \frac{i n_1}{2 n_2} L_0\equiv Z_0 L_0
\quad
\mathrm{and}
\quad R^D_{m-2} = -\frac{i n_{m-1}}{2n_{m-2}}R_m\equiv Z_{m}R_m\,.
\ee
Here one can see that the electric fields in the cavity are larger by one power of the large factors $n_{1,m-1}$ with respect to the fields outside. Indeed, the factors $Z_{0,m}^2\propto n_{1,m-1}^2$ play a role similar to the quality factor of the cavity (in absence of dissipation).

We can now substitute equation~\eqref{ff} into \eqref{eq:powersik} and apply \eqref{outa} to find the emerging power in the form
\begin{subequations}\label{eq:sikans}
\bea
\frac{P_m}{A}
&=&Z_0^2\frac{R_m^2L_0^2}{(R_m^2+L_0^2+\varGamma)^2}\left|\vv{1}{-1}^T\[\sum_{r=2}^{m-2}\frac{b_r}{in_r} (\PP_r-\mathbb{1})\TT^r_2 \] \vv{1}{-1}\right|^2\frac{E_0^2}{2} \label{eq:sikpm}\,,\\[2ex]
\frac{P_0}{A}
&=&Z_0^2\frac{L_0^4}{(R_m^2+L_0^2+\varGamma)^2}\left|\vv{1}{-1}^T\[\sum_{r=2}^{m-2}\frac{b_r}{in_r} (\PP_r-\mathbb{1})\TT^r_2 \] \vv{1}{-1}\right|^2\frac{E_0^2}{2}\,.
\eea
\end{subequations}
We have successfully expressed the overlap integral in terms of a sum over transfer matrices, which will allow us to make a direct comparison to our transfer matrix formalism.

\subsection{Generalised overlap integral formalism}
\label{Generalised_Overlap_Integral_Formalism}

Now that we have an idea of what an overlap integral formalism looks like in a 1D context, we will show that our transfer matrix formalism can be rewritten in a similar form. This can be done by performing some matrix transformations and then reinterpreting each interface as contributing two boundary terms to an integral in each region. Note that this overlap integral formalism is a generalisation of the one derived in section~\ref{Leakycavity}---it will be well defined also for non-resonant configurations.

\subsubsection{Derivation from transfer matrices}

In our transfer matrix formalism, the boost amplitudes are given by the matrix equation~\eqref{transfereq2} in the absence of incoming waves, $R_0=L_m=0$,
\be
\vv{\mathcal{B}_R}{0}=\TT\vv{0}{\mathcal{B}_L}+\MM \vv{1}{1} ,
\ee
which solves as
\be
\vv{\mathcal{B}_R}{\mathcal{B}_L}=\frac{1}{\TT[2,2]}\(\begin{array}{cc}
\TT[2,2]& -\TT[1,2] \\
0  & -1 \\
\end{array}\) \MM \vv{1}{1}.
\ee

Let us focus first on $\mathcal{B}_R$. Note that the top two entries of the matrix multiplying $\MM$ are the same as those of $\text{Det}[\TT]\TT^{-1}$. Thus,
\be
\mathcal{B}_R =\frac{\text{Det}[\TT]}{\TT[2,2]}\vv{1}{0}^T\TT^{-1}\MM\vv{1}{1}
\quad\hbox{with}\quad
\TT^{-1}\MM = \sum^{m}_{s=1} \Ss_{s-1} (\TT^s_0)^{-1},
\ee
which is obtained from equation~\eqref{eq:T-M-matrices} using $\TT= \TT^m_s \TT^s_0$ and its inverse $\TT^{-1}= (\TT^s_0)^{-1}(\TT^m_s)^{-1}$.
Using the Pauli matrix $\sigma_2$ that satisfies $\sigma_2^2=\mathbb{1}$
and $\sigma_2 {\sf A}^T \sigma_2 = {\rm Det}[{\sf A}]{\sf A}^{-1}$ for any $2{\times}2$
matrix~${\sf  A}$,
\be
\mathcal{B}_R =\frac{\text{Det}[\TT]}{\TT[2,2]}\vv{i}{-i}^T\sigma_2 \[\TT^{-1}\MM\]^T\sigma_2 \vv{0}{i},
\ee
where
\be
\sigma_2\[\TT^{-1}\MM\]^T\sigma_2 = \sum^{m}_{s=1} \Ss_{s-1} \sigma_2\[(\TT^s_0)^{-1}\]^T\sigma_2
=  \sum^{m}_{s=1} \Ss_{s-1} \frac{n_s}{n_0}\TT^s_0
\ee
with $\text{Det}[(\TT_0^s)^{-1}]=1/\text{Det}[\TT_0^s]=n_s/n_0$ according to equation~\eqref{eq:DetT}.
This gives an expression for the boost amplitude that includes a sum over the interfaces,
\be
\mathcal{B}_R= \frac{-\text{Det}[\TT]}{\TT[2,2]}\vv{1}{-1}^T\sum^{m}_{s=1} \Ss_{s-1} \frac{n_s}{n_0}\TT^s_0 \vv{0}{1}\,,
\label{eq:BRinterfaces}
\ee
We will now rearrange equation~\eqref{eq:BRinterfaces} as a sum over the different regions~$r$:
\begin{subequations}
\bea
\mathcal{B}_R
&=&
\frac{-(n_0/n_m)}{\TT[2,2]}\vv{1}{-1}^T\sum^{m}_{s=1}  \frac{1}{2}\(\frac{b_s}{n_s^2}-\frac{b_{s-1}}{n_{s-1}^2}\) \frac{n_s}{n_0}\TT^s_0 \vv{0}{1}	
\\[2ex]
&=&\frac{-1}{n_m\TT[2,2]}\frac{1}{2}\vv{1}{-1}^T \Bigg[ \(\frac{b_m}{n_m^2}-\frac{b_{m-1}}{n_{m-1}^2}\) n_m\TT^m_0
+\(\frac{b_{m-1}}{n_{m-1}^2}-\frac{b_{m-2}}{n_{m-2}^2}\) n_{m-1}\TT^{m-1}_0
\nonumber\\[2ex]
&&
\quad\quad\quad\quad \quad\quad\quad\quad \quad\quad
+ \,\, ... \,\, +\(\frac{b_{1}}{n_{1}^2}-\frac{b_{0}}{n_{0}^2}\) n_{1}\TT^{1}_0 \Bigg]\vv{0}{1}
\\[2ex]
&=&
\frac{-1}{n_m\TT[2,2]}\frac{1}{2}\vv{1}{-1}^T\sum^{m-1}_{r=1} \frac{b_r}{n_r} \(\mathbb{1}- \frac{n_{r+1}}{n_r}\GG_r\PP_r\)\TT^r_0 \vv{0}{1}
\nonumber \\[2ex]
&&
- \frac{1}{\TT[2,2]}\frac{1}{2}\vv{1}{-1}^T \frac{b_m}{n_m^2}\TT_0^m \vv{0}{1}
+\frac{1}{\TT[2,2]}\frac{1}{2}\vv{1}{-1}^T\frac{b_0n_1}{n_m n_0^2}\TT_0^1 \vv{0}{1}\,,
\label{eq:BRregions}
\eea
\end{subequations}
where we have used equation~\eqref{Sdef} with~\eqref{eq:Ardef} and~\eqref{eq:brdef},
$\TT^{s+1}_0=\GG_s\PP_s \TT^s_0$, and~\eqref{eq:DetT}. The two unpaired terms in equation~\eqref{eq:BRregions} that are not included in the sum over regions $r$
are of order one and can thus be neglected for
configurations with a large boost factor $\beta$:
\be
\mathcal{B}_R
\simeq
\frac{1}{n_m\TT[2,2]}\frac{1}{2}\vv{1}{-1}^T\sum^{m-1}_{r=1} \frac{b_r}{n_r}
 \( \PP_r-\mathbb{1}\)\TT^r_0 \vv{0}{1}\,,
\label{eq:transferint1}
\ee
where we have used
\be
\vv{1}{-1}^T\GG_r = \frac{n_r}{n_{r+1}}\vv{1}{-1}^T\,.
\ee
By comparing with equation~\eqref{eq:edotb}
we see that~\eqref{eq:transferint1} can be written as
an overlap integral.

Note that equation~\eqref{eq:transferint1} constructs the electric field required for the overlap integral formalism---the condition of a purely left moving wave on the lhs of the device defines
a unique $E$-field configuration. As from \eqref{eq:transmisivityb} the transmissivity is  $1/\TT[2,2]$, we see that this electric field can be found by sending in a left moving wave on the rhs (from medium $m$ to $m-1$) with an amplitude of $1/n_m$. Note that the unpaired terms cannot in general be included due to the ambiguity in defining the terminals of the integral. For certain situations the ambiguity can be lifted. When the magnetic field is of finite extent then one can naturally terminate the integral with the magnetic field. Alternatively, if $|\mathcal R|=1$ (i.e., for standing waves) it is possible to show that the unpaired term in the $m$th interface can be included by extending the integral to the first antinode. So when one of the external media is a mirror, equation~\eqref{eq:transferint1} becomes exact, as the electric field vanishes inside the mirror and so the other unpaired term vanishes.

To get the boost amplitude on the other side we can simply substitute $0$ and $m$ and rearrange the terms
\be
\mathcal{B}_L \simeq  \frac{1}{2n_0}\vv{1}{-1}^T\sum^{m-1}_{r=1} \frac{b_r}{n_r} \( \PP_r-\mathbb{1}\)\TT^r_0 \vv{1}{-\frac{\TT[1,2]}{\TT[2,2]}}\label{eq:transferint2}.
\ee
While one might be concerned with phase issues (only for a standing wave can one always choose a phase to make our generally complex electric field to be purely real), note that this equation applies to the boost amplitude, and so one must remember that we are considering the electric field to be complex. For the boost factor we can more compactly write
\be
\beta_{L,R}\simeq\frac{\omega}{2n_{0,m}|E_{\rm in}|}\left|\int dx\, E_{L,R}b\right|\,, \label{eq:toverlap}
\ee
where $E_{L,R}$ is the complex $E$-field amplitude given by shining a wave of magnitude $|E_{\rm in}|$ in from either the 0 or $m$th layer (left and right hand sides, respectively) in the absence of axions. This formalism is fully general for large $\beta$---it can be applied to a fully open system, even when there is no resonant behaviour.

\begin{figure}[t]
\centering
\includegraphics[width=10cm]{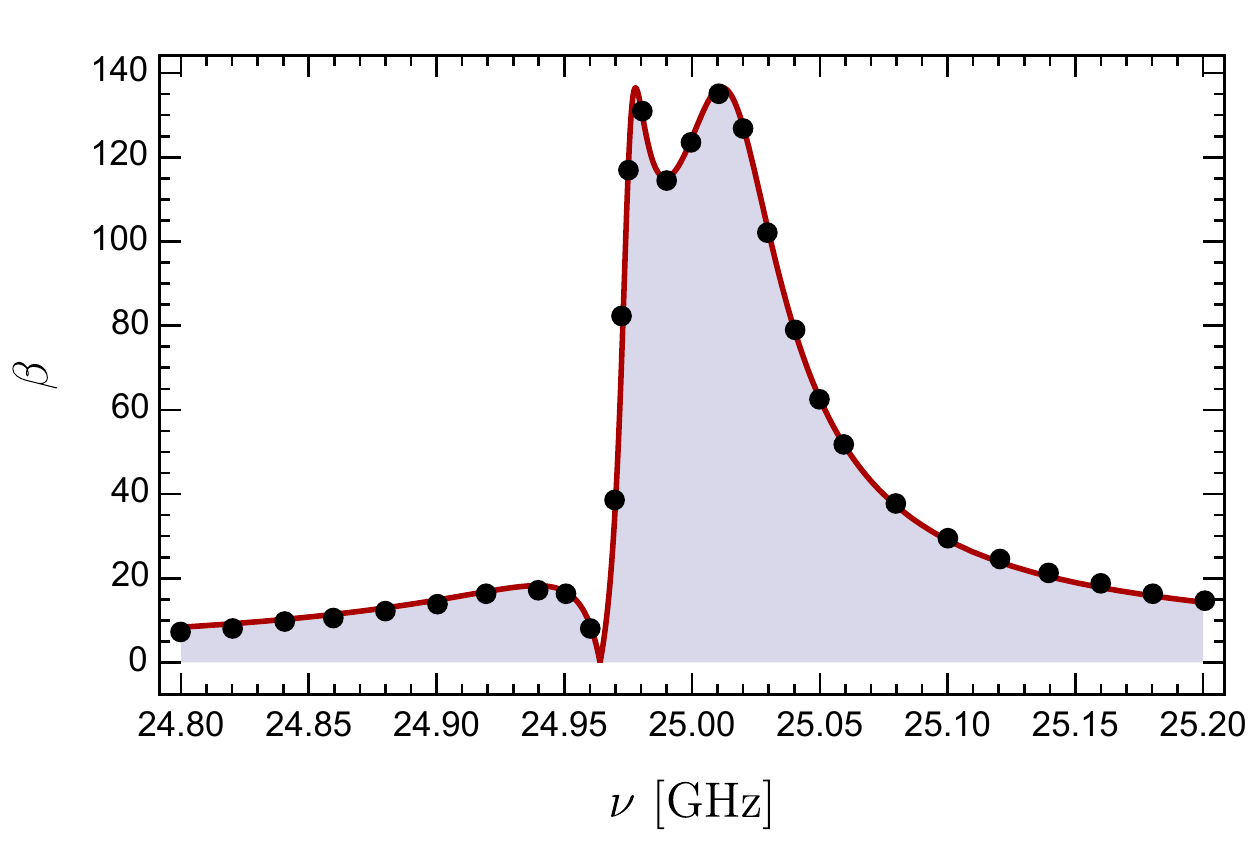}
\caption{Comparison of the boost factor $\beta$ calculated by the generalised overlap integral formalism (black dots) and the transfer matrix formalism (red curve). We plot $\beta(\nu)$ for the B50 configuration of section~\ref{50mhzop}, i.e., a dielectric haloscope consisting of 20~disks, 1~mm thick, refractive index $n=5$, mirror on one side, optimised for bandwidth of 50~MHz centred on 25~GHz (as shown in figure~\ref{fig:50mhz}).}
\label{fig:overlap}
\end{figure}

We have successfully transformed the transfer matrix formalism into an overlap integral formalism. To see this correspondence in practice, we compare the two formalisms in figure~\ref{fig:overlap} for the B50 configuration of section~\ref{50mhzop}. The two methods give very good agreement for the boost factor even off resonance. However, the overlap integral method is more computationally involved as one must first find $E_{L,R}$ and then integrate them.

\subsubsection{Direct comparison with Sikivie's formalism}

While we have developed a generalised overlap integral formalism, it is not immediately obvious that it is the same as that of Sikivie in the resonance limit. In particular, we must check that one integrates the same electric field in both cases and that the normalisations of the integrals agree. In our generalised overlap integral formalism, the normalisation comes from the transmissivity, whereas in the original formalism this comes from, in part, the quality factor and coupling factor.

Instead of considering a situation where the cavity has already had some power input from an unspecified (axion) source, in this transfer matrix inspired picture we are inserting the power externally. However, due to the presence of the cavity walls these are equivalent: a resonant standing wave looks the same regardless of how it is excited. To recover exactly the same formula, assuming a cavity setup as in section~\ref{Leakycavity} we can use $\TT^r_0=\TT^r_2 \TT_0^2$, so
\be
\TT_0^2 \vv{a}{b}=(a+b)Z_0\vv{1}{-1}
\ee
and we can write
\begin{subequations}\label{eq:betacav}
\bea
\mathcal{B}_R
&\simeq&
Z_0\,\frac{1}{\TT[2,2]}\frac{1}{2}\vv{1}{-1}^T\sum^{m-2}_{r=2} \frac{1}{n_r} \(\PP_r-\mathbb{1}\)\TT^r_2\vv{1}{-1}\,,
\label{eq:betacava}\\[2ex]
\mathcal{B}_L
&\simeq&
Z_0\(1-\frac{\TT[1,2]}{\TT[2,2]}\)\frac{1}{2}\vv{1}{-1}^T\sum^{m-2}_{r=2} \frac{1}{n_r} \(\PP_r-\mathbb{1}\)\TT^r_2\vv{1}{-1}\,,
\label{eq:betacavb}
\eea
\end{subequations}
which has exactly the same structure we found in the cavity case, cf.~equation~\eqref{eq:sikans}: the same electric fields are integrated in both cases. Again we have neglected the fields inside the cavity walls. To complete the comparison we must confirm that the normalisations of the integrals agree, i.e., that
\begin{subequations}\label{eq:norm}
\bea
\frac{1}{\TT[2,2]} &=& \frac{2 R_mL_0}{R_m^2+L_0^2+\varGamma},\label{eq:norma}\\[2ex]
1-\frac{\TT[1,2]}{\TT[2,2]} &=& \frac{2 L_0^2}{R_m^2+L_0^2+\varGamma},\label{eq:normb}
\eea
\end{subequations}
where we are implicitly assuming the absolute values of all the quantities.
To show the validity of equation~\eqref{eq:norm}, we must restrict ourselves to the case of a cavity on resonance, without which \eqref{eq:sikivieoverlap} does not apply. From equation~\eqref{eq:betacav} we know~$R_m$ and~$L_0$. They are the same up to factors of~$1/\TT[2,2]$ and~$1-\TT[1,2]/\TT[2,2]$. Further, as we assume a resonance forming a standing wave, $\TT[1,2]/\TT[2,2]$ is real, i.e., a standing wave does not experience a phase shift. Because of this we have
\bea
\frac{2 R_mL_0}{R_m^2+L_0^2+\varGamma}&=&2\frac{\frac{1}{\TT[2,2]}\(1-\frac{\TT[1,2]}{\TT[2,2]}\)}{\(\frac{1}{\TT[2,2]}\)^2+\(1-\frac{\TT[1,2]}{\TT[2,2]}\)^2+\varGamma}\nonumber \\
&=&\frac{1}{\TT[2,2]}\frac{2\(1-\frac{\TT[1,2]}{\TT[2,2]}\)}{\(\frac{1}{\TT[2,2]}\)^2+\(\frac{\TT[1,2]}{\TT[2,2]}\)^2+1-2\frac{\TT[1,2]}{\TT[2,2]}+\varGamma}
=\frac{1}{\TT[2,2]},
\eea
where we have used that $1/\TT[2,2]^2+\(\TT[1,2]/\TT[2,2]\)^2=1-\varGamma$ (conservation of energy). As $L_0/R_m=\TT[2,2]-\TT[1,2]$ we can similarly prove equation~\eqref{eq:normb}.

So the generalised overlap integral formalism developed here agrees with that of Sikivie for a resonant cavity, as one would expect. Indeed, we can make the connection more apparent by rearranging expression~\eqref{eq:toverlap} to get
\begin{subequations}
\begin{eqnarray}
P_{L,R}&\simeq &  {\cal G}_{{\rm d},L,R}V\frac{Q_{{\rm d},L,R}}{m_a}\rho_a g_{a\gamma}^2B_{\rm e}^2, 
\\
{\cal G}_{{\rm d},L,R}&=&\frac{\left|\int dx\, E_{L,R} B_{{\rm e}}\right |^2}{L B^2_{{\rm e}} \int dx\, |E_{\rm in}|^2}, 
\\ 
Q_{{\rm d},L,R}&=&\frac{1}{4}\frac{\int dx\, |E_{L,R}|^2}{E_0^2/m_a},
\end{eqnarray}
\end{subequations}
where $L$ is the length of the haloscope and $V=AL$. As it factors out of $P$, the actual value of $L$ is irrelevant.
Note that $Q_{{\rm d},L,R}$ are not true quality factors: rather than giving the generic response of the system to any input,
it depends on the input wave (i.e., $Q_{{\rm d},L}\neq Q_{{\rm d},R}$ in general). The cavity coupling $\kappa$ (where applicable) is built into our choice of electric fields and the loss tangents of the media.

\begin{figure}[t]
\centering
\includegraphics[width=14cm]{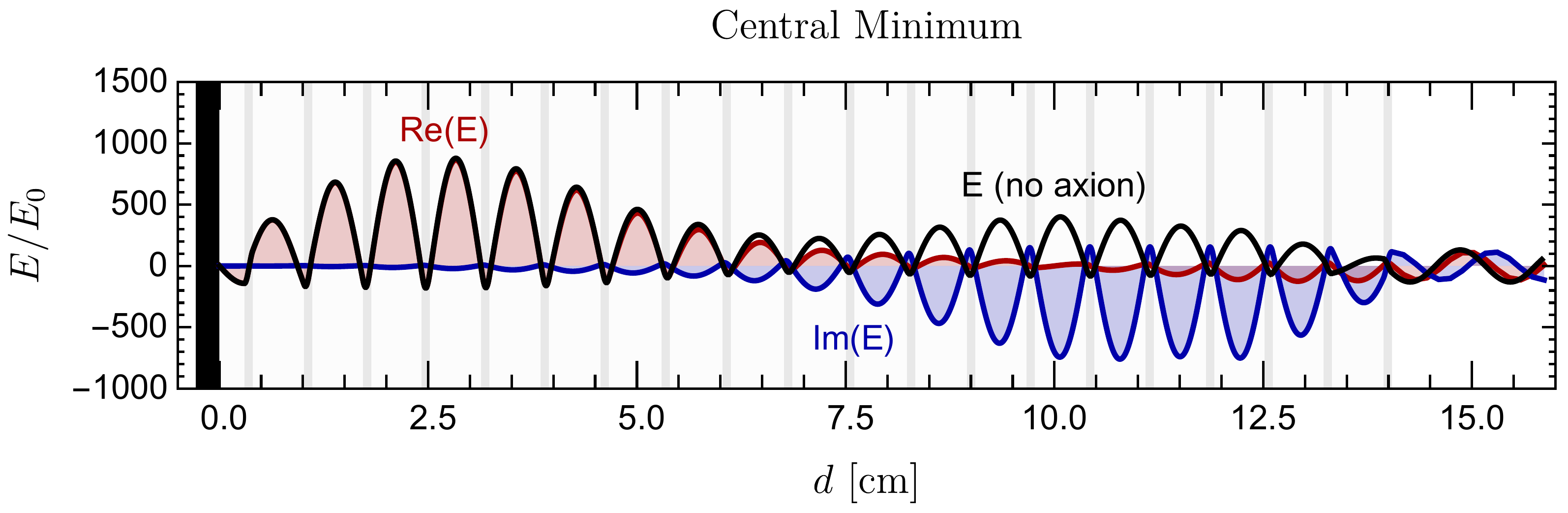}
\caption{Real and imaginary parts of the electric field produced by the axion (red and blue) and the electric field in the absence of axions (black) as a function of distance from the mirror $d$ inside a dielectric haloscope for our B50 configuration  (20~disks, 1~mm thick, refractive index~$n=5$, mirror on one side, bandwidth of 50~MHz centred on 25~GHz). The electric field in the absence of axions has been scaled so that it matches the one produced by axions near the mirror. While there is a correlation between the two, they are different in general. The agreement generally seems to be better on resonance. We consider the frequency associated with the central minimum in figure~\ref{fig:50mhz}.}
\label{fig:E-fields3}
\end{figure}

While the overlap integral formalism can be extended to non-resonant cases, the usual physical interpretation cannot. For a resonant cavity, the axion excites the same resonant mode that is integrated over. For a general dielectric haloscope, the integrated electric field is different from the one generated by the axion field. For example, when one side is a mirror, our integrated electric field is a standing wave. However, when the axion field is included, traveling waves are also present, as can seen in, for example, figure~\ref{fig:E-fields3}. Thus, the generalised overlap integral formalism is more of a mathematical tool than a physical picture.

In this language, one sees that the main concept behind a dielectric haloscope is to use dielectrics to increase the volume and geometry factor of a haloscope while maintaining the flexibility to use resonant and non-resonant setups. This flexibility distinguishes dielectric haloscopes from the more
limited cases studied in Ref.~\cite{Rybka:2014cya,Morris:1984nu}, which propose enhancing the volume and geometry factor of a strictly resonant cavity by either modifying the magnetic field or by using dielectrics.

\section{Scan optimisation}
\label{scanoptimisation}

Any experiment to search for dark-matter axions needs to scan over a large frequency range. For a dielectric haloscope, this means to re-adjust the disk spacings for every search channel $i$ centered on some frequency $\nu_i$, an operation that requires a re-adjustment time~$t_{\rm R}$. For a given set of disks, we seek the optimal channel width $\Delta\nu_i$ to minimise the overall search time
\be
\label{eq:total-time}
t = \sum_i \(\Delta t_i + t_{\rm R}\)\,,
\ee
where we take $t_{\rm R}$ to be independent of channel~$i$.
In section~\ref{discoverypotential} we assumed that $\Delta\nu_i$ should be chosen such that $\Delta t_i=t_{\rm R}$. We here prove this intuitive assumption to be exact.

A given set of dielectric disks allows us to cover a certain frequency range $\nu_1$ to $\nu_2$ by re-adjusting the spacings alone. We need to find the optimal frequency channels $\nu_i$ (width
$\Delta\nu_i$ and boost factor $\beta_i$) to cover the search range $\sum_i\Delta\nu_i=\nu_2-\nu_1$. In order to achieve the desired signal-to-noise in a given channel, the relation between measurement time and boost factor is given by equation~\eqref{eq:scan} to be $\Delta t_i = a_i/\beta_i^4$ with
\be
a_i\sim 400^4\times 1.3\,{\rm days}\(\frac{{\rm S}/{\rm N}}{5}\)^2
\(\frac{1\,\rm m^2}{A}\)^2\(\frac{m_a}{100~\mu{\rm eV}}\)
\(\frac{T_{\rm sys}}{8\,\rm K}\)^2
\(\frac{10\,\rm T}{B_{\rm e}}\)^4\(\frac{0.8}{\eta}\)^{2} C_{a\gamma}^{-4} f_{\rm DM}^{-2} \,,
\ee
where $m_a=2\pi\,\nu_i$ in terms of the search frequency $\nu_i$.
The scanning speed is then
\be
\frac{d\nu}{d t} = \frac{\Delta\nu_i}{\Delta t_i}=\frac{K_i/\beta_i^2}{ a_i/\beta_i^4}\,.
\ee
We have here used the Area Law in the form $\Delta \nu_i \beta_i^2 = K_i$, where $K_i$ is approximately constant for a given set of disks, but in detail depends weakly on frequency $\nu_i$ and boost factor $\beta_i$.

The width and required boost factor varies slowly with $i$ because of the large number of channels, so instead of summing over all channels $i$, we integrate over frequency $\nu$ to obtain the overall measurement time. So for the first term in equation~\eqref{eq:total-time}
we can write
\be
\sum_i \Delta t_i \to \int_{\nu_1}^{\nu_2} \frac{d\nu}{d\nu/dt} = \int_{\nu_1}^{\nu_2}
\frac{d\nu}{K \beta^2 /a}
\ee
and for the second term
\be
\sum_i  t_{\rm R} \to \int_{\nu_1}^{\nu_2} \frac{d\nu}{ \Delta \nu} t_{\rm R} = \int_{\nu_1}^{\nu_2} \frac{d\nu}{K/\beta^2}\,t_{\rm R}\,,
\ee
where $K$ depends weakly on $\nu$ and $\beta$, whereas $a$ can depend more strongly on $\nu$ (at least a linear dependence coming from $m_a$). Thus the quantity we need to minimise with an appropriate choice of $\beta_\nu$ is
\be
\label{eq:t-to-minimize}
t = \int_{\nu_1}^{\nu_2}\frac{d\nu}{K_{\nu,\beta}} \(\frac{a_\nu}{\beta^2_\nu}
+t_{\rm R}\beta^2_\nu\)\,,
\ee
where we show the dependencies on parameters as subscripts. To minimise the integral we
differentiate the integrand with respect to $\beta$
\be
\frac{d }{d\beta}\[ \frac{1}{K_{\nu,\beta}} \(\frac{a_\nu}{\beta^2_\nu}
 +t_{\rm R}\beta^2_\nu\)\] =
 \frac{2}{K}\(- \frac{a}{\beta^3} + t_{\rm R}\beta \)+\frac{1}{K^2}\frac{dK}{d\beta}\(\frac{a}{\beta^2} +t_{\rm R}\beta^2\)= 0 \,.\label{eq:optimum1}
\ee
As the Area Law is generally a good approximation, $dK/d\beta$ is small
which implies an optimal boost factor
\be
\beta_{\rm o}(\nu) \simeq \(\frac{a_\nu}{t_{\rm R}}\)^{1/4}\label{eq:betaop}
\ee
or explicitly
\bea
\label{optimised}
\beta_{\rm o}(\nu)&=&
430 \(\frac{ \rm day}{t_{\rm R}} \)^{1/4}  \biggl(\frac{\nu}{25~{\rm GHz}}\biggr)^{1/4}
\(\frac{{\rm S}/{\rm N}}{5}\)^{1/2}
\nonumber\\[1ex]
&&\kern3em{}\times\(\frac{1\,\rm m^2}{A}\)^{1/2} \(\frac{T_{\rm sys}}{8\,\rm K}\)^{1/2}
\(\frac{10\,\rm T}{B_{\rm e}}\) \(\frac{0.8}{\eta}\)^{1/2} C_{a\gamma}^{-1} f_{\rm DM}^{-1/2} \, .
\eea
The $\nu$ dependence is relatively weak, but generally a higher frequency requires a larger boost factor. $T_{\rm sys}$ can also depend on $\nu$, typically increasing for higher $\nu$.

Inserting the optimal boost factor \eqref{eq:optimum1} in the expression for the search time in
equation~\eqref{eq:t-to-minimize} reveals that the two terms
in brackets are equal, confirming $\Delta t=t_{\rm R}$ for each frequency channel. The minimal scanning time becomes
\be
t_{\rm o}=2 t_{\rm R} \int_{\nu_1}^{\nu_2}\frac{d\nu}{K_{\nu}}\,\beta_{\rm o}^2(\nu)
=2\sqrt{t_{\rm R}} \int_{\nu_1}^{\nu_2}\frac{d\nu}{K_{\nu}}\,\sqrt{a_\nu}\,,
\ee
depending only on $\sqrt{t_{\rm R}}$. If one improves the re-adjustment time, the optimal strategy requires also a shorter measurement time $\Delta t$. To achieve the same sensitivity, one needs a larger boost factor and, according to the Area Law, a narrower channel width and thus more measurement channels.
By the same token, $t_{\rm o}$ decreases only with $C_{a\gamma}^{-2}$ rather than the fourth power. 

The cases used in section~\ref{discoverypotential} are both examples of the boost factor obeying a power law $\beta\propto \nu^p$ for some $p$. The idealised HEMT amplifier with constant $T_{\rm sys}$ 
had $p=1/4$, whereas the quantum-limited amplifiers ($T_{\rm sys}=\omega$) had $p=3/4$. In this case the integrals are easy to evaluate explicitly if we take $K$ to be constant. We write 
$\beta_{\rm o}(\nu)=\beta_2(\nu/\nu_2)^{p}$, where $\beta_2\equiv\beta_{\rm o}(\nu_2)$ and find
\be
t_{\rm o}=\frac{2t_{\rm R}}{K}\int_{\nu_1}^{\nu_2}d\nu\beta_{\rm o}^2(\nu)=
\frac{2t_{\rm R}}{2p+1}\,\frac{\beta_2^2}{K\nu_2^{2p}}\(\nu_2^{2p+1}-\nu_1^{2p+1}\)
\sim \frac{2t_{\rm R}}{2p+1}\,\frac{\beta_2^2 }{K}\nu_2
=\frac{2t_{\rm R}}{2p+1}\,\frac{\nu_2}{\Delta\nu|_{\nu_2}}\,,
\ee
where $\Delta\nu|_{\nu_2}$ is the bandwidth of $\beta_2$ at $\nu_2$. Notice that this bandwidth for the optimal search contains a factor $1/\sqrt{t_{\rm R}}$, i.e., $t_{\rm o}\propto\sqrt{t_{\rm R}}$ as discussed earlier. Interestingly, to first approximation the measurement time depends only on $\nu_2$ and scales in a manner that is invariant under the exact choice of $p$. As $K$ is a linear function of the number of disks, we see that the frequency range measured in a given amount of time is also linear.

\clearpage


\end{document}